\documentclass[iop,twocolappendix]{emulateapj}
\pdfoutput=1
\usepackage{comment}
\usepackage{enumerate}
\usepackage{amsmath}

\usepackage{rotating}
\usepackage{array}
\usepackage{multirow}
\usepackage{threeparttable}
\usepackage{longtable}
\usepackage{float}
\usepackage{color}

\newcommand{\myemail}{mgrootes@cosmos.esa.int;meiert.grootes@mpi-hd.mpg.de;}

\shorttitle{Star Formation of  Spiral Galaxies in the Group Environment}
\shortauthors{Grootes et al.}

\begin{document}


\title{Galaxy And Mass Assembly (GAMA): Gas Fuelling of Spiral Galaxies in the
Local Universe I. - The Effect of the Group Environment
on Star Formation in Spiral Galaxies}


\author{M.~W.~Grootes\altaffilmark{1,2,* },
 R.~J.~Tuffs\altaffilmark{1}, 
 C.~C.~Popescu\altaffilmark{3,4,1}, 
 P.~Norberg\altaffilmark{5}, 
 A.~S.~G.~Robotham\altaffilmark{6,7}, 
 J.~Liske\altaffilmark{8}, 
  E.~Andrae\altaffilmark{1}, 
	I.~K.~Baldry\altaffilmark{9}, 
  M.~Gunawardhana\altaffilmark{5}, 
   L.~S.~Kelvin\altaffilmark{9}, 
   	B.~F.~Madore\altaffilmark{10}, 
   M.~Seibert\altaffilmark{10}, 
    E.~N.~Taylor\altaffilmark{11}, 
    M.~Alpaslan\altaffilmark{12}, 
    M.~J.~I.~Brown\altaffilmark{13}, 
    M.~E.~Cluver\altaffilmark{14}, 
	 S.~P.~Driver\altaffilmark{7,15}, 
	  J.~Bland-Hawthorn\altaffilmark{16}, 
	   B.~W.~Holwerda\altaffilmark{17}, 
	 A.~M.~Hopkins\altaffilmark{18}, 
	 A.~R.~Lopez-Sanchez\altaffilmark{18}, 
	 J.~Loveday \altaffilmark{19}, 
	 M.~Rushton\altaffilmark{4,1,3} 	  
	  }        
        
           
                 

\altaffiltext{1}{Max-Planck-Institut f\"ur Kernphysik, Saupfercheckweg 1, 69117 Heidelberg, Germany}
\altaffiltext{2}{ESA/ESTEC SCI-S, Keplerlaan 1, 2201 AZ, Noordwijk, The Netherlands}
\altaffiltext{*}{\myemail}
\altaffiltext{3}{Jeremiah Horrocks Institute, University of Central Lancashire, Preston PR1 2HE, UK}
\altaffiltext{4}{The Astronomical Institute of the Romanian Academy, Str. Cutitul de Argint 5, Bucharest, Romania}
\altaffiltext{5}{Institute for Computational Cosmology, Department of Physics, Durham University, Durham DH1 3LE, UK}
\altaffiltext{6}{University of Western Australia, Stirling Highway Crawley, WA 6009, Australia}
\altaffiltext{7}{International Centre for Radio Astronomy Research (ICRAR), University of Western Australia, Stirling Highway Crawley, WA 6009, Australia}
\altaffiltext{8}{Hamburger Sternwarte, Universit\"at Hamburg, Gojenbergsweg 112, 21029 Hamburg,  Germany}
\altaffiltext{9}{Astrophysics Research Institute, Liverpool John Moores University, Twelve Quays House, Egerton Wharf, Birkenhead, CH41 1LD, UK}
\altaffiltext{10}{Observatories of the Carnegie Institution for Science, 813 Santa Barbara Street, Pasadena, CA 91101, USA}
\altaffiltext{11}{School of Physics, the University of Melbourne, Parkville, VIC 3010, Australia}
\altaffiltext{12}{NASA Ames Research Center, N232, Moffett Field, Mountain View, CA 94035, USA}
\altaffiltext{13}{School of Physics and Astronomy, Monash University, Clayton, Victoria 3800, Australia}
\altaffiltext{14}{University of the Western Cape, Robert Sobukwe Road, Bellville, 7535, South Africa}
\altaffiltext{15}{Scottish Universities' Physics Alliance (SUPA), School of Physics and Astronomy, University of St Andrews, North Haugh, St Andrews, KY16 9SS, UK}
\altaffiltext{16}{Sydney Institute for Astronomy, School of Physics, University of Sydney NSW 206, Australia}
\altaffiltext{17}{Leiden Observatory, University of Leiden, Niels Bohrweg 2, 2333 CA Leiden, The Netherlands}
\altaffiltext{18}{Australian Astronomical Observatory, PO Box 915, North Ryde, NSW 1670, Australia}
\altaffiltext{19}{Astronomy Centre, University of Sussex, Falmer, Brighton BN1 9QH; UK}

\begin{abstract}
We quantify the effect of the galaxy group environment (for group masses of $10^{12.5}$ to $10^{14.0} M_{\odot}$) on the current star formation rate (SFR) of a pure, morphologically-selected, sample of disk-dominated (i.e. late-type spiral) galaxies with redshift $\le0.13$. The sample embraces a full representation of quiescent and star-forming disks with stellar mass $M_*\ge 10^{9.5} M_{\odot}$. We focus on the effects on SFR of interactions between grouped galaxies and the putative intra-halo medium (IHM) of their host group dark matter halos, isolating these effects from those induced through galaxy--galaxy interactions, and utilising a radiation transfer analysis to remove the inclination-dependence of derived SFRs. The dependence of SFR on $M_*$ is controlled for by measuring offsets $\Delta\mathrm{log}(\psi_*)$ of grouped galaxies about a single power-law relation in specific SFR, $\psi_* \propto M_*^{-0.45\pm0.01}$, exhibited by non-grouped ``field'' galaxies in the sample.

While a small minority of the group satellites are strongly quenched, the group centrals, and the large majority of satellites, exhibit levels of $\psi_*$ statistically indistinguishable from their field counterparts, for all $M_*$, albeit with a higher scatter of $0.44\,$dex about the field reference relation (vs. $0.27\,$dex for the field). Modelling the distributions in $\Delta\mathrm{log}(\psi_*)$, we find that: (i) after infall into groups, disk-dominated galaxies continue to be characterized by a similar rapid cycling of gas into and out of their ISM shown prior to infall, with inflows and outflows of respectively $\sim1.5 - 5$\,x\,SFR and $\sim1\,-\,4$\,x\,SFR, and (ii) that the independence of the continuity of these gas flow cycles on M$_*$ appears inconsistent with the required fuelling being sourced from gas in the circum-galactic medium on scales of $\sim100\,$kpc. Instead, our data favor on-going fuelling of satellites from the IHM of the host group halo on $\sim\,$Mpc scales, i.e. from gas not initially associated with the galaxies upon infall.

Consequently, the color--density relation of the galaxy population as a whole would appear to be primarily due to a change in the mix of disk- and spheroid-dominated morphologies in the denser group environment compared to the field, rather than to a reduced propensity of the IHM in higher mass structures to cool and accrete onto galaxies. We also suggest that the required substantial accretion of IHM gas by satellite disk-dominated galaxies will lead to a progressive reduction in the specific angular momentum of these systems, thereby representing an efficient secular mechanism to transform morphology from star-forming disk-dominated types to more passive spheroid-dominated types.
\end{abstract}





\section{Introduction}\label{INTRODUCTION}
The current paradigm of galaxy formation \citep[e.g.][]{REES1977,WHITE1978,FALL1980,WHITE1991,MO1998} holds that luminous galaxies form and initially evolve as disk galaxies at the center of isolated dark matter halos (DMH). Under this paradigm, as dark matter over-densities decouple from the large scale flow and collapse, the baryons of the ambient intergalactic medium bound to the potential well of the nascent DM halo will collapse and shock-heat at some radius comparable or interior to the virial radius of the halo, giving rise to a pressure supported (and thereby dynamically decoupled from the DM) intra-halo medium (IHM). Subsequently, radiative cooling of the baryons of the IHM will precipitate the further infall of some fraction of the gas towards the center of the DMH. The angular momentum of the cooling baryons, built up from the torques exerted by the tidal shear in the earlier large-scale flow of dark matter\citep[e.g.][]{FALL1980}, is thereby transported from the IHM into a rotationally supported disk of cold gas on some smaller scale related to the specific momentum of the halo \citep[e.g.][]{VANDENBOSCH2002,BETT2010,HAHN2010}. The surface density of gas in the disk increases as gas from the IHM continues to be accreted, until it becomes sufficient for the formation of dense, self-gravitating clouds which rapidly collapse to form stars, which then trace the disk as a visible galaxy. The rotationally supported gas in the disk therefore constitutes an interstellar medium (ISM), at least interior to some radius where the surface density of gas exceeds the threshold for star-formation.\newline

In the subsequent evolution, the galaxy will continue to accrete gas, thus fuelling ongoing star formation in its disk; including gas from the initial IHM, but mainly from secondary infall, i.e. baryons from the ambient IGM of the surrounding large-scale structure in-falling onto the DMH \citep{FILLMORE1984,BERTSCHINGER1985,PICHON2011}. Accordingly the net rate of accretion from the IHM into the ISM of the galaxy, and thereby the availability of fuel for star-formation, will be determined by (i) the \textit{maximum achievable} rate at which accretable, i.e., sufficiently cool, gas can be delivered to the galaxy, as determined by the properties of the large-scale environment, in particular the DMH, and (ii) feedback from processes in the galaxy predicted to regulate the accretion of IHM onto the galaxy.\newline 

A generic expectation of the accretion of IGM onto a DMH is the formation of an accretion shock \citep{BINNEY1977}. However, the formation and radial location of a stable shock depend on the cooling timescale in the post shock gas being longer than the free-fall timescale in order to establish and maintain a pressure supported atmosphere/IHM supporting the shock \citep{REES1977,WHITE1991,BIRNBOIM2003}. In halos where this is not the case at any radius exterior to the galaxy at the center, the IGM being accreted onto the halo will continue to the galaxy on the free-fall timescale, resulting in a highly efficient maximum achievable fuelling rate limited by the cosmological accretion rate onto the DMH, commonly referred to as 'cold mode' accretion \citep{KERES2005,DEKEL2006}. Conversely, in halos capable of supporting a shock, the infalling IGM will be shock heated and remain hot until it radiatively cools on the cooling timescale, resulting in a less efficient maximum fuelling rate determined by the cooling timescale; so-called 'hot mode' accretion \citep{KERES2005,DEKEL2006}. As the cooling timescale depends, inter alia, on the temperature of the post shock gas, and as such on the depth of the potential well of the DMH, i.e. its mass, this introduces an environmental dependence into the process of gas-fuelling in form of a transition between fuelling modes at a certain halo mass and an additional halo mass dependence within the 'hot mode' fuelling \citep{REES1977,WHITE1991,BENSON2001,BIRNBOIM2003,KERES2005,DEKEL2006,BENSON2011,VANDEVOORT2011}. For cosmological DMH detailed thermodynamic considerations of this process find a transition mass between these two modes, i.e. where the free-fall timescale equals the cooling timescale at the virial radius of $\sim 10^{11 - 12} M_{\odot}$ \citep{KERES2005,DEKEL2006}\footnote{It should be noted that the inflow of ambient IGM onto the halo will be anisotropic with preferential inflow along the filaments of the large-scale DM structure \citep{KERES2005,DEKEL2009,BROOKS2009,KERES2009a,PICHON2011}. Thus, the transition between cold and hot modes will not be sharp, as the filamentary flows of cold IGM may penetrate hot atmospheres. The degree to which this is the case is not yet clear, however, although penetration decreases with temperature and extent of the hot halo \citep{NELSON2013}.} As such the accretion in low mass halos, and thus predominantly in the early universe is dominated by cold-mode accretion while hot-mode accretion becomes increasingly relevant at lower redshifts and in the present universe \citep[e.g.][]{DEKEL2013}.\newline  

The net rate of accretion from the IHM into the ISM of the galaxy, and thereby the availability of fuel for star-formation, however, will not be determined by the maximum achievable fuelling rate alone. Rather the accretion of IHM into the ISM is predicted to be subject to regulation by galaxy specific feedback linked to energetic processes in the galaxy, e.g star formation and AGN activity. This feedback includes the mechanical removal of gas from the ISM as well as the heating of the IHM preventing it from cooling\footnote{Feedback from the galaxy may also impact the cooling timescale in the post shock gas by heating the IHM and/or enriching it with metals from the ISM.}. Feedback from star formation, i.e. from super novae, is predicted to remove gas from the ISM of the galaxy, most efficiently for low mass galaxies. While star formation and thus stellar feedback is an intrinsically stochastic process, the feedback will evolve into a near steady-state relation as galaxies grow large enough to support widespread star formation activity, albeit that the efficiency of stellar feedback in removing ISM from the galaxy will decrease with increasing mass of the galaxy/depth of the potential well \citep[e.g.][and references therein]{FAUCHER-GIGUERE2011,HOPKINS2013}, leading to a self-regulated level of accretion of gas from the IHM into the ISM. For the most massive galaxies, residing in massive DMH, AGN feedback from the black hole at the center of the galaxy, heating the IHM and preventing it from cooling and being accreted, is predicted to dominate the feedback from the galaxy \citep[e.g.][and references therein]{FABIAN2012}. Unlike star formation driven feedback, where a quasi steady-state relation is expected, AGN feedback, which is still a major subject of investigation, may always be stochastic in nature \citep[e.g.][]{POPE2007,PAVLOVSKI2009,HICKOX2014,WERNER2014}. Overall, the growth of the galaxy will thus continue until the supply of gas from the IHM is interrupted, e.g. by the galaxy being shifted away from the center of the potential well by a merging event with another halo of comparable or larger mass, or by the activity of an AGN efficiently heating the IHM.\newline

In summary, for galaxies at the center of their DMH - so-called centrals - basic physical considerations founded on the current paradigm of galaxy formation predict the rate at which gas from the IHM is accreted into the ISM of the galaxy, i.e. its gas-fuelling, to be determined by a balance between the possible rate of accretion as set by the DMH and galaxy specific feedback, thus displaying a dependence on both environmental and galaxy specific properties. This picture is consistent with work on the abundance matching of galaxies with halos from DM simulations, which suggests that the efficiency of the conversion of baryons to stars is greatest in DMH of $\sim 10^{12} M_{\odot}$ \citep{MOSTER2010,BEHROOZI2013}. Thus, regardless of the exact underlying cause, $10^{12}M_{\odot}$ represents a critical mass in understanding environment dependent galaxy evolution and gas-fuelling in particular.\newline

In addition to centrals, the hierarchical formation of large-scale structure expected for a $\Lambda$CDM universe gives rise to a population of so-called satellite galaxies, i.e. galaxies which are bound to their host DMH but are not at rest with respect to its center of mass, having been captured during the merging of two smaller DMH. In the context of the flow of gas from the IHM into the ISM it is essential to distinguish between these two types of galaxy group members. While for centrals, the physical processes -- driven by galaxy--IHM interactions -- which determine gas-fuelling can reasonably be expected to to be similar to those of isolated field central galaxies, this is not the case for satellites. For satellites, their motion relative to a putative virialized hot IHM introduces further galaxy--IHM interactions which may affect the rate of accretion of gas from the IHM into the ISM of the galaxy, as well as the gas content of the ISM and of any circum-galactic reservoirs of gas bound to the galaxy (CGM; circum-galactic medium). This includes ram-pressure stripping of the ISM of galaxies in the environment of galaxy clusters (and groups) \citep[e.g.][]{GUNN1972,ABADI1999, HESTER2006,BAHE2015}, as well as ram-pressure stripping of the CGM of a galaxy in the galaxy group and low mass cluster environment, a process often referred to as 'strangulation' \citep[e.g.][]{LARSON1980,KIMM2009}, as it is thought to slowly limit star formation in the galaxy by removing the gas reservoirs from which star formation is fuelled. Thus, satellite galaxies are expected to display suppressed star formation activity with respect to comparable field galaxies.\newline

On the scale of massive clusters of galaxies the predicted processes and trends have been observed, both directly by observations of ram-pressure stripped tails of gas emanating from galaxies \citep{FUMAGALLI2014}, as well as indirectly by the frequent occurrence of galaxies in these massive clusters truncated in H$\alpha$ and by a prevalence of galaxies with red colors and suppressed SFR \citep[e.g.][] {KOOPMANN1998,KOOPMANN2004,GAVAZZI2013}.\newline

An empirical quantification of the predictions on the scale of lower mass galaxy groups, however, has proven challenging, as the ram-pressure stripping of the ISM and CGM is expected to be less severe and potentially even limited to the CGM (\citealt[e.g.][]{KAWATA2008,MCCARTHY2008}, but see also \citealt{HESTER2006,BAHE2015} for a contrasting view), necessitating a statistical consideration of the galaxy group population to discern the impact of the group environment on these galaxies. In order to observationally identify galaxy groups many works \citep[e.g.][]{SMAIL1998,PIMBBLET2002,BALOGH2002,JELTEMA2006,PIMBBLET2006,URQUHART2010,ERFANIANFAR2014} have made use of X-ray selected samples for which properties of the DMH such as mass may be deduced from the X-ray emission. However, samples selected in this manner  may be biased towards more massive DMH and against the more ubiquitous (and therefore arguably more important) loose, low- and intermediate mass galaxy groups. To circumvent this potential bias, with the onset of wide-field spectroscopic galaxy surveys, many studies have made use of optically defined spectroscopic galaxy group catalogs \citep[e.g.][]{GOMEZ2003,BALOGH2004,
KAUFFMANN2004,COLLISTER2005,WEINMANN2006,ROBOTHAM2006,VANDENBOSCH2008,PASQUALI2009,WEINMANN2010,HESTER2010,VANDERWEL2010,
PENG2012,WHEELER2014,BALOGH2015}. These, however, suffer from the relatively low spectroscopic completeness in dense regions achieved by most spectroscopic surveys, such that the majority of galaxies in lower mass halos are central galaxies rather than satellites, and the halo masses assigned to each group depend on the shape of the assumed halo mass function. An alternative approach, pursued by a number of authors, has been to consider the (marked) correlation functions of galaxy samples drawn from spectroscopic surveys and to consider the clustering properties of red and blue galaxies \citep[e.g.][]{BLANTONBERLIND2007,SKIBBA2009,ZEHAVI2011}. While largely model independent, this approach makes linking observations of galaxy properties to the properties of their host group difficult.  Nevertheless, in general, all these works have found the fraction of red and quiescent galaxies to be larger in galaxy groups than in the field, in line with expectations, leading to the general assumption that galaxies are cut off from gas-fuelling upon becoming satellites, although the exact combination, importance and effectivity of the processes assumed to be responsible remain a subject of debate \citep[e.g.][]{BLANTONBERLIND2007,KIMM2009,VANDENBOSCH2008,PASQUALI2009,HESTER2010,WETZEL2013,WETZEL2014,MCGEE2014,PENG2015}.\newline        

Interpreting such observations in terms of the gas-fuelling and ISM content of galaxies and its relation to the group environment, however, is subject to a number of compounding problems, foremost amongst which is that of galaxy morphology. Empirically, the abundance of spheroidal galaxies is known to be higher in denser environments, corresponding to galaxy clusters (and to a lesser extent galaxy groups), than amongst largely isolated field galaxies \citep[e.g.][]{DRESSLER1980,GOTO2003,BAMFORD2009}, i.e it is higher in the higher mass DMH of these objects. However, it is not clear to what extent spheroidal galaxies are capable of retaining cold gas and sustaining significant star formation for any prolonged period \citep[e.g.][]{OOSTERLOO2010,SMITH2012}. In other words, while copious amounts of cold gas are observed in rotationally supported disk/spiral galaxies, the virial temperature of spheroidal pressure supported systems is well above that conducive to forming and maintaining giant molecular clouds. Thus, the prevalence of red, low sSFR galaxies may actually be more indicative of transformative processes affecting the morphology of satellite galaxies, than of effects linked to the supply of gas, making a control of the galaxy morphology paramount to any empirically driven investigation of gas-fuelling. \newline

 Finally, the ability to interpret observations of the properties of group galaxies in the context of the gas-fuelling of these objects requires the ability to control for degeneracies in the observables arising from galaxy--IHM and galaxy--galaxy interactions, as well as that the observables considered be sensitive to changes on timescales $\lesssim 1\,$Gyr, i.e. comparable to the typical dynamical timescale of galaxy groups and shorter than that to which properties such as red and blue fractions, stellar metallicity, and optical colors are sensitive. Accordingly, empirically probing gas-fuelling and its environmental dependencies requires a sample of known morphology probing the environment down to the scale of low mass groups of $\lesssim10^{12} M_{\odot}$, for which the measurement of the gas content (or its proxy tracer) is sensitive to changes on the scale of $10^8\,$yr and for which the effects of galaxy--IHM interactions can be isolated.
Thus, although a number of works have accounted for the morphology of their samples \citep[e.g.][]{HASHIMOTO1998,BAMFORD2009,HESTER2010}, the fundamental process of gas-fuelling in the group environment currently lacks a direct incisive empirical reference with which to compare and constrain theoretical predictions.\newline

In this paper (Paper I) and its companion papers in this series (Grootes et al., in prep) we focus on remedying this situation and providing a direct empirical reference with which to compare predictions of gas-fuelling as a function of environment with a focus on galaxy groups. This work makes use of a sample of galaxies of known uniform disk-dominated morphology (which we will refer to as spirals for simplicity), probing the full ranges of group environmental properties (e.g. DMH mass), galaxy specific properties (e.g. stellar mass $M_*$, SFR), and galaxy properties related to the group environment (e.g. central or satellite, distance from group center). In identifying and selecting galaxy groups we make use of the spectroscopic galaxy group catalogue (the G3C, \citealt{ROBOTHAM2011b}) of the Galaxy And Mass Assembly survey (GAMA, \citealt{DRIVER2011,LISKE2015}). This catalog samples the full mass range of galaxy groups (down to DMH masses of $\sim 10^{12}M_{\odot}$) with high completeness, enabling the determination of robust dynamical mass estimates, and represents the only resource of a statistically significant number of spectroscopic galaxy groups with kinematic determinations of the DMH mass down to low masses currently available.\newline

Given the scarcity of direct measurements of the ISM content of galaxies in wide-field spectroscopic surveys\footnote{For wide-field spectroscopic surveys, direct measurements of the ISM content of the majority of surveyed galaxies are generally not available given the  very long exposure time radio observations that would be required to obtain the necessary data. While this is currently also the GAMA survey, upcoming surveys using pathfinder facilities for the Square Kilometer Array (ASKAP DINGO, PI: M.Meyer), are striving to remedy this situation in the GAMA fields.}
, in our analysis we make use of the SFR of a galaxy derived from its NUV emission, tracing star formation activity on timescales of $\sim 10^8\,$yr (as shown in Fig.~\ref{fig_NUVTRACER}), as a proxy measurement of its ISM content.\newline 

In addition to ensuring that the relation between ISM and star formation is as consistent as possible over the range of environments for the galaxies considered, controlling for galaxy morphology also aids in isolating the effects of galaxy--IHM interactions from those of galaxy--galaxy interactions which may severely impact the SFR of galaxies \citep[e.g.][]{ROBOTHAM2014,DAVIES2015,ALATALO2015,BITSAKIS2016}. As major galaxy--galaxy interactions can strongly perturb disk galaxies and lead to a morphological transformation, focussing on disk dominated galaxies ensures that no major merger has taken place, effectively enabling us, in combination with the de-selection of close pairs of galaxies based on the G3C, to isolate the effects of galaxy--IHM interactions.\newline

The plan of this paper is then as follows. In Section~\ref{DATA} we briefly describe the GAMA survey as well as the relevant raw data products, followed by a description in Section~\ref{DATA_ANCILLARY} of the relevant derived physical properties. We then detail our sample selection and the resulting samples of disk-dominated/spiral galaxies in Section~\ref{SAMPLESELECTION}. Subsequently we present our core empirical results on the sSFR--stellar mass relation and the distribution of sSFR for field and group spirals in Sections~\ref{sSFRFIELD} \& \ref{sSFRGROUP}, as well as for central and satellite (group) spiral galaxies in Section~\ref{SATCENT}. Making use of our samples and the relations derived we investigate the star formation activity and star formation history of group satellite spiral galaxies in Section~\ref{SFR_SAT}, contrasting a range of simple parameterized star formation histories (SFH) with our observations to identify relevant elements of the SFH. In Section~\ref{GASFUELLING} we then consider our results on the star formation activity and history of spiral satellite galaxies in the context of the gas-fuelling cycle of these objects including the implications of our results in terms of the gas reservoirs from which the gas-fuelling may be sourced . Finally, in Section~\ref{DISCUSSION}, we discuss the broader implications of our results, and summarize our results and conclude in Section~\ref{CONCLUSION}.\newline

In subsequent papers (Grootes et al., in prep.) we will focus on the gas-fuelling of central spiral galaxies and proceed with a detailed investigation of the impact of the group environment, as characterized e.g. by the mass of the DMH, the mean galaxy density in the galaxy group, and the presence/absence of an AGN, on the gas-fuelling of our samples of satellite and central spiral galaxies in galaxy groups, again using the field spiral galaxies as a reference.\newline

Throughout the paper, except where stated otherwise, we make use of magnitudes on the AB scale \citep{OKEGUNN1983} and an $\Omega_M = 0.3$, $\Omega_{\lambda} = 0.7$, $H_0 = 70\,\mathrm{km}\mathrm{s}^{-1}\mathrm{Mpc}^{-1}$ cosmology
\citep{SPERGEL2003}.\newline

\section{Data: The GAMA Survey}\label{DATA}
Our analysis of the effect of environment on the SFR and gas-fuelling of spiral galaxies is based on the Galaxy and Mass Assembly survey (GAMA;\citealt{DRIVER2011}). GAMA consists of a highly complete spectroscopic survey covering 286 deg$^2$ to a main survey limit of $r_{\mathrm{AB}} \le 19.8\,$mag in three
equatorial (G09, G12, and G15) and two southern (G02 and G23) regions using the 2dF instrument and the AAOmega spectrograph on the Anglo-Australian Telescope. Uniquely, the spectroscopic survey is accompanied by an associated multi-wavelength database spanning the full UV-optical-FIR/submm-radio spectrum.   
A full description of the survey is 
given in \citet{DRIVER2011} and \citet{LISKE2015}  with details of the spectroscopy provided in \citet{HOPKINS2013}, and details of the input catalogue and tiling algorithm provided in 
\citet{BALDRY2010} and \citet{ROBOTHAM2010}, respectively. Importantly in the context of our investigation, GAMA has obtained science quality redshifts\footnote{GAMA assigns each redshift determined from a spectrum a quality metric $nQ$, the details of which are described in \citet{LISKE2015}. Briefly, however, redshifts used for science purposes should fulfill $nQ \ge 3$.} for $263,719$ target galaxies covering $0<z\lesssim0.5$ with a 
median redshift of $z \sim 0.2\,$ and an overall completeness of $>98$\% \footnote{In the equatorial regions.} to its limiting depth. Due to its multi-pass nature and tiling strategy this completeness remains constant even on small scales, i.e, is unaffected by the density of neighboring galaxies, enabling the construction of a high fidelity galaxy group catalogue extending to low mass, low multiplicity groups of $\lesssim 10^{12} M_{\odot}$ \citep{ROBOTHAM2011b}. For the work presented here we have made use of the first three years of GAMA data - frozen and referred to as GAMA I - consisting of the three equatorial fields to a 
homogeneous depth of $r_{\mathrm{AB}} \le 19.4\,$mag\footnote{the $r$-band magnitude limit for the GAMA survey is defined as the SDSS Petrosian foreground extinction corrected $r$-band magnitude.}(for both galaxies and galaxy groups). In the following we briefly present the GAMA data products relevant to this work.\newline

\subsection{GAMA Spectroscopy:Redshifts \& Emission Line Measurements}\label{DATA_SPECTROSCOPY}
Our main use of the spectroscopic data of the GAMA survey is in the form of redshift measurements which have enabled the construction of the galaxy group catalogue \citep{ROBOTHAM2011b}. However, we also make use of the emission line measurements to identify AGN (as detailed in section~\ref{DATA_EML}). Spiral galaxies hosting AGN are not used, since the UV emission of such objects may no longer be a reliable tracer of their  star formation activity. A full description of the GAMA spectroscopy is given in \citet{HOPKINS2013}, along with details of the quantitative measurement of emission lines, while the determination of redshifts from the spectra is described in \citet{LISKE2015}. \newline

\subsection{GAMA Photometry: Optical}\label{DATA_PHOTOMETRY}
Our analysis makes use of optical photometry for the determination of  the sizes, inclinations, and morphologies of galaxies as well as in determining their stellar masses.
The GAMA optical photometry ($u,g,r,i,z,$) is based on archival imaging data of SDSS\footnote{This is now being replaced by KiDS imaging.}. As outlined in \citet{DRIVER2011} and detailed in \citet{HILL2011} and 
\citet{KELVIN2012}, the archival imaging data is scaled to a common zeropoint on the AB magnitude system and
convolved using a Gaussian kernel to obtain a common FWHM of the PSF of 2". The resulting data frames are combined using the \texttt{SWARP} software developed by the 
TERAPIX group \citep{BERTIN2002}, which performs background subtraction using the method described for \texttt{SExtractor} \citep{BERTIN1996}. From these 'SWARPS' 
aperture matched Kron photometry is extracted as detailed in\citet{HILL2011} and S\'ersic photometry is extracted by fitting the light profiles using single S\'ersic profiles 
as detailed in \citet{KELVIN2012}. Along with the value of the fit profile integrated to 10 effective radii, the index of the profile $n$, the half-light angular size, and the ratio 
of semi-minor to semi-major axis are also reported together with quality control information regarding the fit. \newline

Foreground extinction corrections in all optical bands have been calculated following \citet{SCHLEGEL1998} and k-corrections to $z = 0$ have been calculated using \verb kcorrrect_v4.2 \citep{BLANTON2007}.\newline

\subsection{GAMA Photometry: UV}\label{DATA_PHOTOMETRY_UV}
Critical to our investigation is the use of space-borne spatially integrated UV photometry to measure SFR.
Coverage of the GAMA fields in the ultraviolet (FUV and NUV) is provided by GALEX in the context of GALEX MIS \citep{MARTIN2005,MORRISSEY2007} and by a dedicated guest investigator 
program \textit{GALEX-GAMA} providing a largely homogeneous coverage to $\sim23\,\mathrm{mag}$. 
Details of the GAMA UV photometry are provided in \citet{LISKE2015}, Andrae et al. (in prep.), and on the GALEX-GAMA website\footnote{www.mpi-hd.mpg.de/galex-gama/}. In summary extraction of UV photometry proceeds as follows. GAMA provides a total of three measurements of UV fluxes. First, all GALEX data is processed using the GALEX pipeline v7 to obtain a uniform blind source catalog \footnote{The band merged GALEX blind catalog is $NUV$-centric, i.e. $FUV$ fluxes have been extracted in $NUV$ defined apertures, entailing that no cataloged source can be detected only in the $FUV$.} with a signal-to-noise ($S/N$) cut at $2.5\,\sigma$ in the $NUV$. This catalog has subsequently been matched to the GAMA optical catalog using an advanced matching technique which accounts for the possibility of multiple matches between 
optical and UV sources, redistributing flux between the matches as described in Andrae et al. and on the GALEX-GAMA website. Additionally, FUV and NUV photometry at the positions of all GAMA target galaxies is extracted using a curve-of-growth algorithm, as well as in apertures defined based on the measured size of the source in the $r$-band. For one-to-one matches preference is given to the pipeline photometry, while for extended sources and multiple matches, the curve-of-growth and aperture photometry is preferred, since it provides better de-blending and better integrated fluxes in these cases.The resulting best estimates of the total FUV and NUV flux of the galaxy are reported as \texttt{BEST\_FLUX\_NUV}, respectively \texttt{BEST\_FLUX\_FUV}, in the UV photometric catalog and used in the work presented. \newline

Foreground extinction corrections and k-corrections have been applied as in the optical bands. In calculating foreground extinctions in the NUV we make use of $A_{\mathrm{NUV}} = 8.2\,E(B-V)$ as provided by \citet{WYDER2007}. \newline

\section{Derived Physical Properties}\label{DATA_ANCILLARY}
Additionally, we make use of some of the more advanced data products of the GAMA survey. 
Notably, we have made use of the GAMA galaxy group catalog \citep{ROBOTHAM2011b}, as well as the GAMA stellar mass measurements \citep{TAYLOR2011}, and have derived AGN classifications from the emission line measurements and star formation rates from the UV photometry. In the following we provide details on the derived physical properties used in our analysis.\newline

\subsection{The GAMA Galaxy Group Catalog G$^3$Cv5}\label{DATA_G3CV1}  
In order to identify galaxies in groups and to characterize their environment, we make use of the GAMA Galaxy Group Catalog v5 (G$^3$Cv5, \citealt{ROBOTHAM2011b}). Due to the multi-pass nature of the GAMA survey and the resulting high spectroscopic completeness even in dense regions, this unique galaxy group catalog extends the halo mass function down to the range of low mass, low multiplicity galaxy groups, providing measurements of the dynamical mass of the groups over the whole range in mass. The G$^3$Cv5 encompasses the GAMA I 
region extending to a homogeneous depth of $r_{\mathrm{AB}} \le 19.4$, and spans a large range in group multiplicity, i.e. the number of detected group members ($2 \le 
N_{\mathrm{FoF}} \le 264$), as well as an unprecedented range in estimated dynamical mass ($ 5\cdot10^{11} M_{\odot} \lesssim M_{\mathrm{dyn}} \lesssim 10^{15} 
M_{\odot}$).
This catalogue has been constructed using a friends-of-friends (FoF) algorithm to identify galaxy groups in $\alpha/\delta-z$ space. The catalogue contains 12200 
(4487) groups with 2 (3) or more members,  totalling 37576 (22150) of 93325 possible galaxies, i.e. $\sim 40$\% of all galaxies are grouped.\newline

As discussed in \citet{ROBOTHAM2011b} the most accurate recovery of the dynamical center of the group is obtained using the so-called iterative group center. Using this method, the center always coincides with a group member galaxy. For the purposes of our analysis we have defined this galaxy as the central galaxy of the group, and consider all other group member galaxies to be satellite galaxies. We note that \citet{ROBOTHAM2011b} have calibrated the group finder on mock survey light-cones, finding no bias in the recovery of groups, respectively of the center of groups, as a function of larger scale structure. Furthermore, \citet{ALPASLAN2014} have shown that observed galaxy groups from the group catalogue trace out a large scale structure of filaments and tendrils in the GAMA survey volume, so that overall we hold our identification of central and satellite galaxies to be robust.\newline

\subsection{AGN Classification based on Emission Line Measurements}\label{DATA_EML}
In converting UV luminosity to SFR it is essential to ensure, that the measured UV luminosity indeed originates from the star formation activity of the galaxy, and is not dominated by emission from a central AGN. Accordingly, in this work, we have made use of the GAMA emission line database, as detailed in \citet{HOPKINS2013}, to identify AGN. In order to classify a galaxy as hosting an AGN we impose the requirement of line measurements with $\mathrm{S}/\mathrm{N} > 3$ in all four lines required for the BPT classification (H$\alpha$, NII, H$\beta$, and OIII) and that the galaxy lie in the AGN dominated region of parameter space as defined by \citet{KEWLEY2001}.\newline

\subsection{Stellar Mass Estimates}\label{DATA_SM}
In order to control for the effect of intrinsic galaxy properties on the SFR of galaxies, and separate this from environmental effects, we characterize our galaxy sample by stellar mass $M_*$, using the GAMA stellar mass estimates of \citet{TAYLOR2011}, which are derived from the GAMA aperture matched broadband photometry\footnote{Following Taylor, priv. comm., we scale the stellar mass estimates by the ratio of the S\'ersic $r$-band magnitude to the Kron $r$-band magnitude to account for flux missed by the fixed aperture.}. We note, that \citet{TAYLOR2011} make use of a \citet{CHABRIER2003} IMF and the \citet{BRUZUAL2003} stellar population library, and that hence,
any systematic variations due to the choice of IMF or the stellar population library are not taken into account. Furthermore, stellar masses predicted by \citeauthor{TAYLOR2011} 
incorporate a single fixed prediction of the reddening and
attenuation due to dust derived from \citet{CALZETTI2000}.
Thus, expected systematic variations in reddening and attenuation
with inclination, disk opacity and bulge-to-disk ratio
are not taken into account in the determination of $M_*$. 
However, as discussed by \citeauthor{TAYLOR2011}
(see also Figs.~12 \& 15 of \citealt{DRIVER2007}), the resulting shifts in
estimated stellar mass are much smaller than the individual effects
on color and luminosity. Finally, as we have constructed a morphologically selected sample, we are largely robust against possible morphology dependent biases in the stellar mass estimates arising from different stellar populations associated with different galaxy morphologies. Overall, \citet{TAYLOR2011} determine the formal random uncertainties on the derived stellar masses to be $\sim0.1-0.15\,$dex on average, and the precision of the determined mass-to-light ratios to be better than $0.1\,$dex.\newline

\subsection{Star Formation Rates}\label{DATA_SFR}
\begin{figure}
\plotone{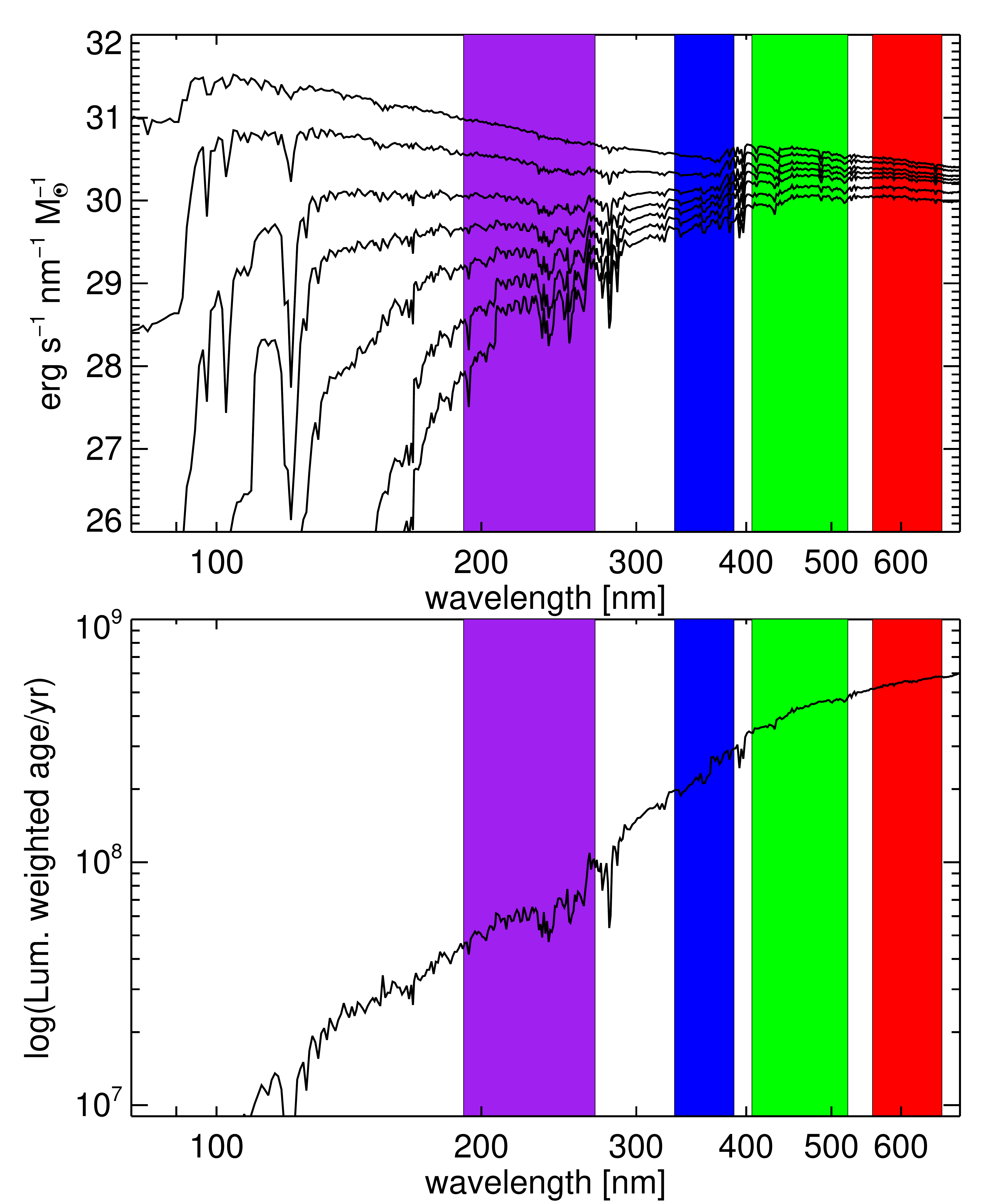}
\caption{\textbf{Top:} Mass normalized spectral luminosity density as a function of wavelength. The spectra correspond to that of a galaxy which has been constantly forming stars until 0, 10, 100, 250, 500, 1000, 2000 Myr ago (from top to bottom). \textbf{Bottom:} Luminosity weighted mean age of the emission of a galaxy with constant SFR as a function of wavelength. The shaded regions correspond to the GALEX NUV filter (purple) and the SDSS u,g,r bands (blue,green,red).}
\label{fig_NUVTRACER}
\end{figure}

Making use of the SFR of late-type galaxies as a tracer of their gas content and its dependence on the galaxies' environment requires a tracer which is sensitive to changes on timescales significantly shorter than the typical dynamical timescale of $\sim1\,$Gyr of galaxy groups. On the other hand, the tracer must reliably trace the spatially integrated star formation of the galaxy and be robust against individual bursts of SF. As shown in Fig.~\ref{fig_NUVTRACER}, which shows the spectral luminosity density of a galaxy as a function of wavelength for a range of times after the cessation of star formation, as well as the luminosity weighted mean age as a function of wavelength for a galaxy with a constant SFR, the NUV emission ideally fulfills these requirements. Probing timescales of order $10^8\,$yr, it can resolve (in time) changes on the typical dynamical timescale of galaxy groups while being robust against individual stochastic bursts of SF, unlike H$\alpha$ emission line based tracers and to a lesser extent the FUV, which trace star formation on timescales of $\sim 10^7\,$yr. Furthermore, the GAMA NUV photometry provides a robust estimate of the total spatially integrated NUV flux of the galaxy, and hence of the total SFR as desired, in contrast to emission line based tracers which require more or less sizeable aperture corrections due to the size of the fibre, depending on the distance of the source. Finally, the conversion of NUV luminosity to a SFR may depend on the age of the stellar population, i.e. the star formation history, and on the metallicity. Using the spectral synthesis code \texttt{Starburst99} \citep{LEITHERER2014}, we find the derived SFR to vary by $\lesssim 10\%$ over a range of $0.008 < Z < 0.05$ (a large range compared to that expected based on the evolution of the average metallicity of star forming galaxies over the redshift range $0 < z < 0.8$ \citealt{YUAN2013}), and between a constant SFH and a declining SFH following the SFMS. \newline

It is, however, essential to make use of the \textit{intrinsic} NUV emission of the galaxies, i.e. to correct for the attenuation of the stellar emission due to the dust in the galaxy which is particularly severe at short (UV) wavelengths \citep[e.g.][]{TUFFS2004}.\newline

In the context of the work presented here, it is important that these corrections be as precise and accurate as possible for two main reasons:

\begin{enumerate}[i.]
\item{With the analysis relying on the identification of systematic effects of the SFR and sSFR, all scatter in the values of $M_{NUV}$ used in determining these quantities will reduce the sensitivity of the analysis.}\label{scatter}

\item{In order to provide a quantitative analysis which can eventually be used in constraining structure formation calculations, an accurate treatment of systematic effects influencing the determination of intrinsic SFR is required.}\label{systematic}
\end{enumerate}

For our purposes we have adopted the method of \citet{GROOTES2013} which uses the radiation transfer model of \citet{POPESCU2011} and supplies attenuation corrections on an object-by-object basis for spiral galaxies, taking into account the orientation of the galaxy in question and estimating the disk opacity from the stellar mass surface density. A recent quantitative comparison of this method with other methods of deriving attenuation corrections, including the UV-slope, has shown it to have a higher fidelity, with smaller scatter and systematics in measuring SFR compared to other commonly used methods \citep[see Figs.~4 \& 9 of][]{DAVIES2016}.\newline

\begin{figure}
\plotone{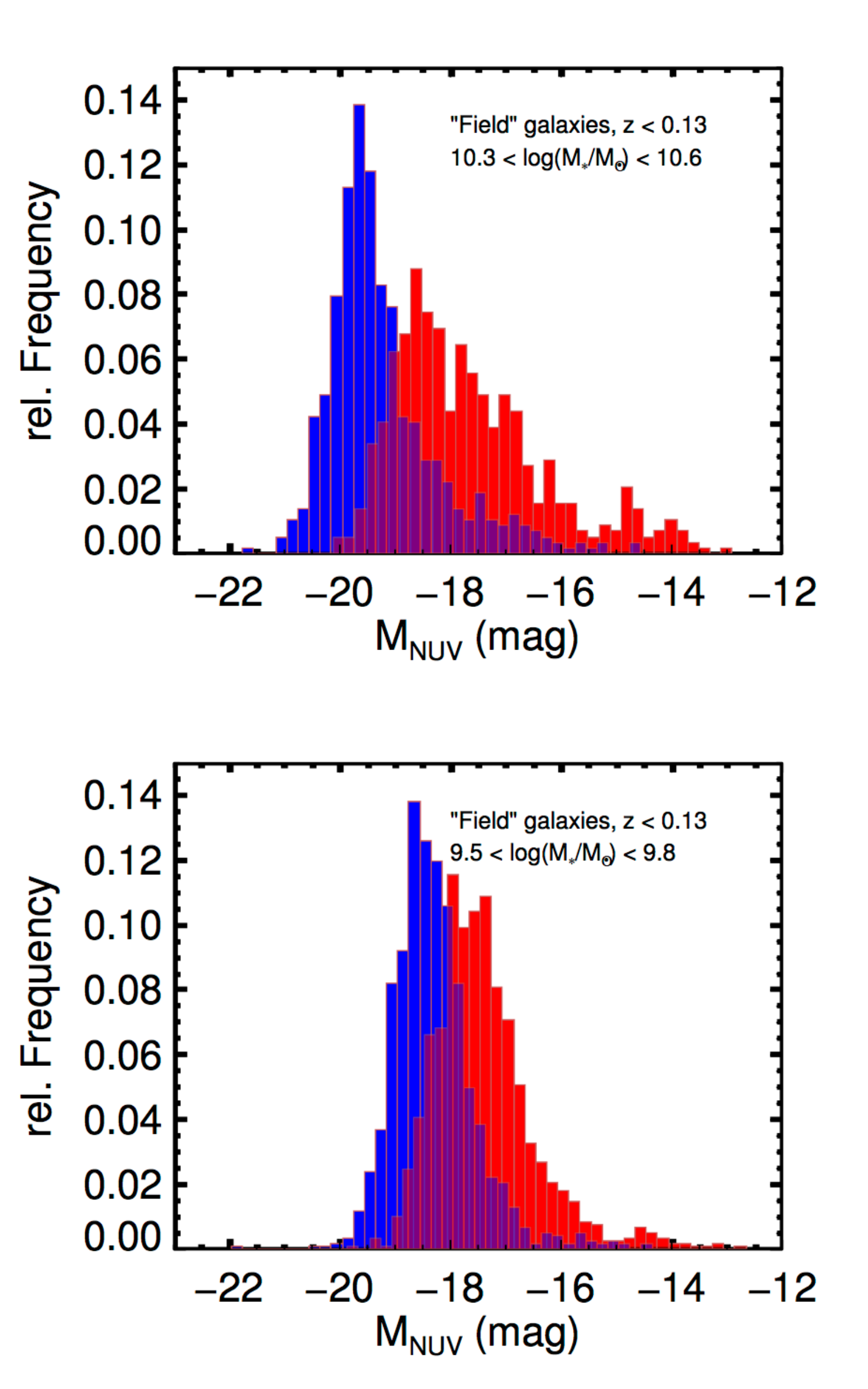}
\caption{Distributions of NUV absolute magnitude $M_{\mathrm{NUV}}$ in two ranges of stellar mass before (red) and after (blue) applying attenuation corrections as prescribed by \citet{GROOTES2013}. Notice the reduction of the scatter and the removal of the tail towards faint values of $M_{\mathrm{NUV}}$, especially in the high stellar mass range.}
\label{fig_NUVDIST}
\end{figure}

The geometry on which the RT model relies has been empirically calibrated on a sample of near-by edge-on spirals galaxies. Details of the derivation of attenuation corrections are provided in Appendix~\ref{APPEND_ATTCOR}. Corrections are typically $\sim1.4\,$mag for high stellar mass galaxies ($M \simeq 10^{10.5} M_{\odot}$)	and	lower	($\sim0.74\,$mag)	for	lower mass galaxies ($M_* \simeq 10^{9.5}M_{\odot}$). To illustrate the impact of the attenuation corrections, in Fig.~\ref{fig_NUVDIST} we show the distribution of	NUV	absolute	magnitudes	for	largely	isolated	spiral galaxies	in two ranges of stellar mass ($10^{9.5}M_{\odot}	<	M_*	<	10^{9.8} M_{\odot}$ and $10^{10.3}M_{\odot}	<	M_*	<	10^{10.6}M_{\odot}$) drawn from our \textsl{\textsc{fieldgalaxy}} sample (see Section~\ref{SAMPLESELECTION_SPIRALS_FIELD} for a definition of the \textsl{\textsc{fieldgalaxy}} sample) before (red) and after (blue) applying attenuation corrections to the observed NUV emission. The tail in the distribution due to dust for more edge-on systems is effectively removed. This tail would otherwise have been confused with galaxy quenching in a way not depending on the color of the galaxy.\newline

Using the intrinsic absolute foreground extinction corrected NUV magnitudes derived in this manner we estimate the SFR $\Phi_*$ using the conversion given in \citet{KENNICUTT1998b} scaled from a \citet{SALPETER1955} IMF to a \citet{CHABRIER2003} IMF as in \citet{SALIM2007}. For ease of comparison we explicitly supply our conversion from NUV luminosity to SFR in Eq.~\ref{eq_SFR}. It is then also simple to derive the sSFR $\psi_*$ following Eq.~\ref{eq_sSFR}.

\begin{eqnarray}
\Phi_* [M_{\odot} \mathrm{yr}^{-1}] &=& \frac{L_{\mathrm{NUV}} [\mathrm{Js}^{-1}\mathrm{Hz}^{-1}]}{1.58\times 7.14\cdot 10^{20}} \label{eq_SFR} \\
\nonumber \\
\psi_* &=& \Phi_*/M_* \label{eq_sSFR} 
\end{eqnarray}

\section{Sample Selection}\label{SAMPLESELECTION}
For our purpose of using disk-dominated/spiral galaxies as test particles to probe the influence of environment on the star formation and gas-fuelling of galaxies, we require a morphologically selected sample of spiral galaxies in galaxy groups. However, in order to separate environmental effects from the effects of secular evolution, we also require a morphologically selected sample of non-grouped spiral galaxies as a reference sample. For reasons of brevity we will refer to non-grouped galaxies as field galaxies. Additionally, as the probability of the morphological transformation of a galaxy may/will vary with environment, we require a well-defined uniform parent sample from which to select the morphologically defined samples for our analysis, which will allow us to quantify the evolution in the morphological fractions between different environments.
Furthermore, these requirements entail that the sample must be of homogeneous depth, must have been observed by GALEX\footnote{GALEX coverage of the GAMA equatorial footprint is high but not complete. See Andrae et al. (in prep) \& \citet{LISKE2015} for details.}, and have available stellar mass measurements as well as structural information in the form of S\'ersic photometry.    
In the following, we describe the sample selection process, beginning with the definition of a uniform parent sample. A synoptic overview of the sample selection is provided in Table~\ref{tab_SAMPLE}.\newline

\begin{table}[htb]
\begin{threeparttable}
\centering
\caption{Summary of sample selection process\label{tab_SAMPLE}}
\begin{tabular}{llcc}
Sample & Criteria & No. of Gal. & No. of Groups \\
\hline
Parent Sample & i - vi & 16791 & 2734\\
Spiral Galaxies & see Section~\ref{SAMPLESELECTION_SPIRALS} &7988 & 1861 \\
\textsl{\textsc{fieldgalaxy}} & see Section~\ref{SAMPLESELECTION_SPIRALS_FIELD}& 5202 & - \\
\textsl{\textsc{groupgalaxy}} & see Section~\ref{SAMPLESELECTION_SPIRALS_GROUP}& 971 & 532 \\
satellites & see Section~\ref{SAMPLESELECTION_SPIRALS_SATCENT}& 892 &  502\\
centrals & see Section~\ref{SAMPLESELECTION_SPIRALS_SATCENT}& 79 &  79\\
\hline
\hline
\end{tabular}
\end{threeparttable}

\end{table}

\subsection{The Parent Sample}
As the basis of our analysis we have constructed a uniform sample of galaxies from the GAMA data-base by selecting those which fulfill the following criteria:

\begin{enumerate}[i.]
\item{$r_{\mathrm{AB}} \le 19.4$}\label{it1}
\item{science quality redshift available from the GAMA dataset. 
}\label{it2}
\item{GALEX $NUV$ coverage of the galaxy position. Not affected by artifacts (de-selection of window and dichroic reflection artifacts).}\label{it3}
\item{redshift $z\le0.13$.}\label{it4}
\item{successful S\'ersic profile photometry in the GAMA dataset 
($r$-band quality flag = 0).}\label{it5}
\item{GAMA stellar mass estimate with $M_* \ge 10^{9} M_{\odot}$ 
.}\label{it6} 
\end{enumerate}  

resulting in a sample of 16791 galaxies.
Criteria (\ref{it1}) \& (\ref{it2}) ensure a balanced comparison of group and field galaxies by restricting the selection to the galaxies used in the construction of the galaxy group catalogue G$^3$Cv5. This work makes use of NUV photometry in estimating SFR of galaxies, and (\ref{it3}) ensures that a source has either been detected or that an upper limit can be derived. The redshift limit given by (\ref{it4}) ensures that the resolution of the imaging data is sufficient to allow reasonable determinations of galaxy morphology, while (\ref{it5}) ensures that the necessary structural information for the morphological classification, as discussed in detail below,  and the attenuation corrections is indeed available. Finally, in combination criteria (i), (iv), and (vi) ensure that our sample selection is robust against the effects of cosmological surface brightness dimming over the volume considered. \newline

As shown by \citeauthor{TAYLOR2011} (see Fig.~6 of \citealt{TAYLOR2011}), the GAMA survey (limited to $r_{\mathrm{AB}} = 19.4$), in the redshift range of $z\le0.13$ is largely stellar mass complete, i.e. volume limited, to $M_* \gtrsim 10^{9.5} M_{\odot}$ ($\gtrsim80\,$\% complete to $M_* \gtrsim 10^{9.5} M_{\odot}$ at $z \approx 0.13$).  Thus, choosing a stellar mass limit as specified in (\ref{it6}) in combination with (\ref{it4}) leads to a nearly volume-limited sample of galaxies. It must be noted, however,  that below $M_* = 10^{9.5}$, the galaxy samples selected will suffer from a Malmquist bias towards blue galaxies. Quantitatively, for a mass of $M_* = 10^{9} M_{\odot}$, the survey will only be largely mass complete to $z=0.08$. By introducing a color bias to the galaxy population, the Malmquist bias affecting the stellar mass completeness of the GAMA survey at $M_*\le10^{9.5}M_{\odot}$ may also give rise to a bias in the SFR and $\psi_*$ properties of the galaxy samples in that range of stellar mass. Nevertheless, in order to at least provide an indication of the behavior of galaxies with $M_* < 10^{9.5} M_{\odot}$, we extend our sample down to $M_* = 10^{9}M_{\odot}$ and have taken the bias into account appropriately. A detailed quantification and discussion of the bias is provided in Sections~\ref{SAMPLESELECTION_SPIRALS_FIELD},\ref{SAMPLESELECTION_SPIRALS_GROUP}, \& \ref{sSFRFIELD}.\newline

\subsection{Selection of Disk/Spiral Galaxies}\label{SAMPLESELECTION_SPIRALS} 
A key element in our approach is the selection of a morphologically defined pure sample of disk/spiral galaxies, unbiased in their SFR distribution. This requirement entails that no selection method which makes use of information linked to ongoing star formation activity (e.g. galaxy colors or clumpiness) can be used. For the purpose of selecting our sample we have therefore adopted the method of \citet{GROOTES2014}. This method, which has been trained using the GALAXY ZOO DR1 \citep{LINTOTT2011}, provides the user with a number of selection parameter combinations some of which are optimized to recover samples with an unbiased SFR distribution. \newline

In particular we have chosen to use the parameter combination (log($n$),log($r_e$),$M_i$) , where $n$ is the index of the single S\'ersic profile fit to the galaxy in the $r$-band, $r_e$ is the $r$-band effective (half-light) radius, and $M_i$ is the total $i$-band absolute magnitude.
As shown in \cite{GROOTES2014}, this particular parameter combination selects $\gtrsim77$\%\footnote{As shown in \citet{GROOTES2014} the rate of recovery decreases for very small and very bulge dominated systems.}  of SDSS galaxies classified as spiral/disk galaxies in GALAXY ZOO DR1 ( $\gtrsim70$\% of visual spiral/disk galaxies extending to types S0/Sa based on the classifications of \citet{NAIR2010}), with a contamination of $\lesssim2$\% by elliptical galaxies. Nevertheless, as demonstrated in \citet{GROOTES2014}, the use of this parameter combination results in samples which are representative of the SFR distribution of visual spiral/disk galaxies, as the recovered samples are largely unbiased with respect to the H$\alpha$ equivalent width distribution (indicative of the sSFR distribution). We also find support for this representative recovery of the parent sSFR distribution when considering the $z<0.06$ subsample of GAMA galaxies with visual morphological classifications presented in \citet{KELVIN2014}. For these sources we find the overall distribution of SFR at fixed stellar mass to be statistically indistinguishable for a sample selected by our adopted proxy, as well as by the available visual classifications (even under the inclusion of S0/Sa galaxies).

Although the performance of the selection method has been demonstrated on the parent population of spiral/disk galaxies, the use in this work of galaxy samples differentiated by environment requires the consideration of a further difficulty. For a spiral galaxy consisting of a predominantly old\footnote{In terms of its stellar population.} bulge component, and a younger star-forming disk, a quenching of the star formation will lead to a secular passive fading of the disk with respect to the bulge which might cause a spiral galaxy to be shifted out of the selection by changing the resulting value of $n$ or $r_e$, although its actual morphology remains unaltered. Thus, although the recovery of the SFR distribution appears largely unbiased, the possibility of slight remaining bias against quenched systems remains. As the group environment may cause a cessation or decline of star formation in member galaxies, the possibility of the environment exacerbating the possible small bias induced by fading, and impacting the recovery of the group spiral/disk population arises. However, as we show in detail in Appendix~\ref{APPEND_MORPHSEL}, even for the higher range of bulge-to-disk ratios represented by the higher stellar mass range of our samples ($B/T \approx 0,3$) we only expect shifts of $\sim0.1$ over timescales of several Gyr, i.e. not out of the range of $B/T$ values encompassed by the selection method of \citet{GROOTES2014}, so that passive fading will not significantly bias our sample selections.\newline 

In our analysis we have used the parameter combination (log($n$),log($r_e$),$M_i$) to provide a sample of morphologically late-type/disk galaxies. A detailed discussion of the morphological selection is provided in Appendix~\ref{APPEND_MORPHSEL}. Applying the morphological selection to the sample of 16791 galaxies previously selected we obtain a sample of 7988 disk/spiral galaxies.\newline

 \subsection{The Field Galaxy Sample}\label{SAMPLESELECTION_SPIRALS_FIELD}
From the sample of disk/spiral galaxies we select a so-called 'field' sample for reference purposes in this paper (and the following papers in this series) by selecting those galaxies which have not been grouped together with any other spectroscopic GAMA galaxy in the G$^3$Cv5 to the apparent magnitude limit of $r_\mathrm{AB} \le 19.4\,$mag. Furthermore, we impose the requirement, that the galaxy not host an AGN. This results in a sample of 5202 galaxies, referred to as the \textsl{\textsc{fieldgalaxy}} sample. As a comparison, a total of 9606 galaxies from the parent sample are non-grouped in the G$^3$Cv5, and we refer to these as the field galaxy parent sample.\newline

It should be emphasized that the \textsl{\textsc{fieldgalaxy}} sample does not strictly represent a sample of truly isolated galaxies, as potentially galaxies below the magnitude limit may be associated with its constituent galaxies (i.e. rendering them grouped). However, given the stellar mass completeness of GAMA to $M_* \approx 10^{9.5} M_{\odot}$ at $z=0.13$ as well as the high spectroscopic completeness achieved by the GAMA survey it is nevertheless very likely that the \textsl{\textsc{fieldgalaxy}} sample galaxies lie at the center of their DMH and are the dominant galaxy in it, as for normal mass-to-light ratios it is unlikely that they are actually the satellite of a more massive but $r$-band faint galaxy. 
As such, the galaxies in the \textsl{\textsc{fieldgalaxy}} sample can be thought of as representing a highly pure sample of largely isolated spiral central galaxies.\newline 

\begin{figure}
\plotone{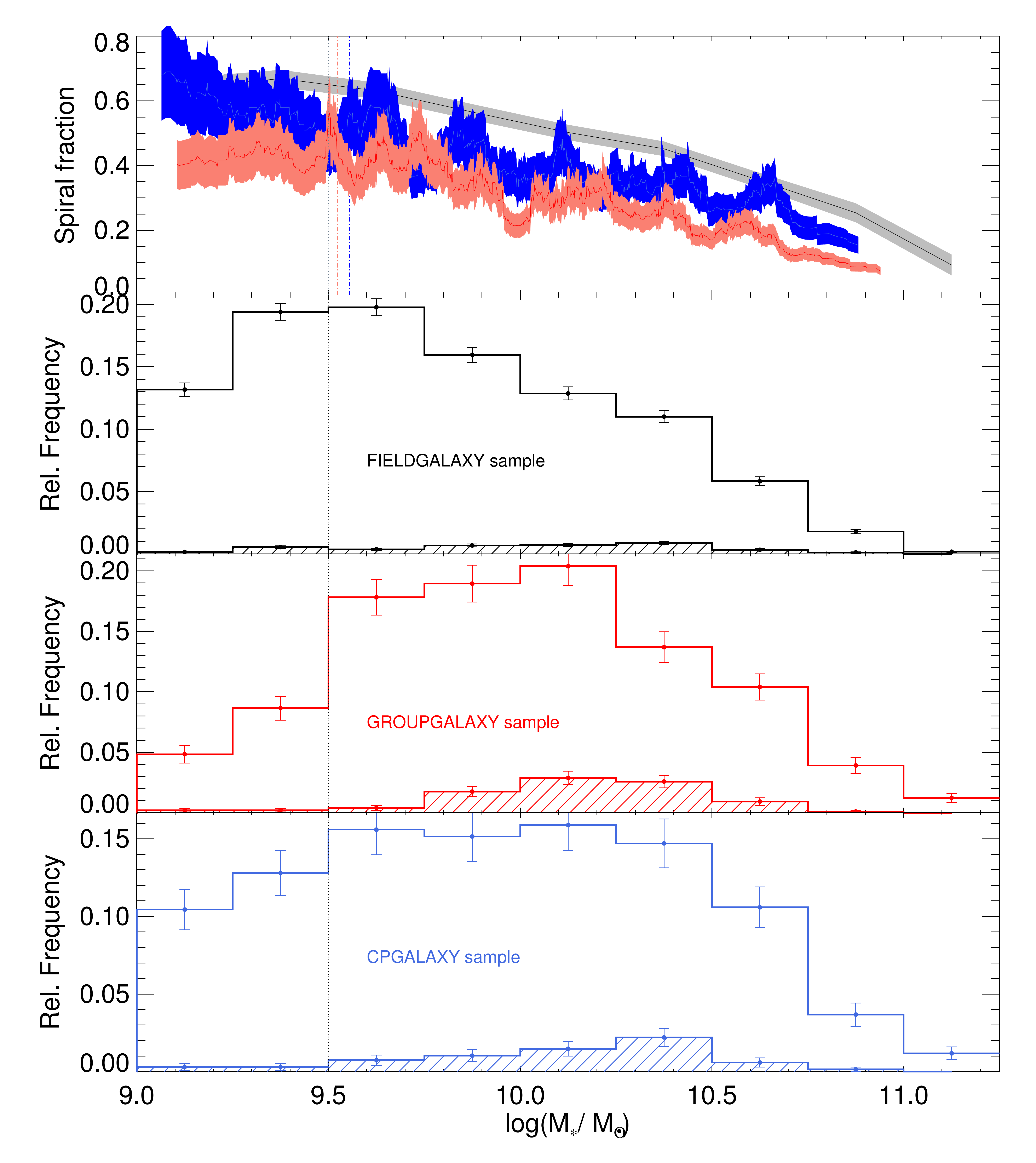}
\caption{Spiral fraction (i.e. fractional contribution of the spiral sample in question to the relevant parent sample) for the \textsl{\textsc{fieldgalaxy}}, \textsl{\textsc{groupgalaxy}}, and \textsl{\textsc{cpgalaxy}} samples in sliding tophat bins containing 40 galaxies as detailed in Section~\ref{sSFRGROUP}. The shaded area indicates the (Poisson) uncertainty in each bin. Colored dashed-dotted lines indicate the stellar mass above which the bins can be considered to be complete. The lower panels show the distribution of stellar mass $M_*$ for the \textsl{\textsc{fieldgalaxy}}, \textsl{\textsc{groupgalaxy}}, and \textsl{\textsc{cpgalaxy}} samples. Line-filled histograms show the stellar mass distributions of sources with NUV upper limits. The spiral fraction as function of stellar mass and the stellar mass distributions are available online as 'data behind the figure'.}
\label{fig_GFCP_SPFRAC}  
\end{figure}

The fraction of the field galaxy parent sample included in the \textsl{\textsc{fieldgalaxy}} sample, i.e. the field spiral fraction, varies as a function of stellar mass $M_*$, as shown in Fig.~\ref{fig_GFCP_SPFRAC}. We find the spiral fraction to decrease from $\sim65$\% at $M_* \approx 10^{9.5}M_{\odot}$ to $\lesssim30$\% at $M_* \approx 10^{10.75}M_{\odot}$. In terms of frequency, the distribution of $M_*$ for the \textsl{\textsc{fieldgalaxy}} sample is peaked at the lower bound of the volume limited mass range (see Fig.~\ref{fig_GFCP_SPFRAC}) , with the frequency gradually declining towards higher values of $M_*$ and only $\sim2$\% of the sample being more massive than $10^{10.75} M_{\odot}$.\newline

\begin{figure*}
\plotone{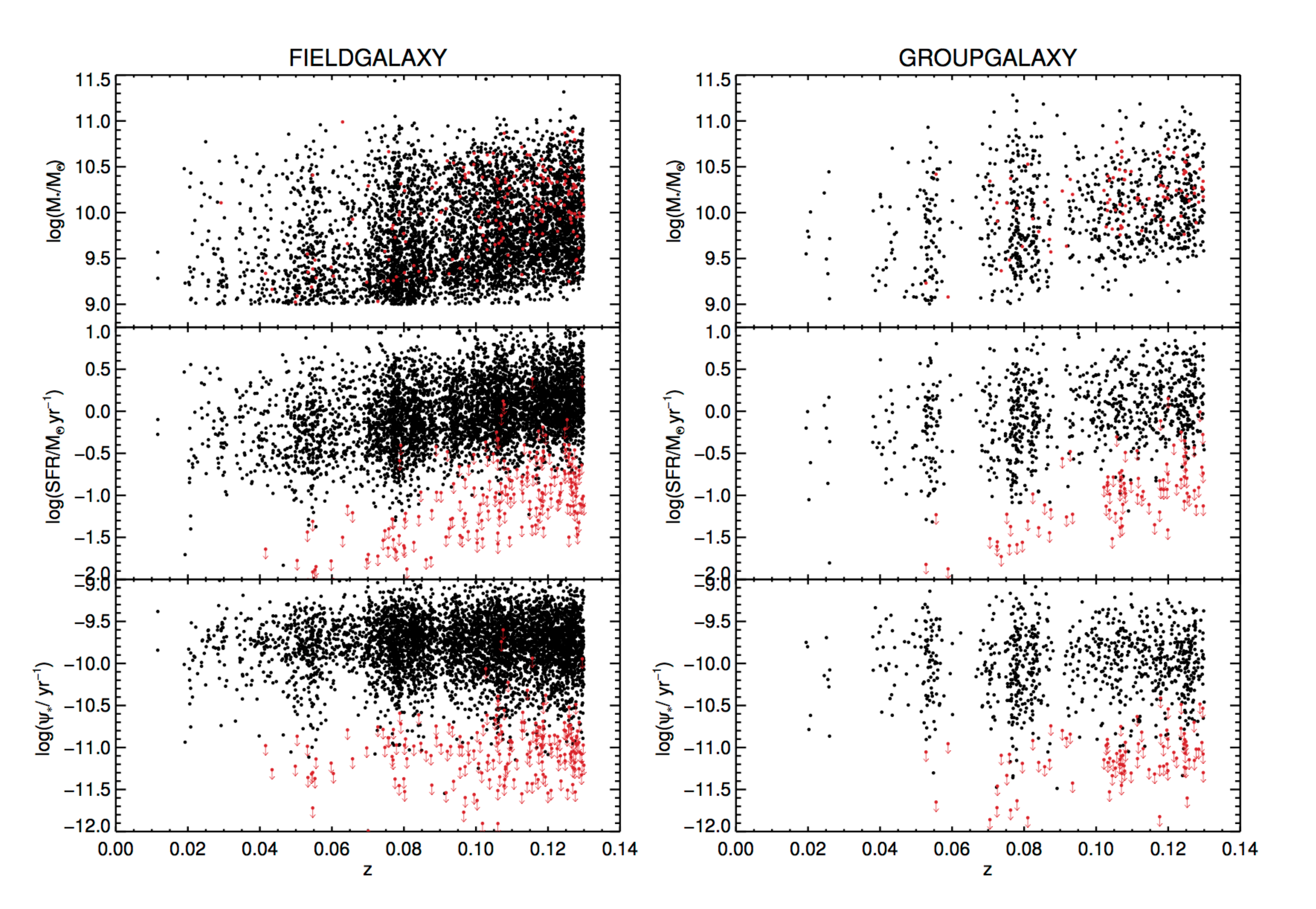}
\caption{Distributions of stellar mass $M_*$ (top), SFR (middle), and sSFR $\psi_*$ (bottom), as a function of redshift z for the \textsl{\textsc{fieldgalaxy}} sample (left) and the \textsl{\textsc{groupgalaxy}} sample (right). Galaxies for which only 2.5-$\sigma$ upper limits in the NUV are available are shown in red. The effects of the Malmquist bias on the population of galaxies with $M_* < 10^{9.5}$ are clearly visible. Above this mass no indication of a bias is present. The vast majority of sources are detected by GALEX in the NUV, ensuring that the median and quartiles of the distributions are defined by detections rather than upper limits. }
\label{fig_sampdist}
\end{figure*}

Finally, the distributions of stellar mass $M_*$, SFR and sSFR $\psi_*$ as a function of redshift $z$ for the \textsl{\textsc{fieldgalaxy}} sample are shown in Fig.~\ref{fig_sampdist} and evidence the presence of the previously discussed Malmquist bias. Figs.~\ref{fig_GFCP_SPFRAC} \&\ref{fig_sampdist} also demonstrate, that the GALEX NUV coverage is sufficiently deep so as to ensure that the median, and quartiles of the distributions of SFR and sSFR are defined by actual detections, rather than by upper limits. \newline

\subsection{The Group Galaxy Sample}\label{SAMPLESELECTION_SPIRALS_GROUP}
The sample of field spiral galaxies is complemented by our sample of spiral/disk-dominated galaxies within galaxy groups as characterized by the G$^3$Cv5 of \citet{ROBOTHAM2011b} (see also Section~\ref{DATA_ANCILLARY}), referred to as the \textsl{\textsc{groupgalaxy}} sample. In constructing this sample, we proceed by selecting from the sample of 7988 disk/spiral galaxies all those which are assigned to a group with 3 or more members (of any morphology), each with $M_* \ge 10^{9.5} M_{\odot}$. This selection ensures, that the groups considered in our analysis can be selected over the full redshift range considered, thus avoiding any implicit bias in $\psi_*$ as a function of group properties, which could result from the Malmquist bias in the galaxy sample. Obviously, groups selected in this manner thus may actually consist of more members, some having $M_* \le 10^{9.5} M_{\odot}$, due to the flux-limited nature of the GAMA survey. \newline

From this selection, we discard all galaxies residing in groups in which the velocity dispersion is dominated by the total error on the velocity dispersion \footnote{As discussed in \citet{ROBOTHAM2011b} no estimate of the dynamical mass is possible in groups in which the total error on the velocity dispersion, composed of the uncertainties on the individual redshifts, is comparable to the measured velocity dispersion.}, and furthermore impose the requirement that the galaxy not host an AGN.\newline

In using the galaxies in our samples as test-particles and their SFR as a probe of gas-fuelling, it is essential to exclude close pairs of galaxies, as the effects of galaxy--galaxy interactions are known to boost the rate at which galaxies convert their ISM into stars \citep[e.g.][]{BARTON2000,ROBOTHAM2013,DAVIES2015}. Although these galaxy--galaxy interactions, which are likely to be present in close pairs, are an important and interesting aspect of galaxy evolution in the group environment, they will be superimposed on the galaxy--IHM effects which are the focus of this work. We therefore discard galaxies which are a member of a close pair, i.e. have a neighbor galaxy within $1000\,\mathrm{km}\,\mathrm{s}^{-1}$ and a projected separation $\le 50\,\mathrm{kpc}\,h^{-1}$.\newline

To verify that the minimal separation chosen in the exclusion of close pairs of galaxies is sufficient to isolate galaxy--IHM interactions from galaxy--galaxy interactions we consider the offset of the sSFR $\psi_*$ of the galaxies in the \textsl{\textsc{groupgalaxy}} sample from the median value of $\psi_*$ for \textsl{\textsc{fieldgalaxy}} sample galaxies of the same mass, $\Delta \mathrm{log}(\psi_*)$ as defined in Eq.~\ref{eq_dlogpsi} of Section~\ref{sSFRGROUP}, as a function of stellar mass $M_*$, and of the projected distance to the nearest group member galaxy $r_{\mathrm{proj,NN}}$, as shown in Fig.~\ref{fig_SATCENTDISTSSFR}. No systematic  dependence of  $\Delta \mathrm{log}(\psi_*)$ on $r_{\mathrm{proj,NN}}$ is visible for $r_{\mathrm{proj,NN}} \ge 50\,\mathrm{kpc}h^{-1}$, implying that environmental effects on $\psi_*$ as a function of group parameters are unlikely to be contaminated by the effects of recent interactions. For galaxies within our exclusion limit, we do see signs of an enhanced star formation at low projected distances, in particular for $M_*\gtrsim 10^{10} M_{\odot}$, in line with the results of the dedicated investigation of star formation in close pairs presented by \citet{DAVIES2015}.\newline
 
Applying this selection, the resulting \textsl{\textsc{groupgalaxy}} sample consists of 971 galaxies drawn from 532 distinct galaxy groups as identified by the G$^3$Cv5. As a comparison, a total of 4419 galaxies from the parent sample reside in galaxy groups with 3 or more members (this number includes close pair galaxies, as well as AGN host galaxies). We refer to these galaxies as the group galaxy parent sample.\newline

In terms of the spiral fraction as a function of  stellar mass for group galaxies, we find that the trend found for the \textsl{\textsc{fieldgalaxy}} sample is approximately mirrored by the \textsl{\textsc{groupgalaxy}} sample over the full range in $M_*$. However, the actual fraction of spiral galaxies embodied by the \textsl{\textsc{groupgalaxy}} sample is lower by $30 - 40$\% over the entire range in $M_*$ considered. Furthermore, the distribution of $M_*$ for the \textsl{\textsc{groupgalaxy}} sample, as shown in Fig.~\ref{fig_GFCP_SPFRAC}, is more skewed towards intermediate and higher mass galaxies than that of the \textsl{\textsc{fieldgalaxy}} sample, displaying a peak in the relative frequency distribution at $M_* \approx 10^{10} - 10^{10.25} M_{\odot}$ and increased relative weight above this stellar mass with respect to the \textsl{\textsc{fieldgalaxy}} sample.  This increase in relative
weight can be attributed to the requirement that a group contain $\ge3$ galaxies with $M_* \ge 10^{9.5}M_{\odot}$ in order to enter the \textsl{\textsc{groupgalaxy}} sample, leading to a selection of more massive halos than those sampled by the largely isolated central galaxies of the \textsl{\textsc{fieldgalaxy}} sample. These relatively more massive halos can and do host relatively more massive galaxies, skewing the stellar mass distribution.\newline

Finally, as for the \textsl{\textsc{fieldgalaxy}} sample, the distributions of $M_*$, SFR, and sSFR for the \textsl{\textsc{groupgalaxy}} sample are depicted in Fig~\ref{fig_sampdist} and show evidence of the expected Malmquist bias below $M_∗ = 10^{9.5}M_{\odot}$.\newline

\subsection{Group Central \& Satellite Spiral Galaxies}\label{SAMPLESELECTION_SPIRALS_SATCENT}
As outlined in the Introduction the distinction between a galaxy being a central or a satellite galaxy may be fundamental to its SFR and star formation history. For the purpose of our detailed investigation of the impact of the group environment on the SFR of spiral/disk galaxies, we further divide the \textsl{\textsc{groupgalaxy}} sample into satellite and central spiral group galaxies, which we will refer to as satellites and centrals, respectively. This distinction is based on the identification of the group central galaxy supplied by the G$^3$Cv5.\newline

\begin{figure}
\plotone{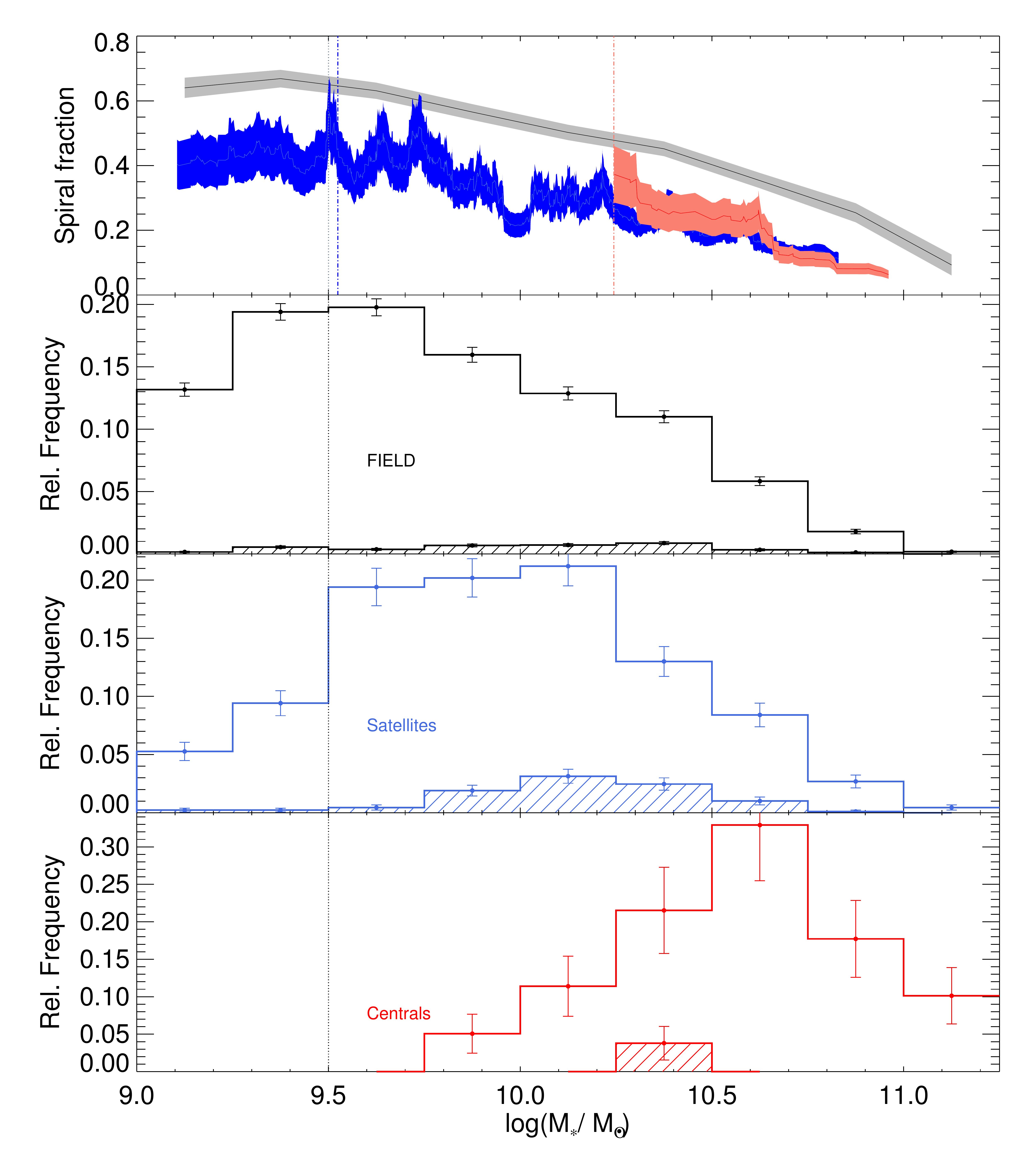}
\caption{The top panel shows the fraction of galaxies classified as spirals as a function of $M_*$ for the \textsl{\textsc{fieldgalaxy}} sample (black), the satellite galaxies in the \textsl{\textsc{groupgalaxy}} sample (blue), and the central galaxies in the \textsl{\textsc{groupgalaxy}} sample (red). Fractions have been determined in the sliding tophat bins containing 40 and 25 galaxies in the case of the satellites and centrals, respectively, as described in Section~\ref{sSFRGROUP}. The shaded areas indicate the (Poisson) uncertainties. The dash-dotted colored lines indicate the mass above which the population of the tophat bin can be considered mass complete. The bottom panels show the distribution of $M_*$ for each galaxy category, with the distribution of the sources with upper limits in $\psi_*$ shown as a line-filled histogram. The dotted vertical gray line indicates the mass limit beyond which the samples considered represent a volume limited sample. The spiral fraction as function of stellar mass and the stellar mass distributions are available online as 'data behind the figure'.}
\label{fig_SATCENTSPF}  
\end{figure}

\begin{figure}
\plotone{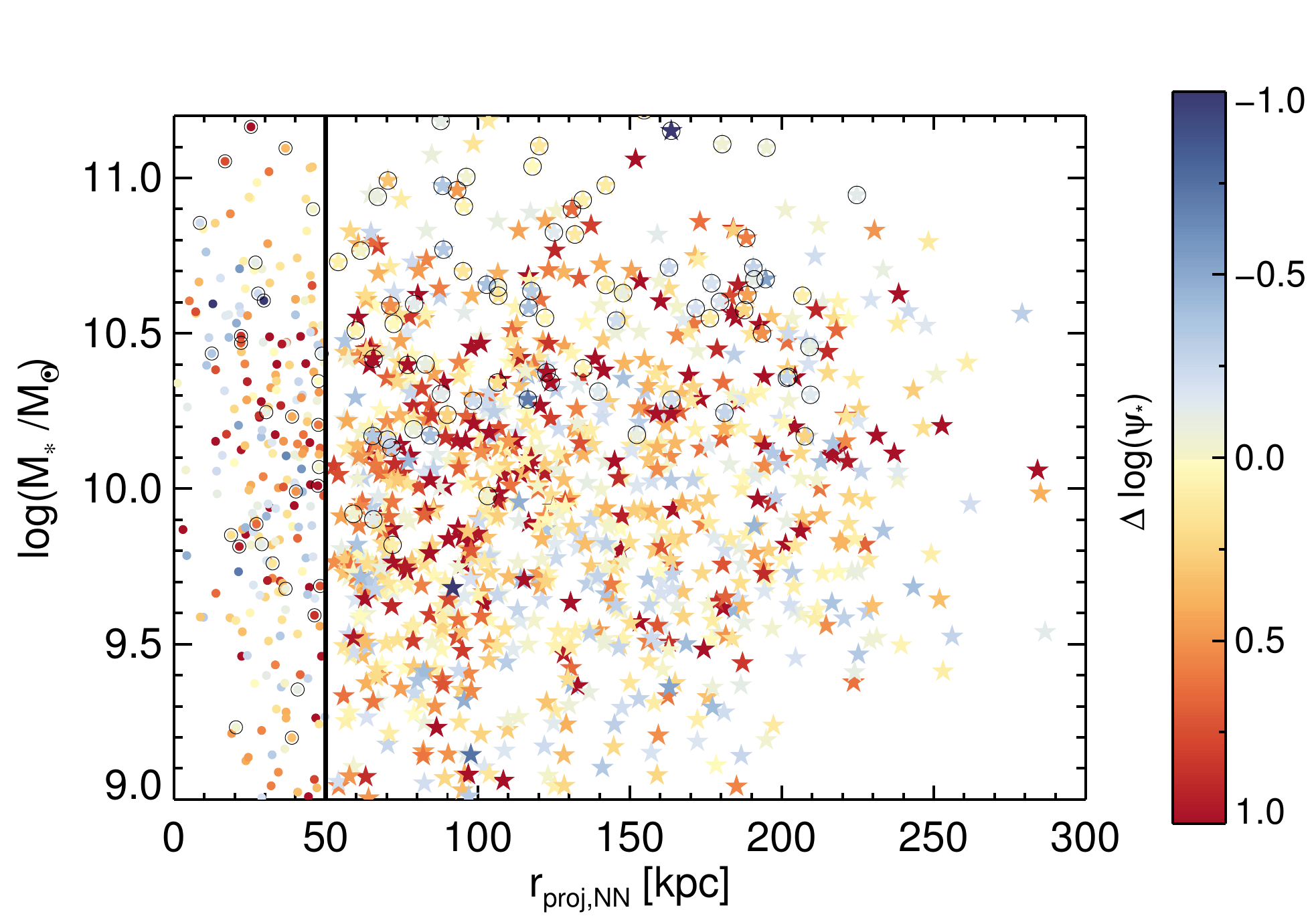}
\caption{$M_*$ as a function of the projected distance to the nearest group neighbor, $r_{\mathrm{proj,NN}}$. Galaxies in the \textsl{\textsc{groupgalaxy}} sample as shown as stars, with galaxies which are the central galaxy of their respective group marked by a circle. The offset from the median value of $\psi_*$ for field sample galaxies of the same stellar mass as the satellite ($\Delta \mathrm{log}(\psi_*)$) is color coded from blue (enhanced) to red (suppressed) as shown in the figure. The vertical solid black line indicates the minimum projected separation required for inclusion in the \textsl{\textsc{groupgalaxy}} sample. Galaxies which would have been included in the \textsl{\textsc{groupgalaxy}} sample save for having a (group member) neighbor within $50\,h^{-1}\,$kpc are shown as colored filled circles, with centrals again indicated by black circles.}   
\label{fig_SATCENTDISTSSFR}  
\end{figure}

In total, we find 892 group satellite spiral galaxies, and 79 group central spiral galaxies. The spiral fractions of the satellites and centrals\footnote{ Spiral fractions for centrals and satellites are defined as the fraction of group central/satellite galaxies in the relevant parent sample which are in the central/satellite subsample of the \textsl{\textsc{groupgalaxy}} sample.} as a function of $M_*$ are shown in Fig.~\ref{fig_SATCENTSPF}. With the satellites constituting $\sim91$\% of the \textsl{\textsc{groupgalaxy}} sample, the spiral fraction of the satellites unsurprisingly mirrors that of the \textsl{\textsc{groupgalaxy}} sample both in trend and absolute values. However, we find that the spiral fraction of the centrals not only also displays the same trend with stellar mass as the satellites and the \textsl{\textsc{fieldgalaxy}} sample, but that the actual spiral fraction of centrals is comparable to that of satellites over the full mutual mass range, i.e. $30-40$\% suppressed with respect to the \textsl{\textsc{fieldgalaxy}} sample.\newline  

Fig.~\ref{fig_SATCENTSPF} also shows the distributions of $M_*$ for the group satellite and group central spiral galaxies. As for the \textsl{\textsc{groupgalaxy}} sample, the sample of satellite spiral galaxies is more skewed towards intermediate values of stellar mass with respect to the \textsl{\textsc{fieldgalaxy}} sample (cf. panel 2 of Fig.~\ref{fig_SATCENTSPF}), with the distribution of $M_*$ peaking at $M_* = 10^{10} - 10^{10.25} M_{\odot}$, as shown in the third panel from the top in Fig.~\ref{fig_SATCENTSPF}. The relative frequencies of the satellites and the \textsl{\textsc{groupgalaxy}} sample agree within their uncertainties, although the satellites appear slightly more weighted towards lower values of $M_*$ (see Figs.~\ref{fig_GFCP_SPFRAC} \& \ref{fig_SATCENTSPF}), as expected given the distribution of $M_*$ for centrals (see bottom panel of Fig.~\ref{fig_SATCENTSPF}). The latter, namely, is skewed toward high mass galaxies, with the distribution of $M_*$ for the sample of centrals peaking at $M_* = 10^{10.5} - 10^{10.75} M_{\odot}$.\newline

\subsection{The Close Pair and Merging Galaxy Samples}
We have constructed our \textsl{\textsc{fieldgalaxy}} and \textsl{\textsc{groupgalaxy}} samples to exclude close pairs of galaxies, defined as being grouped with $\Delta v \le 1000 \mathrm{km}\,\mathrm{s}^{-1}$ and a projected separation less than $50 h^{-1} \mathrm{kpc}$, and merging galaxies, in order to safe-guard against contamination by the effects of galaxy--galaxy interactions on the SFR of our sample galaxies. To qualitatively understand whether this exclusion is necessary/justified, we have constructed a sample of close pair galaxies, referred to as the \textsl{\textsc{cpgalaxy}} sample, which meet all the morphological and photometric requirements of the \textsl{\textsc{groupgalaxy}} sample (including not hosting an AGN) but are a member of a close pair of galaxies \footnote{Unlike for the \textsl{\textsc{groupgalaxy}} sample no requirement on the multiplicity of the group has been applied, i.e galaxies in groups of two galaxies have been included.}. In total, this \textsl{\textsc{cpgalaxy}} sample contains 680 galaxies. Of these, 50 have been visually classified as merging galaxies. We have removed these from the \textsl{\textsc{cpgalaxy}} sample, and designate these 50 galaxies to be the \textsl{\textsc{merger}} sample. At this point we emphasize that for these samples 
derived properties such as SFR may suffer from systematic physical effects impacting our ability to recover intrinsic SFR. The purpose of these samples is solely to gain a qualitative measure of the potential impact of galaxy--galaxy interactions on the SFR of spiral galaxies.\newline

In terms of the distribution of stellar mass and the spiral fraction of the \textsl{\textsc{cpgalaxy}} sample Fig.~\ref{fig_GFCP_SPFRAC} shows that the \textsl{\textsc{cpgalaxy}} sample is more skewed towards intermediate and high stellar masses than the \textsl{\textsc{fieldgalaxy}} sample, similar to the \textsl{\textsc{groupgalaxy}} sample. However, the peak at intermediate values of $M_*$ is less pronounced. While, the spiral fraction of the \textsl{\textsc{cpgalaxy}} sample also approximately mirrors the trend with stellar mass found for the \textsl{\textsc{fieldgalaxy}} and \textsl{\textsc{groupgalaxy}} samples, in terms of the actual fraction of spirals, at a given stellar mass the \textsl{\textsc{cpgalaxy}} sample fraction lies between the \textsl{\textsc{fieldgalaxy}} and \textsl{\textsc{groupgalaxy}} samples, being lower than that of the  \textsl{\textsc{fieldgalaxy}}  by $10-30$\% over the full range in $M_*$. These trends are in line with expectations, given the \textsl{\textsc{cpgalaxy}} being composed of galaxies in groups as well as in very low multiplicity systems.\newline

\subsection{Selection of Groups}
No explicit selection of the galaxy groups on the basis of group properties has been made in this analysis. Instead, groups and their member galaxies have been included by virtue of their hosting a spiral/disk-dominated galaxy of the \textsl{\textsc{groupgalaxy}} sample as detailed in Sect.~\ref{SAMPLESELECTION_SPIRALS_GROUP}, with the only requirement being that the group consist of at least three member galaxies, each with $M_* \ge 10^{9.5} M_{\odot}$, in order to guarantee that the sample of groups be volume limited.\newline

Inside of $z=0.13$ the G$^3$Cv5 contains 824 galaxy groups for which an estimate of the dynamical mass is possible and which consist of at least three member galaxies (of any morphology), each with $M_* \ge 10^{9.5} M_{\odot}$ and science quality redshifts. Of these, a total of 631 groups ($77$\%) contain at least one spiral galaxy. In 99 of these the (only) spiral galaxy is a member of a close pair, resulting in a total of 532 ($65$\%) of all possible galaxy groups being probed by the \textsl{\textsc{groupgalaxy}} sample. Although \citet{GROOTES2014} cite a completeness of $\sim 70$\% for the morphological selection criteria we have adopted, and it is accordingly possible that some of the groups not sampled by the \textsl{\textsc{groupgalaxy}} sample do in fact contain a spiral galaxy (in addition to those in which the spiral is in a close pair), it is nevertheless conceivable that the selection applied to define the \textsl{\textsc{groupgalaxy}} sample (in particular the morphological selection) might introduce a bias into the population of galaxy groups considered. In the following we will briefly consider potential biases in the distribution of group parameters of groups included in the \textsl{\textsc{groupgalaxy}} sample, compared to the full volume-limited sample of galaxy groups.\newline     

\begin{figure}
\plotone{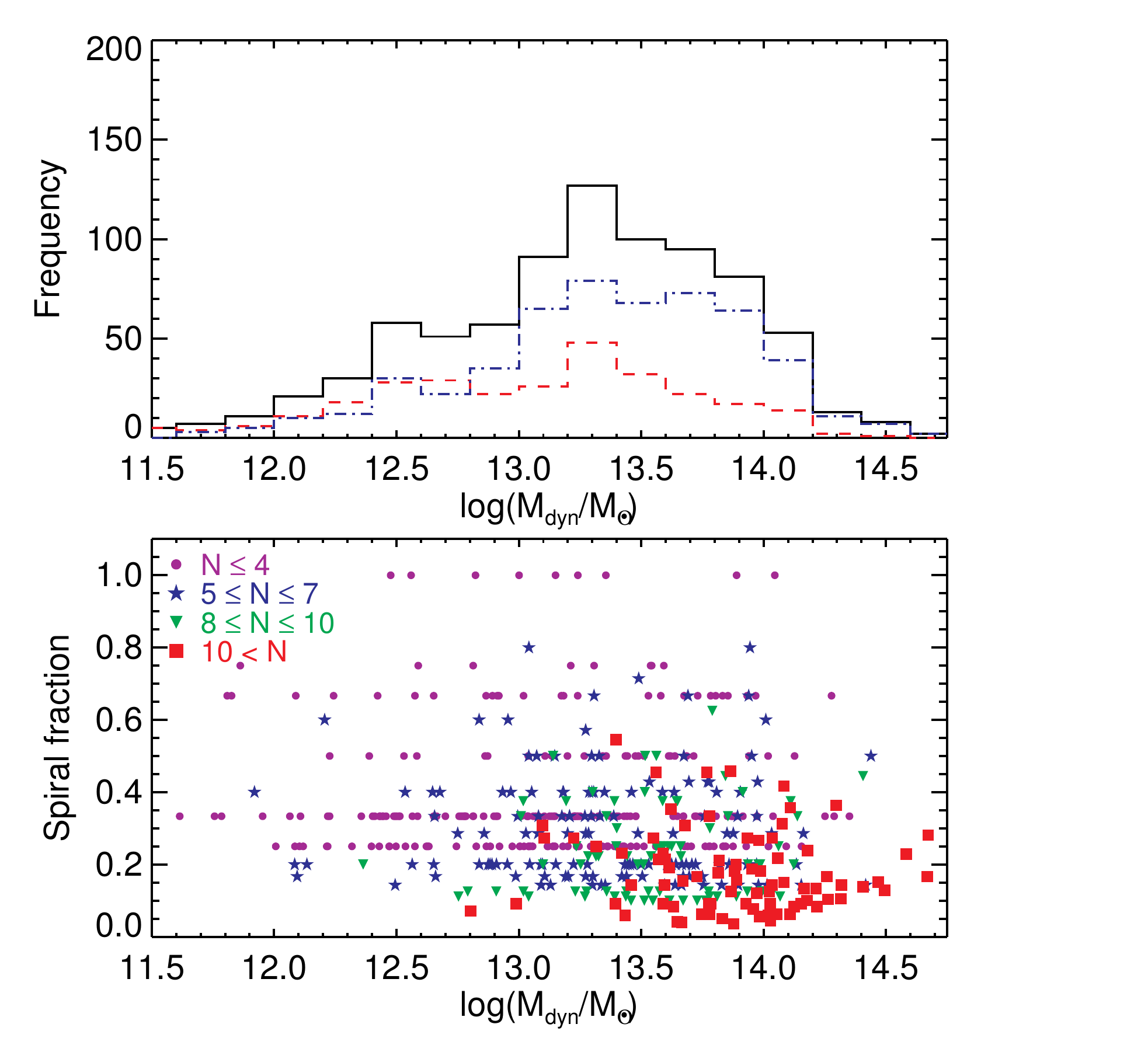}
\caption{\textbf{Top:} Distribution of the estimated dynamical mass of the galaxy group $M_{\mathrm{dyn}}$ as provided by the G$^3$Cv5 for all 824 galaxy groups with $\ge 3$ member galaxies with $M_* \ge 10^{9.5} M_{\odot}$ and science quality redshifts for which an estimate of the dynamical mass is possible (black solid line). The distribution of $M_{\mathrm{dyn}}$ for the 532 groups sampled by the \textsl{\textsc{groupgalaxy}} sample (i.e containing a spiral) is shown in blue, while that of the 292 groups not sampled (i.e not containing a spiral  galaxy) is shown in red. \textbf{Bottom}: Spiral fraction (i.e. fraction of group members in the \textsl{\textsc{groupgalaxy}} sample) as a function of $M_{\mathrm{dyn}}$ for the galaxy groups sampled by the \textsl{\textsc{groupgalaxy}} sample. The number of group members is encoded in the color and plotting symbol used (purple circle: $N\le4$; blue star: $5\le N\le7$; green inverted triangle: $8\le N \le10$; red square: $N>10$).}
\label{fig_NOSP}
\end{figure}

The top panel of Fig.~\ref{fig_NOSP} shows the distribution of group dynamical mass for all 824 groups (black), as well as for the 532 (blue) and 292 (red) groups probed, respectively not probed, by the \textsl{\textsc{groupgalaxy}} sample (top panel), while the bottom depicts the fraction of group member galaxies which are spirals as function of the dynamical mass of the group (for the 532 groups probed by the \textsl{\textsc{groupgalaxy}} sample. We find the distribution of group dynamical mass for groups sampled by the \textsl{\textsc{groupgalaxy}} sample to be skewed towards more massive groups than that of the groups without a spiral galaxy. This, however, results from the fact that, although the average spiral fraction decreases with group mass, the multiplicity increases. This increase in multiplicity is more rapid than the decrease in the spiral fraction, so that higher mass groups are slightly more likely to host at least one spiral galaxy and thus be included in the \textsl{\textsc{groupgalaxy}} sample. Any bias towards higher mass, however, is mild and the selection of the \textsl{\textsc{groupgalaxy}} sample does not appear to introduce any significant bias into the population of groups considered in our analysis.\newline

\section{The sSFR - $M_*$ Relation for Field Spiral Galaxies}\label{sSFRFIELD}
It is well documented that the main observationally accessible property influencing the sSFR of galaxy is its stellar mass $M_*$ \citep{NOESKE2007,PENG2010,WHITAKER2012}. Therefore, in aiming to identify the influence of environmental effects on the sSFR of a galaxy, we require a means of accounting for this dependence and separating out the effects of the environment from galaxy specific effects. Here, we make use of the relation between the sSFR $\psi_*$ and the stellar mass $M_*$ for spiral/disk-dominated galaxies thought to be free of major environmental influences, i.e. the largely isolated central spiral galaxies of our \textsl{\textsc{fieldgalaxy}} sample, as shown in Fig.~\ref{fig_FIELDSSFRSM}, as a baseline from which to identify environmental effects.\newline

Fig.~\ref{fig_FIELDSSFRSM} also clearly reiterates the need to make use of the \textit{intrinsic} (i.e corrected
for the affects of attenuation by dust) $\psi_{*} - M_{*}$ relation, as a comparison of the corrected and uncorrected relation shows that attenuation due to dust both significantly increases the scatter\footnote{All measurements of scatter were calculated as the difference between the quartiles of the distribution in $\psi_*$, averaged over equal sized bins in $M_*$ of 0.25 dex in width, and weighted by the number of galaxies in each bin.} ($0.53\,$dex vs. $0.36\,$dex for the intrinsic relation), and changes the (high stellar mass) slope of the relation. Furthermore, Fig.~\ref{fig_FIELDSSFRSM} (and also Fig.~\ref{fig_GROUPSSFRSM}), in which the $5$\% most inclined galaxies in each mass bin have been highlighted, demonstrates the importance of accounting for the inclination of the galaxy as discussed in Section~\ref{DATA_SFR}, as it is apparent that a significant fraction of the apparently red high mass galaxies are highly inclined and actually have 'normal' sSFR.\newline

Applying the attenuation corrections obtained using the method of \citet{GROOTES2013} (as detailed in Appendix~\ref{APPEND_ATTCOR}), we find the intrinsic $\psi_{*}$--$M_{*}$ relation for the spiral/disk galaxies of the \textsl{\textsc{fieldgalaxy}} sample, to be well described by a power law
$\psi_*\propto M_*^{\gamma}$ with a slope of $\gamma  = -0.45 \pm 0.01$ over the entire range of $9.0 \le \mathrm{log}(M_* /M_{\odot} ) \le 11.25$. The values of scatter and the coefficients of the power law fits before and after applying attenuation corrections are compiled in Table~\ref{tab_PLfits}. Because of the morphological selection which makes no use of any parameters linked directly to star formation, the relation presented here represents a real unbiased specification of the $\psi_*$--$M_*$ relation for spiral/disk galaxies in the field, including spiral galaxies with little or no SF. We note that this consideration of morphologically selected spirals, rather than a star formation driven selection, leads to our derived slope being considerably steeper than that found by \citet{PENG2010} (who selected visibly blue galaxies) and comparable to that of e.g. \citet{WHITAKER2012} who used a more encompassing selection of star-forming systems. We refer the reader to \citet{GROOTES2014} for a more detailed discussion.\newline

It is quite likely that much of the remaining scatter may be intrinsic and not due to dust, as the accuracy of the systematic dust corrections is supported by the reduction in scatter. Furthermore, we draw attention to the fact that although the scatter in the $\psi_*$-−$M_*$ plane is reduced, due to the rarefication of the population of galaxies with low values of sSFR after the application of the attenuation corrections, a population of quiescent galaxies with very low sSFR remains. These quiescent galaxies are predominantly of intermediate and high stellar mass, as can be seen from a comparison of the middle and top panels of Fig.~\ref{fig_FIELDSSFRSM}. Accordingly, as we have considered largely isolated central spiral galaxies, this result may imply the existence of a secular shut-off mechanism for star formation in spirals in isolated halos, linked only to the properties of the galaxy and its surrounding IHM.\newline

\begin{table*}[tb]
\centering
\begin{threeparttable}
\caption{Compilation of power law fits to the $\psi_* - M_*$ relation\label{tab_PLfits}}
\begin{tabular}{ccccccc}
Sample & (un)corrected (u/c) & scatter [dex] & 1-$\sigma$ Equiv. [dex]$^{\alpha}$ &$\gamma$ & A & section \\
\hline
\textsl{\textsc{fieldgalaxy}} & u & $0.53$ & $0.39$ & $-0.78\pm 0.02$ & $-10.30 \pm 0.03$  & \ref{sSFRFIELD} \\ 
\textsl{\textsc{fieldgalaxy}} & c & $0.36$ & $0.27$ & $-0.45\pm 0.01$ & $-9.86 \pm 0.02$ & \ref{sSFRFIELD} \\
\textsl{\textsc{fieldgalaxy}} ($z<0.06$) & c & $0.35$ & $0.26$ &$-0.42\pm 0.06$ & $-9.94 \pm 0.03$ & \ref{sSFRFIELD} \\ 
\textsl{\textsc{fieldgalaxy}} ($z>0.06$) & c & $0.36$ & $0.27$ & $-0.45\pm 0.01$ & $-9.85 \pm 0.02$ & \ref{sSFRFIELD} \\
\textsl{\textsc{groupgalaxy}} & u & $0.77$ & $0.57$ & $-0.74\pm 0.04$ & $-10.44 \pm 0.02$ & \ref{sSFRGROUP} \\ 
\textsl{\textsc{groupgalaxy}} & c & $0.59$ & $0.44$ & $-0.43\pm 0.03$ & $-9.99 \pm 0.02$ & \ref{sSFRGROUP} \\
\hline 
\hline
\end{tabular}
\begin{tablenotes}[flushleft]
\item{Power law fits of the form $\mathrm{log}(\psi_*) = A + \gamma \cdot \left(\mathrm{log}(M_*) -10\right)$ to the $\psi_* - M_*$ relations for different samples of spiral galaxies. The column (un)corrected signifies whether attenuation corrections have been applied (c) or not (u). Scatter is calculated as detailed in section~\ref{sSFRFIELD}. The uncertainties reflect the formal uncertainties of the fit.}
\item[$^{\alpha}$]{The weighted mean interquartile range is converted to an equivalent 1-$\sigma$ scatter under the assumption of a Gaussian distribution.} 
\end{tablenotes}
\end{threeparttable}
\end{table*}

\subsection{Quantification of the Malquist Bias for Low Mass Galaxies}\label{QUANTMALM}
While our samples of spiral galaxies, i.e. the \textsl{\textsc{fieldgalaxy}} and \textsl{\textsc{groupgalaxy}} sample, are volume limited at $M_* \gtrsim 10^{9.5} M_{\odot}$, below this mass, driven by the relation between galaxy color and the mass-to-light ratio, the distribution of galaxy colors, and hence sSFR, may be subject to Malmquist bias, biasing the distributions towards bluer colors and higher sSFR. To quantify the impact this bias may have below the mass completeness limit of $M_* \gtrsim 10^{9.5} M_{\odot}$, and to verify that there is indeed no bias at greater stellar masses, we consider the \textsl{\textsc{fieldgalaxy}} sample, split into a local and distant sample at $z=0.06$ as shown in Fig.~\ref{fig_SSFRSMFIELD_LOCDIST}. Considering the median sSFR for subsamples of the local and distant samples with $10^9 M_{\odot} \le M_* \le 10^{9.5} M_{\odot}$, we find that in the local sample 
it is $\sim0.16\,$dex lower, and the interquartile range is $\sim 0.05\,$dex smaller, than in the distant sample. A similar consideration of the local and distant sub-samples of the \textsl{\textsc{fieldgalaxy}} sample limited to $10^{10} M_{\odot} \le M_* \le 10^{10.5} M_{\odot}$, and thus expected to be complete for both redshift ranges, displays a shift of $\sim0.06\,$dex towards lower values for the low $z$ sample, with similar interquartile ranges ($\sim0.4\,$dex) in both samples. Fitting power laws to the local and distant sub-sample in the range $M_* > 10^{9.5}M_{\odot}$ (see Table~\ref{tab_PLfits}), we find the fit to the local sample to be offset towards lower sSFR by $0.07 - 0.1\,$dex with respect to the fit to the distant sample, while the slopes agree within their uncertainties. We will discuss this shift in the context of the evolution of the main sequence of star forming galaxies in section~\ref{NGCEN}.  The additional shift of $\lesssim0.09\,$dex in the low mass range, however, can thus be attributed to the Malmquist bias affecting the sample below $M_* = 10^{9.5} M_{\odot}$.\newline 

We complement this test by comparing the distribution of the intrinsic $g-i$ color, derived by \citet{TAYLOR2011} in parallel to the stellar mass estimates and based on stellar population synthesis modelling, for the local and distant subsamples of the \textsl{\textsc{fieldgalaxy}} sample using a Kolmogorov-Smirnoff test. The distributions of the $g-i$ color are considered in a sliding bin of $0.2\,$dex in $M_*$. While the distributions in the bins up to and including $10^{9.3} M_{\odot} \le M_* \le 10^{9.5} M_{\odot}$ are statistically consistent with having been drawn from different parent distributions at above the 95\% confidence level ($p\approx 0.01$), the higher mass bins show no statistical evidence of having been drawn from different parent distributions $p\approx 0.545$.\newline

We have applied analogous tests to the \textsl{\textsc{groupgalaxy}} sample using the same subsample definitions, and find shifts of $0.1\,$dex and $0.05\,$dex towards lower values of $\psi_*$ for the local sub-samples in the low and high mass ranges, respectively. For the range $10^9 M_{\odot} \le M_* \le 10^{9.5} M_{\odot}$ the null hypothesis that the $g-i$ color distributions have been drawn from the same sub-sample can be (marginally) rejected ($p=0.042$), while for the high mass bin there is no significant evidence of the $g-i$ distributions having been drawn from statistically different parent samples ($p=0.52$).\newline

Overall, these results imply that, as expected, the galaxy samples suffer from a mild bias towards blue, star-forming galaxies below $M_* = 10^{9.5} M_{\odot}$, with a systematic upward bias of $\lesssim0.06\,$dex in the median sSFR of these objects, but that above this mass the samples are indeed complete and volume limited.\newline

\subsection{The $\psi_*$--$M_*$ relation for Spiral galaxies vs. the Main Sequence of Star-forming Galaxies}
The $\psi_*$--$M_*$ relation for spiral/disk-dominated galaxies considered here is closely related to the well established so-called main sequence of star-forming galaxies (SFMS) \citep[e.g.][]{NOESKE2007,PENG2010,WHITAKER2012}. As shown by \citet{WUYTS2011}, out to $z\approx 2.5$ the locus of star forming galaxies in the SFR--$M_*$ plane, commonly referred to as the SFMS, is dominated by galaxies whose light profiles can be well described by an exponential disk with a S\'ersic index of $\sim1$. As such, as the majority of our spiral galaxies are star forming (albeit that we observe a non-negligible population of quiescent spirals), the $\psi_*$--$M_*$ relation for spiral galaxies is likely to form the dominant backbone of the SFMS, at least in the local universe. In turn, this implies that the $\psi_*$--$M_*$ relation for spirals, which may arguably be more constrained in terms of the physical drivers \citep{GROOTES2014}, may be used to gain insight into the physical drivers of the SFMS. Here we briefly touch on two such uses.\newline

\subsubsection{Redshift Evolution}\label{NGCEN}
In our above consideration of the Malmquist bias affecting our sample we have spilt the \textsl{\textsc{fieldgalaxy}} sample into two redshift ranges  ($0 < z \le 0.06$ \& $0.06 < z \le 0.13$). For the stellar mass range $M_* > 10^{9.5} M_{\odot}$, i.e. not affected by the Malmquist bias, we have found the normalization of the relation to shift by $\sim0.1\,$dex while the slopes of the power law fits in both redshift ranges agree within uncertainties. This shift can be attributed to a real evolution of the $\psi_*$--$M_*$ relation over this small redshift baseline. Comparing these results with the recently published fit of the redshift evolution of the SFMS provided by \citep{SPEAGLE2014}, we find the shift of $\sim0.1\,$dex to be in line with the evolution predicted for the SFMS over this redshift baseline\footnote{For comparison purposes, we have used the median redshift of the galaxy population in the high stellar mass local and distant bins which are $z=0.05$ and $z=0.1$, respectively. We then calculate the expected shift using Eq.~28 of \citet{SPEAGLE2014}.}. As the evolution over the small redshift  baseline for the isolated central spirals of our sample can only readily be attributed to a smooth evolution of the available amount of fuel for star formation, this lends support to the idea \citep[e.g.][]{KERES2005,DAVE2011,LILLY2013,SAINTONGE2013} that the smooth evolution of the SFMS over much longer redshift baselines, as observed by \citeauthor{SPEAGLE2014}, is also driven by gas supply processes. In this context we will consider the gas-fuelling of central spiral galaxies in more detail in a subsequent paper in this series. \newline

\subsubsection{Impact of Galaxy Morphology on the SFMS} 
Over the last decade, the main sequence of star-forming galaxies has generally been considered to be well described by a single power law with a fixed slope and a normalization evolving with redshift \citep{NOESKE2007,PENG2010}. However, more recently, a number of authors have found that at the high stellar mass end, the slope of the SFMS appears to flatten \citep{KARIM2011,WHITAKER2012, SCHREIBER2015,LEE2015,ERFANIANFAR2016}. This break, located at $M_* \gtrsim10^{10} - 10^{10.5}$, is generally attributed to the increasing contribution of a passive bulge component to the stellar mass of a galaxy \citep[e.g.][]{SCHREIBER2015,LEE2015,ERFANIANFAR2016}, in line with the findings of \citet{WUYTS2011}.\newline

As shown above, the $\psi_*$--$M_*$ relation for our \textsl{\textsc{fieldgalaxy}} sample, consisting of a morphologically selected pure sample of disk dominated systems, shows no indication of a break over its entire stellar mass range of $10^9 M_{\odot} - 10^{11}M_{\odot}$. As we will show in section~\ref{sSFRGROUP} this is also the case for the $\psi_*$--$M_*$ relation of the \textsl{\textsc{groupgalaxy}} sample, with both findings being consistent with previous work on the $\psi_*$--$M_*$ relation of spiral galaxies \citep{GROOTES2014}. As a result, we thus conclude that our findings supply strong evidence that the observed break in the SFMS can be attributed to more bulge-dominated galaxies entering/dominating the sample at higher stellar mass. In essence the occurrence of the break is then linked to bulges and disks differing in the amount of star formation they can support per unit stellar mass, which may potentially be linked to their different kinematics as argued in \citet{GROOTES2014}. We will return to this question in the context of a more detailed analysis of the SFMS, decomposed by morphology, in an upcoming paper (Davies et al., in prep.)\newline

\begin{figure}
\plotone{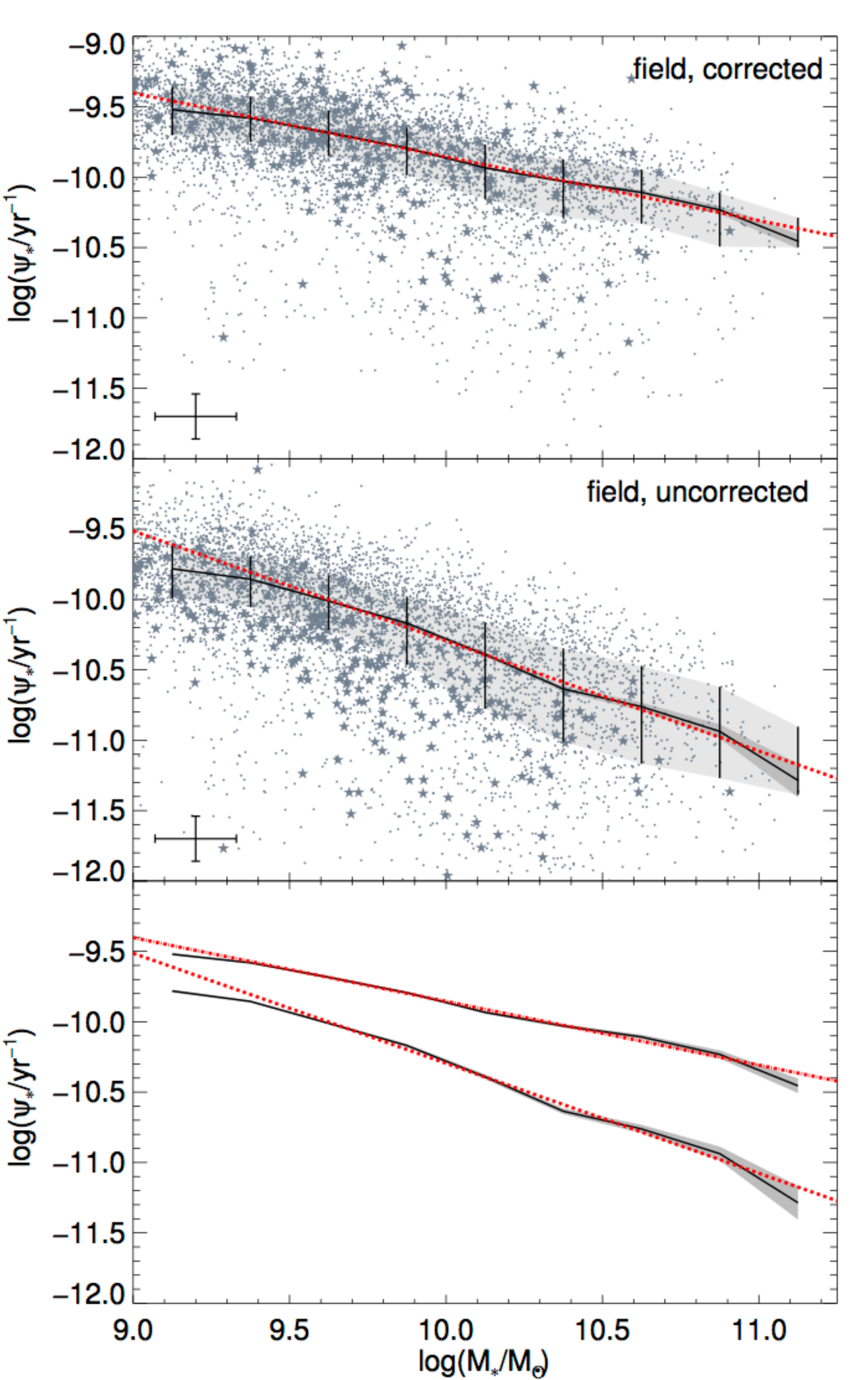}
\caption{$\psi_*$ as a function of $M_*$ for the \textsl{\textsc{fieldgalaxy}} sample, before (middle) and after (top) application of attenuation corrections. The median of the distribution in bins of $0.25\,$dex in $M_*$ is shown as a solid line with the errorbars and the light gray shaded region indicating the interquartile range and the uncertainty in the median (estimated by bootstrapping) shown as a gray shaded region. Median measurement uncertainties are shown at bottom left.
The red dashed lines indicate single power law fits to the binwise median values of $\psi_*$ in the range of $M_* > 10^{9.5} M_{\odot}$. The 5\% most inclined galaxies are denoted by stars. A comparison of the corrected and uncorrected sSFR of these galaxies shows that a significant fraction of red galaxies actually have normal sSFR. 
 For ease of comparison the bottom panel shows both the corrected and uncorrected median relations, as well as the corresponding fitted power laws. The corrected and uncorrected \textsl{\textsc{fieldgalaxy}} $\psi_*$--$M_*$ relations, including interquartile range and uncertainty of the median, are available online as 'data behind the figure'. }
\label{fig_FIELDSSFRSM}
\end{figure}

\begin{figure}
\plotone{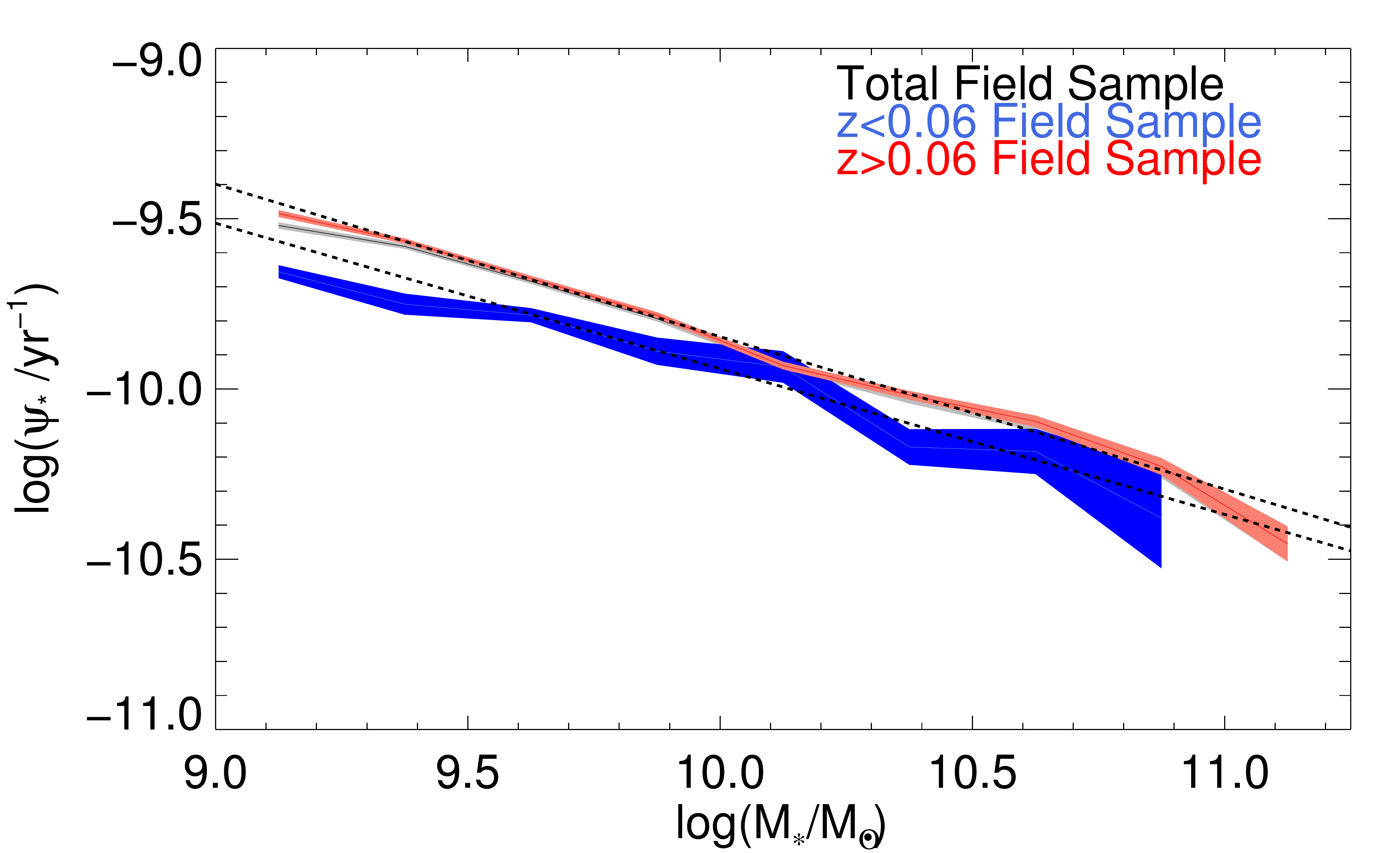}
\caption{$\psi_*$ as a function of $M_*$ for the \textsl{\textsc{fieldgalaxy}} sample (gray), and the local (blue) and distant (red) subsamples. The median of the distribution in bins of $0.25\,$dex in $M_*$ is shown as a solid line with the shaded region indicating the bootstrapped uncertainty in the median. The power law fits tabulated in Table~\ref{tab_PLfits} are plotted as black dashed lines.}
\label{fig_SSFRSMFIELD_LOCDIST}
\end{figure}

\section{The sSFR - $M_*$ Relation for Spiral Galaxies in the Group Environment}\label{sSFRGROUP}  
\begin{figure}
\plotone{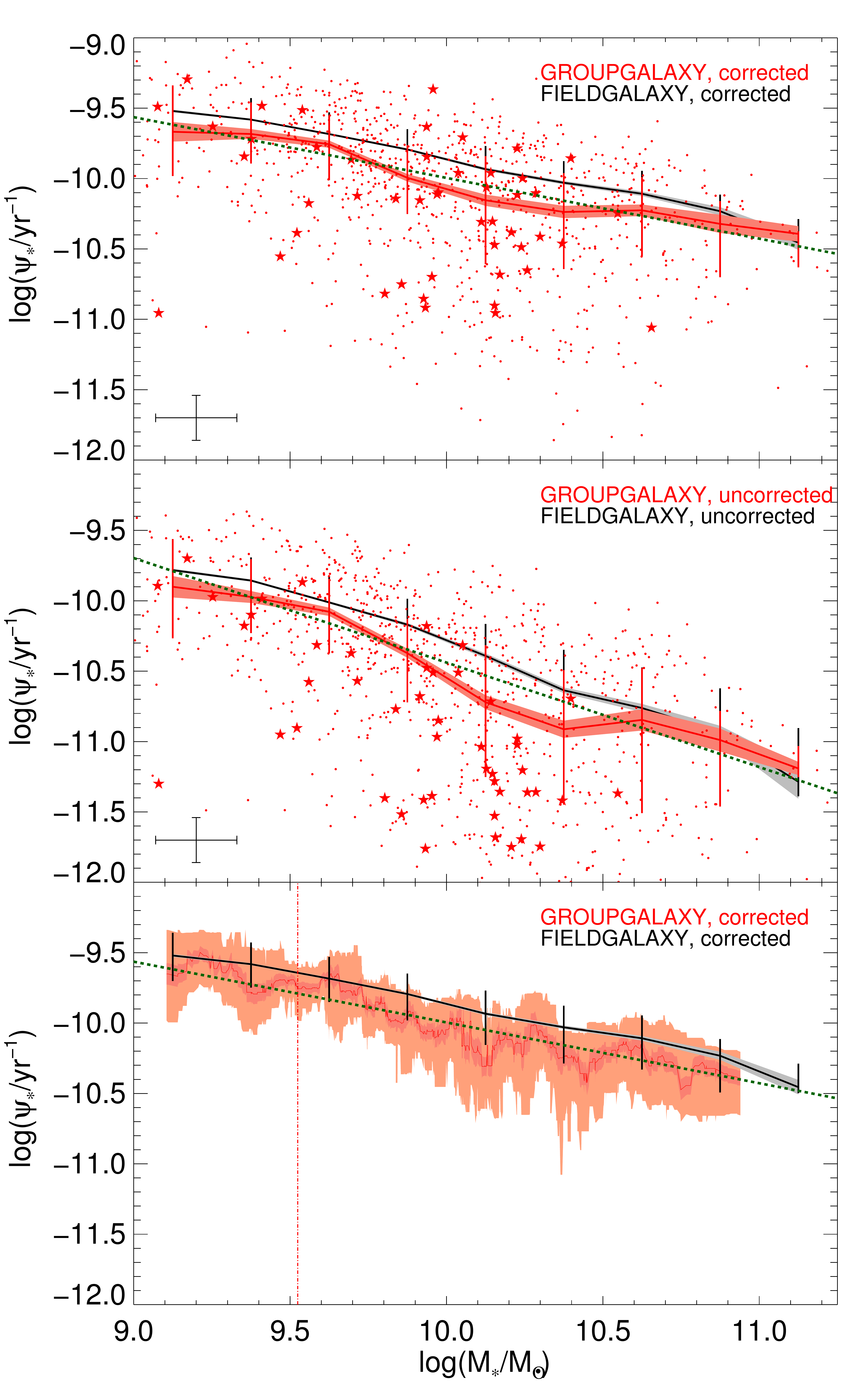}
\caption{$\psi_*$ as a function of $M_*$ for the \textsl{\textsc{groupgalaxy}} sample, before (middle) and after (top) application of attenuation corrections. The median of the distribution in bins of $0.25\,$dex in $M_*$ is shown as a red solid line with the errorbars indicating the interquartile range and the bootstrapped uncertainty in the median shown as a shaded region. For comparison the average measurement uncertainties (1-$\sigma$) for an average galaxy are shown at bottom left.
The relation for the \textsl{\textsc{fieldgalaxy}} sample is over-plotted for comparison. Dark green dashed lines indicate single power law fits to the binwise median values of $\psi_*$ for the \textsl{\textsc{groupgalaxy}} sample in the range of $M_* > 10^{9.5} M_{\odot}$.  As in Fig.~\ref{fig_FIELDSSFRSM} the 5\% most inclined galaxies are denoted by stars, demonstrating that also in the group environment a significant fraction of apparently red low sSFR galaxies actually have normal sSFR.
The bottom panel shows the median $\psi_* - M_*$ relation for the \textsl{\textsc{groupgalaxy}} sample determined in a sliding top hat bin containing 40 galaxies. The resulting median relation is shown as a solid red line, with the uncertainty on the median and the interquartile range shown as dark, respectively light, shaded regions. The $\psi_* - M_*$ relation from Fig.~\ref{fig_FIELDSSFRSM} is over-plotted, as is the power law fit to the attenuation corrected \textsl{\textsc{groupgalaxy}} sample from the top panel (see also Table~\ref{tab_PLfits}. The stellar mass limit above which the tophat bins can be considered mass complete is shown as a vertical red dash-dotted line. The $\psi_*$--$M_*$ relations for the \textsl{\textsc{fieldgalaxy}} and \textsl{\textsc{groupgalaxy}} samples, including interquartile ranges and uncertainties of the median (both for fixed and tophat binning), are available online as 'data behind the figure'.} 
\label{fig_GROUPSSFRSM}
\end{figure}

For the \textsl{\textsc{groupgalaxy}} sample we derive the median $\psi_* - M_*$ relation in exactly the same manner as used for the \textsl{\textsc{fieldgalaxy}} sample. This is illustrated in Fig.~\ref{fig_GROUPSSFRSM} where we plot the relation for the \textsl{\textsc{groupgalaxy}} sample before (top panel) and after (middle panel) applying attenuation corrections, using the same binning in stellar mass $M_*$ as for the \textsl{\textsc{fieldgalaxy}} sample shown in Fig.~\ref{fig_FIELDSSFRSM}. In both cases the scatter in the relation is larger than that for the field sample. 
This may result from  a larger intrinsic scatter in the relation or be due to differential effects of the environment on the dust content of and hence attenuation of emission from galaxies. However, as we show in Appendix~\ref{APPEND_ATTCOR}, the latter explanation appears unlikely, so we attribute the finding to a true increase in the scatter of the SFR of spiral/disk galaxies at fixed $M_*$ in the group environment.\newline

As for the \textsl{\textsc{fieldgalaxy}} sample, we find that the attenuation corrected $\psi_*$--$M_*$ relation of the \textsl{\textsc{groupgalaxy}} sample can be, to first order, well described by a power law $\psi_{*} \propto M_{*}^{\gamma}$. The results of this fit are listed in Table~\ref{tab_PLfits}. After applying attenuation corrections we find a value of $\gamma = -0.43 \pm 0.03$ over the full range in $M_*$ considered, with a scatter of $0.59\,$dex interquartile\footnote{The scatter has been determined as for the \textsl{\textsc{fieldgalaxy}} sample relation.} (see Table~\ref{tab_PLfits}). Thus, the slope of the relation is close to that found for the \textsl{\textsc{fieldgalaxy}} sample, although the scatter is twice as large, indicative of an influence of the group environment on the sSFR of group member spiral galaxies.\newline

A comparison of our results for the power law fits to the $\psi_* - M_*$ relation of the \textsl{\textsc{fieldgalaxy}} and \textsl{\textsc{groupgalaxy}} samples, as listed in table~\ref{tab_PLfits}, with the power law fit by \citet{GROOTES2014} to the $\psi_*$--$M_*$ relation of a morphologically selected sample of galaxies analogous to ours but without distinction between group and field galaxies ($\gamma = -0.5 \pm 0.12$) finds them to be consistent. The slopes of the fits agree within the uncertainties, and while \citet{GROOTES2014} find a scatter of $0.43\,$dex inter-quartile rather than $0.35\,$dex as for the \textsl{\textsc{fieldgalaxy}} sample, as we find $\sim22$\% of spiral galaxies of a given stellar mass $M_*$ are \textit{not} in the \textsl{\textsc{fieldgalaxy}} sample, this is entirely consistent with the expected contribution to the scatter arising from grouped spiral galaxies.\newline

For the purposes of considering the influence of the group environment on the SFR of spiral/disk galaxies, a rigid stellar mass binning such as hitherto employed may blunt the sensitivity of the analysis to differential effects as a function of stellar mass, in particular if the binning is relatively coarse (due to small sample sizes). For this reason, in the rest of the paper, we have chosen to adopt a slightly different approach to identifying the median $\psi_* - M_*$ relation for our samples of group galaxies. Rather than using a fixed binning in stellar mass, we make use of a sliding tophat bin containing 40 galaxies, and determine the median and quartiles of the sSFR $\psi_*$ in this bin, plotting the derived values against the median stellar mass of the galaxies in the bin. The choice of 40 galaxies is dictated by the need for the span in mass to be small enough for the systematic shift of $\psi_*$ arising from gradients in the source density of the sample as a function of stellar mass to be small over the extent stellar mass sampled by the bin.  The result of this approach is shown in the bottom panel of Fig.~\ref{fig_GROUPSSFRSM}, where the median is shown as a solid red line, and the uncertainty on the median and the interquartile range are depicted as dark and light shaded regions, respectively. Its advantages are immediately apparent, as the undulation of the binned $\psi_*$--$M_*$ relation around the power law fit for the attenuation corrected \textsl{\textsc{groupgalaxy}} sample seen in the top panel of Fig.~\ref{fig_GROUPSSFRSM} is revealed as most likely arising from statistical fluctuations.\newline

\subsection{Comparison with the \textsl{\textsc{fieldgalaxy}} sample}\label{sSFRGROUP_COMP_FIELD}
\begin{figure}
\plotone{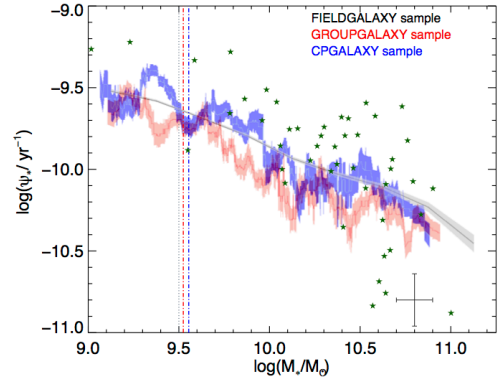}
\caption{$\psi_*$ as a function of $M_*$ for the \textsl{\textsc{fieldgalaxy}} sample (grey), the \textsl{\textsc{groupgalaxy}} sample (red), and the \textsl{\textsc{cpgalaxy}} sample blue. For the \textsl{\textsc{groupgalaxy}} and \textsl{\textsc{cpgalaxy}} samples, the median value of $\psi_*$ in a sliding tophat bin containing 40 galaxies is shown as a solid line. The wider shaded area indicates the bootstrapped uncertainty on the median. For the \textsl{\textsc{fieldgalaxy}} sample the relation is shown in bins of equal size in $M_*$ as in Fig.~\ref{fig_FIELDSSFRSM}, with the shaded area again corresponding to the uncertainty in the median. The stellar mass limit above which the samples can be considered mass complete ($M_* =10^{9.5}M_{\odot}$) is indicated by a dotted gray line. Colored dash-dotted lines indicate the stellar mass above which the galaxies in the moving tophat bins are all above the mass completeness limit. The galaxies in the \textsl{\textsc{merger}} sample have been over-plotted as dark green stars. The $\psi_*$--$M_*$ relations for the \textsl{\textsc{fieldgalaxy}},\textsl{\textsc{groupgalaxy}}, and \textsl{\textsc{cpgalaxy}} samples, including uncertainties of the median, are available online as 'data behind the figure'.} 
\label{fig_GROUPFIELDCP_1}  
\end{figure}

In order to quantify the effect of the group environment on the star formation of spiral/disk galaxies, we begin by comparing the median $\psi_*$--$M_*$ relation of the \textsl{\textsc{groupgalaxy}} and \textsl{\textsc{fieldgalaxy}} samples as shown in Fig.~\ref{fig_GROUPFIELDCP_1} (see also Fig.~\ref{fig_FIELDSSFRSM}). The median $\psi_*$--$M_*$ relation for the \textsl{\textsc{groupgalaxy}} sample in a sliding tophat bin of 40 galaxies in width is shown in red, with the uncertainty on the median indicated by the shaded area, while the reference relation defined by the \textsl{\textsc{fieldgalaxy}} sample is shown in black/gray. Although the scatter is large, it is apparent that the median relation of the \textsl{\textsc{groupgalaxy}} sample is suppressed with respect to the reference relation for galaxies with stellar mass $M_* \gtrsim 10^{9.7} M_{\odot}$. This suppression, while only mild ($0.1 - 0.2\,$dex), is quite marked with no readily discernible dependence on galaxy stellar mass.\newline

Although the median represents a statistically robust characteristic of a sample, very different distributions may result in the same median value. Therefore, to complement our consideration of the median $\psi_*$--$M_*$ relation which collapses the range of $\psi_*$ for a narrow bin of stellar mass $M_*$, we also consider the full distribution of $\psi_*$ in two disjoint ranges of $M_*$ separated at $M_* = 10^{10} M_{\odot}$. Specifically, we investigate the effect of the group environment on the sSFR of spiral galaxies by  considering the offset of a galaxy's sSFR $\psi_*$ from the median value found for galaxies of comparable stellar mass in the \textsl{\textsc{fieldgalaxy}} sample, defined as
\begin{equation}
\Delta \mathrm{log}\psi_* = \mathrm{log}(\psi_*) - \mathrm{log}(\overline{\psi_{*,\mathrm{field}}}(M_{*}))\,,
\label{eq_dlogpsi}
\end{equation}
where $\overline{\psi_{*,\mathrm{field}}}(M_{*})$ is the median value of $\psi_*$ for a field galaxy of mass $M_{*}$. $\overline{\psi_{*,\mathrm{field}}}(M_{*})$ as used in Eq.~\ref{eq_dlogpsi} has been defined as a piecewise continuous function obtained by the linear interpolation of the binned median values of $\psi_*$ as shown in Fig.~\ref{fig_FIELDSSFRSM}.\newline

\begin{figure}
\plotone{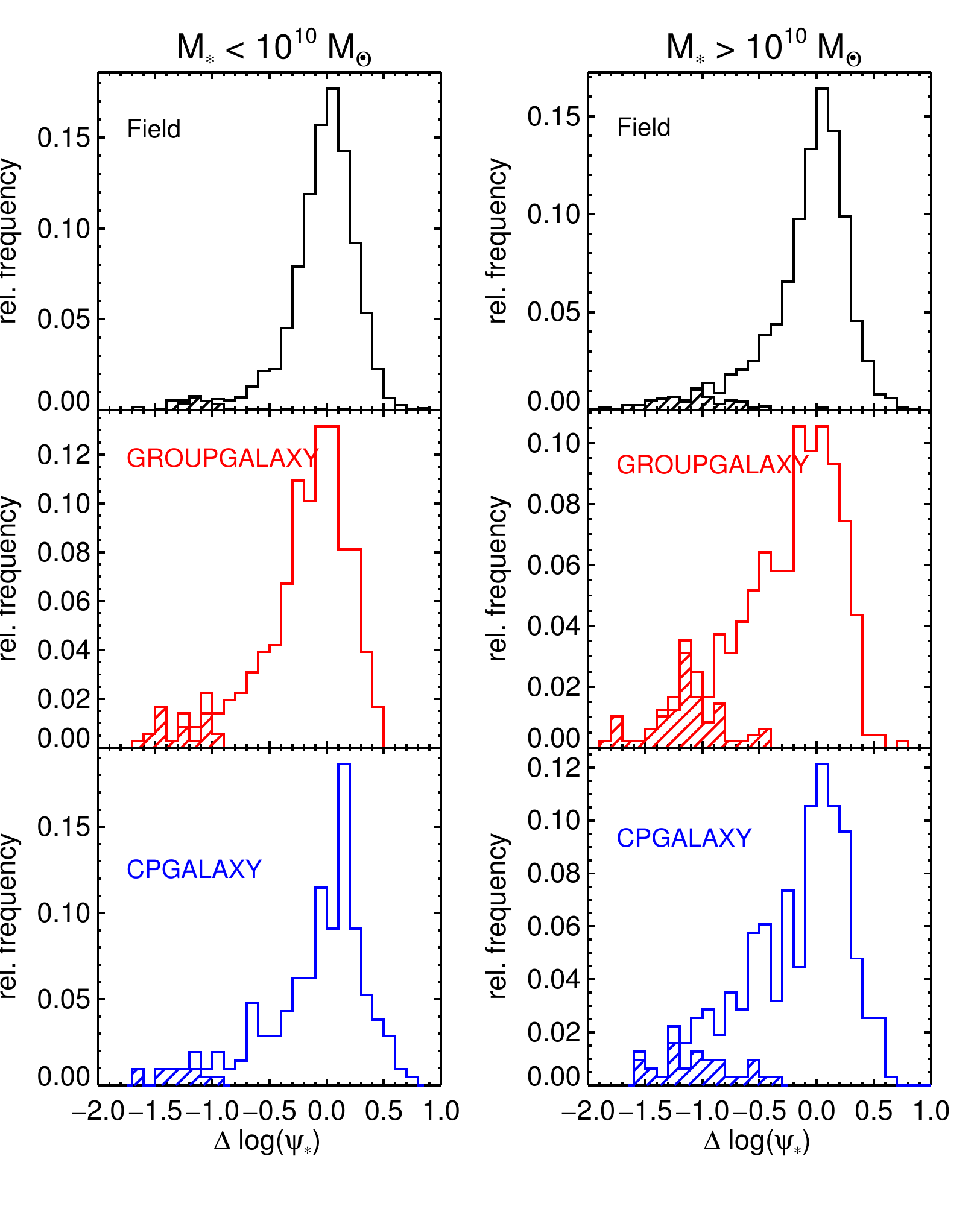}
\caption{The distributions of $\Delta \mathrm{log}(\psi_*)$ for the \textsl{\textsc{fieldgalaxy}}, \textsl{\textsc{groupgalaxy}}, and \textsl{\textsc{cpgalaxy}} samples in the low (left) and high (right) stellar mass range, separated at $M_* = 10^{10} M_{\odot}$. For the low mass range we only consider the range over which the samples can be deemed mass complete ($10^{9.5} M_{\odot} \le M_* < 10^{10} M_{\odot}$) in constructing the histograms.The distribution of upper limits is indicated by the line-filled histograms. The distributions of $\Delta\mathrm{log}(\psi_*)$ of the  \textsl{\textsc{fieldgalaxy}},\textsl{\textsc{groupgalaxy}}, and \textsl{\textsc{cpgalaxy}} samples in both stellar mass ranges are available online as 'data behind the figure'.}
\label{fig_GROUPFIELDCP_2}  
\end{figure}

Fig.~\ref{fig_GROUPFIELDCP_2} shows the distributions of $\Delta \mathrm{log}(\psi_*)$ for the \textsl{\textsc{fieldgalaxy}} (top), \textsl{\textsc{groupgalaxy}} (middle), and \textsl{\textsc{cpgalaxy}} (bottom) samples in the low (left) and high (right) stellar mass range. For the \textsl{\textsc{fieldgalaxy}} sample, we find that the distribution is strongly peaked around its median in both stellar mass ranges, with a small asymmetrical tail extending to values of $\Delta \mathrm{log}\psi_*$ corresponding to very low sSFR. This tail is more populous in the higher stellar mass range ($M_* \ge 10^{10} M_{\odot}$), encompassing $18$\% of the sample in this range of $M_*$ compared to $7$\% in the low stellar mass range, in agreement with Fig.~\ref{fig_FIELDSSFRSM}.\newline

Similarly to the \textsl{\textsc{fieldgalaxy}} sample, the distribution of $\Delta \mathrm{log}(\psi_*)$ for the \textsl{\textsc{groupgalaxy}} sample displays a pronounced peak in both ranges of stellar mass, coinciding with that of the \textsl{\textsc{fieldgalaxy}} sample. However, the population of galaxies in the \textsl{\textsc{groupgalaxy}} sample with very low values of $\Delta \mathrm{log}\psi_*$, i.e. strongly suppressed sSFR with respect to the median of the \textsl{\textsc{fieldgalaxy}} sample, is significantly larger in both stellar mass ranges, with $\gtrsim 20$\% of the population with $10^{9} \le M_*\le 10^{9.5}M_{\odot}$ having $\Delta \mathrm{log}\psi_* < -0.5$ in comparison to $7$\% for the \textsl{\textsc{fieldgalaxy}} sample, and $30$\% of the \textsl{\textsc{groupgalaxy}} having $\Delta \mathrm{log}\psi_* < -0.5$, compared to $18$\% of the \textsl{\textsc{fieldgalaxy}} sample in the range $M_*\ > 10^{10} M_{\odot}$. Thus, the small shift observable in the median $\psi_* - M_*$ relation can be attributed to an increase in the minority population of galaxies with strongly suppressed sSFR.\newline 

A more rigorous quantitative statistical investigation of  of the similarity of the distributions of $\Delta \mathrm{log}(\psi_*)$, and thus, by extension, also of the significance of the observed shift in $\psi_* - M_*$, is complicated by the fact that the measurements of $\psi_*$ include upper limits at the $2.5\,\sigma$ level ($NUV$ upper limits derived for the GALEX-GAMA photometry) in addition to reliable detections \footnote{Given that the distribution of the actual measurements of sSFR of undetected objects is likely to follow a Poisson distribution, the inclusion of these data at the $2.5\,\sigma$ upper limit level may significantly alter the shape of the distribution (with this being of increasing importance for samples with a potentially suppressed sSFR $\psi_*$).}. Accordingly, in quantifying the (lack of) similarity of the samples it is necessary to make use of a non-parametric test capable of accounting for censoring in the data. We have, therefore, adopted the generalized Wilcoxon test as suggested by \citet{PETO1972}, applied to the case of upper limits
 by \citet[e.g.][]{AVNI1980,PFLEIDERER1982,FEIGELSON1985}, and available in the statistical analysis package \texttt{STSDAS}\footnote{The \texttt{STSDAS} is a data analysis package based on the IRAF environment and developed and maintained by the software division of the Space Telescope Science Institute, Baltimore, Maryland, USA.}. In the following we will refer to this test simply as the Peto test. It should be noted that any such test, by necessity, applies a weighting scheme to the upper limits, making the test more or less sensitive to different regions of the distribution, and cannot recover the information discarded by the use of upper limits.\newline

Applying Peto tests to compare the distributions of $\Delta \mathrm{log}(\psi_*)$ for the \textsl{\textsc{fieldgalaxy}} and \textsl{\textsc{groupgalaxy}} samples supports our previous findings in the sense that the distributions are found to differ significantly in both stellar mass ranges ($p < 10^{-5}$). A summary of the Peto tests comparing the distributions of $\Delta \mathrm{log}(\psi_*)$ for the \textsl{\textsc{fieldgalaxy}} and \textsl{\textsc{groupgalaxy}} samples in both stellar mass ranges is presented in Table~\ref{tab_GFCP}.\newline

When comparing the distributions of $\Delta \mathrm{log}(\psi_*)$ in wide bins of $M_*$ one must, in principle, also consider the relative distributions of $M_*$ \textit{within} the relevant bins. However, as the differences in the relative weighting in $M_*$ in each range of $M_*$ between the samples are small and the offset in the median relation is largely uniform over the full stellar mass range, the comparability of the distributions of $\Delta \mathrm{log}(\psi_*)$ for the \textsl{\textsc{fieldgalaxy}} and \textsl{\textsc{groupgalaxy}} samples is not strongly biased.\newline

\begin{table}
\centering
\begin{threeparttable}
\caption{Comparison of the \textsl{\textsc{fieldgalaxy}}, \textsl{\textsc{groupgalaxy}}, and \textsl{\textsc{cpgalaxy}} samples \label{tab_GFCP}}
\begin{tabular}{|cc|c|c|c|}
\cline{3-5}
\multicolumn{2}{c|}{} &\rotatebox{90}{\textsl{\textsc{groupgalaxy}}\ }& \rotatebox{90}{CPGALAXY\ }&\rotatebox{90}{\textsl{\textsc{fieldgalaxy}}\ }\\
\hline
\multirow{2}{*}{\textsl{\textsc{groupgalaxy}}} & \multicolumn{1}{c|}{$\mathrm{log}(M_*)\le 10$} & 1.&$\le10^{-5}$&$\le10^{-5}$\\
 & \multicolumn{1}{c|}{$\mathrm{log}(M_*)> 10$} & 1.&0.024&$\le10^{-5}$ \\
 \hline
\multirow{2}{*}{CPGALAXY} & \multicolumn{1}{c|}{$\mathrm{log}(M_*)\le 10$} & $\le10^{-5}$&1.&0.87\\
 & \multicolumn{1}{c|}{$\mathrm{log}(M_*)> 10$} & 0.024&1.&$\le10^{-5}$ \\
 \hline
 \multirow{2}{*}{\textsl{\textsc{fieldgalaxy}}} & \multicolumn{1}{c|}{$\mathrm{log}(M_*)\le 10$} & $\le10^{-5}$&0.87&1.\\
 & \multicolumn{1}{c|}{$\mathrm{log}(M_*)> 10$} & $\le10^{-5}$&$\le10^{-5}$&1. \\
\hline
\end{tabular}
\begin{tablenotes}
\item{Significance ($p$) values of Peto tests performed between the \textsl{\textsc{fieldgalaxy}}, \textsl{\textsc{groupgalaxy}}, and \textsl{\textsc{cpgalaxy}} samples. For each combination Peto tests have been performed in two disjoint bins of stellar mass spilt at $M_*=10^{10} M_{\odot}$ (the low stellar mass bin has been limited to the mass complete sample, i.e $M_* \ge 10^{9.5} M_{\odot}$. }
\end{tablenotes}
\end{threeparttable}

\end{table}

Finally, the main systematic uncertainty in the absolute shifts in $\psi_*$ found for the members of the \textsl{\textsc{groupgalaxy}} sample is likely to be due to environment dependent effects on the dust content and distribution of spiral galaxies as discussed in Appendix~\ref{APPEND_ATTCOR}. However, shifts of the magnitude required to increase the population of galaxies with strongly suppressed sSFR seem unlikely, as they would require systematic changes in the dust surface density of the spiral galaxies by factors of a few, compared to the relation calibrated by \citet{GROOTES2013}. This is discussed further in Appendix~\ref{APPEND_ATTCOR}.\newline

In summary, we find that for the \textsl{\textsc{groupgalaxy}} sample, i.e. spiral/disk galaxies in galaxy groups, the median sSFR $\psi_*$ at a given stellar mass is suppressed by only $0.1-0.2\,$dex compared to similar objects in the \textsl{\textsc{fieldgalaxy}} sample. Furthermore, this shift is the result of an increase in size of the minority population of galaxies with strongly suppressed sSFR with respect to the median of the \textsl{\textsc{fieldgalaxy}} sample, while the majority of galaxies in the \textsl{\textsc{groupgalaxy}} sample have sSFR comparable to their \textsl{\textsc{fieldgalaxy}} sample counterparts.\newline

\subsection{SFR of Spiral Galaxies in Close Pairs}
While the members of close pairs of galaxies have been excluded from the \textit{GROUPGALXAY} sample for the purposes of our main investigation we briefly consider the \textsl{\textsc{cpgalaxy}} sample and the \textsl{\textsc{merger}} sample in terms of its median $\psi_* - M_*$ relation and distribution of $\Delta \mathrm{log}(\psi_*)$.
as shown in Figs.~\ref{fig_GROUPFIELDCP_1} \& \ref{fig_GROUPFIELDCP_2} 
\footnote{Even though strong perturbative galaxy-galaxy interactions are likely to lead to morphological transformations, a subset of these close pair and merger galaxies will still have a largely spiral/disk structure. In so far as these are identified as spirals, they have been treated analogously to the spiral galaxies in the \textsl{\textsc{fieldgalaxy}} and \textsl{\textsc{groupgalaxy}} samples. The attenuation corrections and SFR estimates may, however, be less accurate for these perturbed systems.}. \newline
 
As expected \citep[e.g.][]{BARTON2000,ROBOTHAM2013,DAVIES2015}, we find that the sSFR of merging systems is, on average, enhanced, even with respect to field spirals as shown by the green stars in Fig.~\ref{fig_GROUPFIELDCP_1}. There seems to be little stellar mass dependence of this enhancement over the range of $9.5 \le \mathrm{log}(M_*/M_{\odot}) \le 10.5$. Above $\mathrm{log}(M_*/M_{\odot}) = 10.5$, however, the sSFR of merging spirals appears to no longer only be enhanced, with some galaxies also showing a strongly suppressed sSFR with respect to the \textsl{\textsc{fieldgalaxy}} reference sample, resulting in a huge spread in the sSFR of merging spiral galaxies at these masses.\newline
  
For spiral galaxies which are members of close pairs of galaxies but not merging, we find that the median $\psi_*$--$M_*$ relation, shown in blue in Fig.~\ref{fig_GROUPFIELDCP_1}, is comparable to the field reference relation, and may even be elevated for galaxies with a stellar mass below $\sim10^{10}M_{\odot}$. Fig.~\ref{fig_GROUPFIELDCP_2} shows the distributions of $\Delta\mathrm{log}(\psi_*)$ for the \textsl{\textsc{cpgalaxy}} (blue) and \textsl{\textsc{fieldgalaxy}} (black) samples in the low (left column) and high (right column) stellar mass ranges.  Performing a Peto test comparing the distributions of $\Delta \mathrm{log}(\psi_*)$ of the \textsl{\textsc{cpgalaxy}} and \textsl{\textsc{fieldgalaxy}} samples in the low stellar mass range, one finds that the null hypothesis is not rejected ($p=0.87$; see Table.~\ref{tab_GFCP} for the results of Peto tests comparing the \textsl{\textsc{cpgalaxy}}, \textsl{\textsc{groupgalaxy}}, and \textsl{\textsc{fieldgalaxy}} samples in both ranges of stellar mass), however, a closer inspection does find the relative weight of the tail of galaxies with suppressed sSFR in the \textsl{\textsc{cpgalaxy}} sample to be greater than for the \textsl{\textsc{fieldgalaxy}} sample, albeit slightly less so than for the \textsl{\textsc{groupgalaxy}} sample, as well as finding the \textsl{\textsc{cpgalaxy}} sample to be more skewed towards high values of $\Delta \mathrm{log}\psi_*$, i.e. increased sSFR, including a slight shift in the position of the peak. In line with these findings, we also find the distributions of $\Delta \mathrm{log}(\psi_*)$ of the \textsl{\textsc{cpgalaxy}} sample and the \textsl{\textsc{groupgalaxy}} sample to differ significantly ($p\lesssim10^{-5}$) in the low stellar mass range.\newline

In the high stellar mass range we find the distribution of $\Delta\mathrm{log}(\psi_*)$ for the \textsl{\textsc{cpgalaxy}} to be peaked at the position of the peak of the \textsl{\textsc{fieldgalaxy}} sample. However, in this mass range, the population of galaxies with strongly suppressed sSFR in the \textsl{\textsc{cpgalaxy}} sample is fully comparable to that of the \textsl{\textsc{groupgalaxy}} sample. Nevertheless, the \textsl{\textsc{cpgalaxy}} sample is slightly skewed towards increased sSFR with respect to the \textsl{\textsc{fieldgalaxy}} sample. As a result, although the distributions of $\Delta \mathrm{log}\psi_*$ for both samples
differ significantly ($p<10^{-5}$), the median $\psi_* - M_*$ relation, lying between that of the \textsl{\textsc{groupgalaxy}} sample and the \textsl{\textsc{fieldgalaxy}} sample, as shown in Fig.~\ref{fig_GROUPFIELDCP_1}, is generally comparable to the field reference relation. Regardless of the increased similarity of the \textsl{\textsc{cpgalaxy}} and \textsl{\textsc{groupgalaxy}} samples in the range of $M_* \ge 10^{10} M_{\odot}$, the distributions of $\Delta\mathrm{log}(\psi_*)$ are found to differ significantly, as in the low mass range, in summary retroactively justifying the exclusion of close pairs from our analysis.\newline

Overall, we find that merging activity has a strong affect on the sSFR of spiral/disk galaxies, leading to a significant enhancement of the sSFR in spiral galaxies with $M_* \le 10^{10.5} M_{\odot}$ and to a very large scatter above this mass. This effect is markedly stronger than any more general environmental impact. The sSFR of galaxies in close pairs appears to be only marginally affected by the fact of having a neighbor in the direct vicinity. However, for the adopted definition of a close pair (a neighbor within $50\,\mathrm{kpc}\,h^{-1}$ projected distance and $1000\,\mathrm{km} s^{-1}$), it is likely that a large fraction of the close pairs identified in this manner are by no means interacting, diluting possible effects.\newline

\section{Spiral Galaxies in the Group Environment: Centrals and Satellites}\label{SATCENT}
Hitherto we have considered the \textsl{\textsc{groupgalaxy}} sample as a whole, i.e. we have considered all spiral/disk-dominated galaxies in galaxy groups, regardless of their being a satellite galaxy in the group, or of being the central galaxy of the group. However, as mentioned in the introduction, this distinction may be fundamental to the ability of galaxies to accrete gas and fuel on-going star-formation. In the following, we therefore separate the \textsl{\textsc{groupgalaxy}} sample into central and satellite spiral group galaxies as described in Section~\ref{SAMPLESELECTION_SPIRALS_SATCENT}.\newline

\subsection{Group Central Spiral Galaxies}\label{CENT}
Using the $\psi_* - M_*$ relation of the \textsl{\textsc{fieldgalaxy}} sample, i.e. largely isolated central spiral galaxies, as a reference, we consider the impact of the group environment on the star formation, respectively the $\psi_* - M_*$ relation, of group central spiral galaxies. As shown in Fig.~\ref{fig_SATCENTSPF}, the sample of group central spiral galaxies is skewed toward high mass galaxies, as expected given the nature of these objects as the central galaxy of a group encompassing at least three galaxies with $M_* \ge 10^{9.5} M_{\odot}$. In terms of the median $\psi_* - M_*$ relation, however, that of the group centrals is very close  to that of our reference sample over the full mutual range in $M_*$, as shown in red in Fig.~\ref{fig_SATCENT_1}. Nevertheless, there is a hint that the slope in the relation may be slightly steeper for group central spiral galaxies than for the \textsl{\textsc{fieldgalaxy}} sample, with galaxies at $M_* \approx 10^{10.4}M_{\odot}$ having slightly higher median sSFR than the reference relation, while those with $M_* \gtrsim 10^{11}$ appear to have a minimally suppressed median value of $\psi_*$.\newline

Considering the distributions of $\Delta \mathrm{log}(\psi_*)$ (shown in red in Fig.~\ref{fig_SATCENT_2} for the low and high stellar mass ranges, respectively), the centrals strongly resemble the \textsl{\textsc{fieldgalaxy}} sample, with a pronounced peak at the position of the peak of the \textsl{\textsc{fieldgalaxy}} sample and a negligible tail of galaxies with strongly suppressed sSFR with respect to the median relation of the \textsl{\textsc{fieldgalaxy}} sample ($13$\% in the stellar mass range $M_* \ge 10^{10} M_{\odot}$). As a total of four central spiral galaxies have masses below $M_*=10^{10} M_{\odot}$ we ignore the low stellar mass range in our comparisons. As summarized in Table~\ref{tab_SATCEN}, the null hypothesis that the distributions of $\Delta \mathrm{log}(\psi_*)$ for the group central spiral galaxies and the \textsl{\textsc{fieldgalaxy}} sample in the mass range $M_* > 10^{10} M_{\odot}$ are statistically similar cannot be discarded.\newline

In summary, our analysis finds that the sSFR of group central spiral galaxies is comparable to that of largely isolated field central spiral galaxies matched in stellar mass, with almost no evidence of any influence of the group environment on the sSFR of central spiral galaxies. 
We will return to this result and our findings on the evolution of the $\psi_* - M_*$ relation of the central spiral galaxies of the \textsl{\textsc{fieldgalaxy}} sample in a subsequent paper in this series considering the gas-fuelling of central spiral galaxies. In this paper, we will continue by focussing on the star formation and gas-fuelling of satellite spiral galaxies.\newline

\subsection{Group Satellite Spiral Galaxies}
\begin{figure}
\plotone{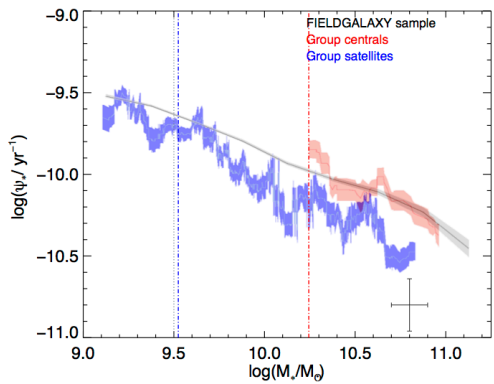}
\caption{$\psi_*$ as a function of $M_*$ for the \textsl{\textsc{fieldgalaxy}} sample (grey), the satellite galaxies in the \textsl{\textsc{groupgalaxy}} sample (blue), and the central galaxies in the \textsl{\textsc{groupgalaxy}} sample (red). For the satellite and central galaxies in the \textsl{\textsc{groupgalaxy}} sample the solid line shows the median value in a sliding tophat bin containing 40, respectively 25, galaxies . The wider shaded area indicates the uncertainty on the median. For the \textsl{\textsc{fieldgalaxy}} sample the relation is shown in bins of equal size in $M_*$ as in Fig.~\ref{fig_GROUPFIELDCP_1}, with the shaded area again corresponding to the uncertainty in the median. The stellar mass limit above which the samples can be considered mass complete ($M_* =10^{9.5}M_{\odot}$) is indicated by a dotted gray line. Colored dash-dotted lines indicate the stellar mass above which the galaxies in the moving tophat bins are all above the mass completeness limit. The $\psi_*$--$M_*$ relations for the \textsl{\textsc{fieldgalaxy}} sample as well as for the central and satellite subsamples of the \textsl{\textsc{groupgalaxy}} sample, including uncertainties of the median, are available online as 'data behind the figure'.}
\label{fig_SATCENT_1}  
\end{figure}

Given the expected differences in the physical circumstances of central spiral galaxies (in general) and satellite spiral galaxies vis \`a vis their ability to accrete gas, we consider the $\psi_* - M_*$ for satellite spiral/disk-dominated galaxies, contrasting it with our reference relation defined by the \textsl{\textsc{fieldgalaxy}} sample. As shown in Fig.~\ref{fig_SATCENT_1} the median $\psi_* - M_*$ relation for satellite spiral galaxies (shown in blue) is suppressed with respect to the reference relation over the full range in stellar mass. The suppression of the median is found to be moderate,  increasing very mildly from $\sim 0.1 - 0.2\,$dex for $M_* < 10^{9.75} M_{\odot}$  to $\sim0.2-0.3\,$dex at a given $M_*$ for $M_* \gtrsim 10^{10} M_{\odot}$. Overall, the offset of the $\psi_* - M_*$ relation for satellite spiral galaxies from the \textsl{\textsc{fieldgalaxy}} sample reference relation appears to be largely independent of stellar mass, albeit possibly with a very weak dependence in the sense that the offset is smaller at lower stellar mass.  \newline

\begin{figure}
\plotone{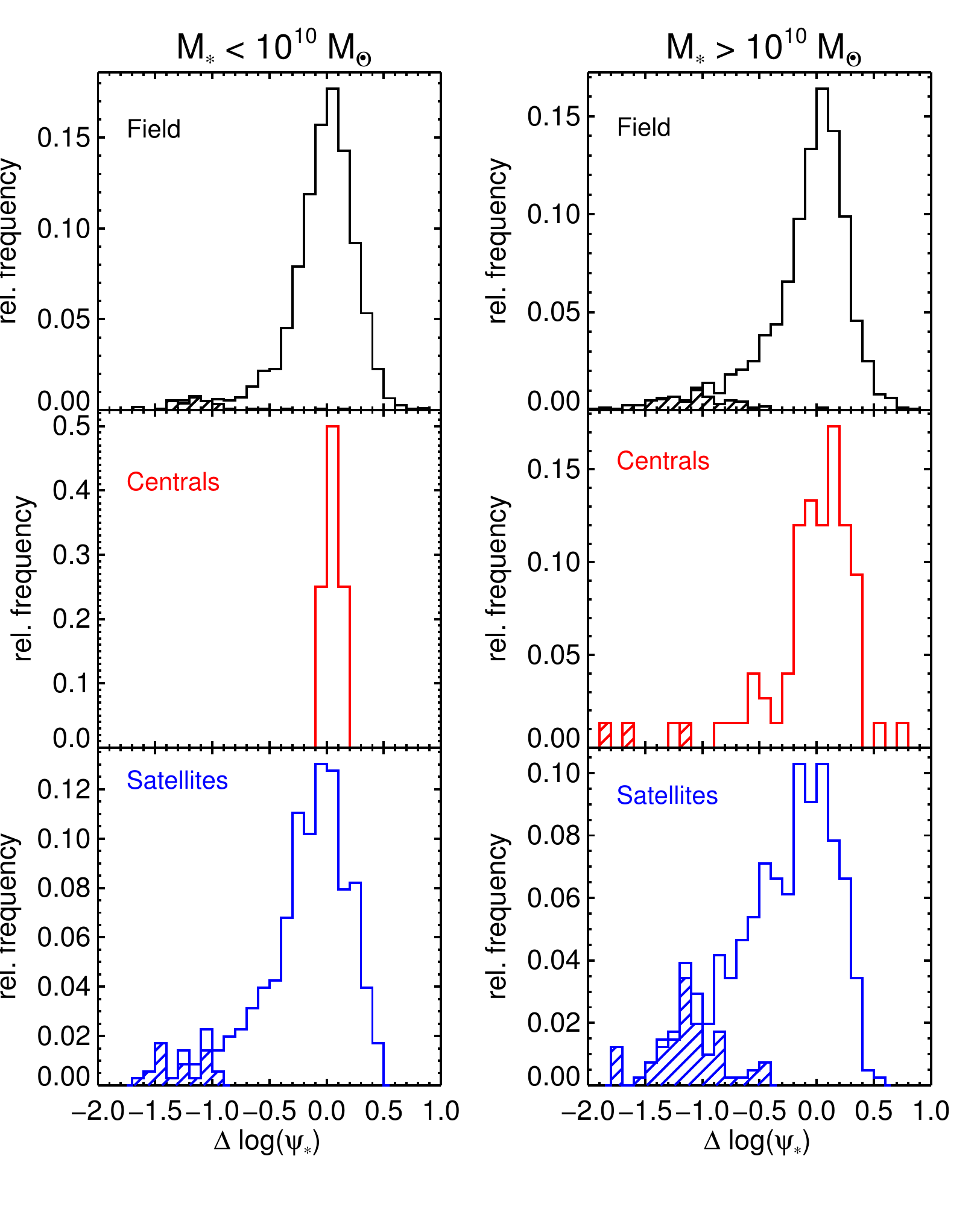}
\caption{Histograms of the distribution of $\Delta \mathrm{log}(\psi_*)$ for field, satellite, and central spiral galaxies with $M_* < 10^{10} M_{\odot}$ (left) and $M_* > 10^{10} M_{\odot}$ (right) respectively. The distribution of upper limits is indicated by the line-filled histograms. The distributions of $\Delta\mathrm{log}(\psi_*)$ for the \textsl{\textsc{fieldgalaxy}} sample as well as for the central and satellite subsamples of the \textsl{\textsc{groupgalaxy}} sample are available online as 'data behind the figure'.}
\label{fig_SATCENT_2}  
\end{figure}

Considering the distributions of $\Delta\mathrm{log}(\psi_*)$ for the satellites (blue) with $M_* < 10^{10}M_{\odot}$ (left column), respectively $M_* \ge 10^{10}M_{\odot}$ (right column) as shown in Fig.~\ref{fig_SATCENT_2}, and comparing with the distributions of the \textsl{\textsc{fieldgalaxy}} sample (black) shown in the same figure, one finds that the satellites' distributions show a strong peak at the position of the peak in the \textsl{\textsc{fieldgalaxy}} sample distribution. However, the population of galaxies with strongly suppressed sSFR is larger in the satellite sample (in both stellar mass ranges) than in the \textsl{\textsc{fieldgalaxy}} sample. In the low stellar mass bin, $18$\% of the satellite galaxies have $\Delta\mathrm{log}(\psi_*) < -0.5$, compared to $7$\% for the \textsl{\textsc{fieldgalaxy}} sample, while the difference in the distributions is even more pronounced in the high stellar mass range, with $32$\% of the satellites having $\Delta\mathrm{log}(\psi_*) < -0.5$, compared to $18$\% for the \textsl{\textsc{fieldgalaxy}} sample. This is mirrored in the results of Peto tests comparing the distributions (the results of Peto tests comparing the distributions of satellites, centrals, and the \textsl{\textsc{fieldgalaxy}} sample are summarized in Table~\ref{tab_SATCEN}), which find them to differ significantly in both ranges of $M_*$\footnote{As for the \textsl{\textsc{fieldgalaxy}} and \textsl{\textsc{groupgalaxy}} samples, the lack of stellar mass dependence of the offset of the median $\psi_*$--$M_*$ for satellite galaxies, combined with the similarity of the relative distributions of $M_*$ within the broad mass ranges considered lends confidence in the comparability of the distributions of $\Delta\mathrm{log}(\psi_*)$ for the \textsl{\textsc{fieldgalaxy}} and satellite spiral samples.}.\newline

Overall, we find the majority ($\gtrsim 70$\%) of satellite spiral galaxies to be forming stars at a rate comparable to their counterparts in the \textsl{\textsc{fieldgalaxy}} sample. The mild suppression of the median $\psi_*$--$M_*$ relation for these group satellite spiral galaxies can be attributed to a minority population ($\lesssim 30$\%) of galaxies with strongly suppressed sSFR with respect to the \textsl{\textsc{fieldgalaxy}} reference.\newline

\begin{table}
\centering
\begin{threeparttable}
\caption{Comparison of Satellites, Centrals, and the \textsl{\textsc{fieldgalaxy}} Sample \label{tab_SATCEN}} 
\begin{tabular}{|cc|c|c|c|}
\cline{3-5}
\multicolumn{2}{c|}{} &\rotatebox{90}{Centrals\ }& \rotatebox{90}{Satellites\ }&\rotatebox{90}{FIELD\ }\\
\hline
\multirow{2}{*}{Centrals} & \multicolumn{1}{c|}{$\mathrm{log}(M_*)\le 10$} & 1.&0.18&0.69\\
 & \multicolumn{1}{c|}{$\mathrm{log}(M_*)> 10$} & 1.&$\le10^{-5}$&0.19 \\
 \hline
\multirow{2}{*}{Satellites} & \multicolumn{1}{c|}{$\mathrm{log}(M_*)\le 10$} & 0.18&1.&$\le10^{-5}$\\
 & \multicolumn{1}{c|}{$\mathrm{log}(M_*)> 10$} & $\le10^{-5}$&1.&$\le10^{-5}$ \\
 \hline
 \multirow{2}{*}{FIELD} & \multicolumn{1}{c|}{$\mathrm{log}(M_*)\le 10$} & 0.69&$\le10^{-5}$&1.\\
 & \multicolumn{1}{c|}{$\mathrm{log}(M_*)> 10$} & 0.19&$\le10^{-5}$&1. \\
\hline
\end{tabular}
\begin{tablenotes}[flushleft]
\item{Significance ($p$) values of Peto tests performed between the central and satellite galaxy sub-samples of the \textsl{\textsc{groupgalaxy}} samples and the reference \textsl{\textsc{fieldgalaxy}} sample. For each combination Peto tests have been performed in two disjoint bins of stellar mass spilt at $M_*=10^{10} M_{\odot}$. }
\end{tablenotes}
\end{threeparttable}

\end{table}

\section{Star-formation and Star-formation histories of Satellite Spiral Galaxies} \label{SFR_SAT}
In the previous section we have shown that the vast majority of satellite spiral/disk galaxies display sSFR comparable to those of their field counterparts, with the observed moderate suppression of the median $\psi_* - M_*$ relation for group satellite spiral galaxies being caused by a minority of galaxies with strongly suppressed sSFR, as shown in Fig.~\ref{fig_SATCENT_2}. Thus, it appears that while the group environment in terms of galaxy--IHM interactions has a strong impact on a minority of satellite spiral/disk galaxies, the majority population remains unaffected and behaves nigh identically to their central counterparts in the field.\newline 

This observed similarity between the sSFR of satellite and field central spiral galaxies is highly surprising, since, as outlined in Sect.~\ref{INTRODUCTION}, satellite galaxies are expected to be largely unable to accrete gas
 to resupply/fuel star formation activity, while field central galaxies are thought to experience on-going gas-fuelling. This is exacerbated by the fact that, as shown in Fig.~\ref{fig_SATCENTSPF}, the spiral fraction as a function of stellar mass for satellite galaxies has only decreased by $30$--$40$\% with respect to galaxies in the field, and accordingly a substantial fraction of spiral satellites have likely resided in the group environment as satellite galaxies for several Gyr. As the majority of  the spiral satellite galaxies ($\gtrsim 70$\%) display sSFR comparable to that of their field counterparts, it seems inevitable that a substantial fraction of these actively star forming spiral satellites have resided in the group environment for an extended period. The question one has to consider is thus whether gas-fuelling is on-going for group satellite spiral galaxies. \newline

As a first step to answering this question we consider the gas exhaustion timescales for spiral galaxies with stellar masses in the range of $9.5 \le \mathrm{log}(M_*/M_{\odot}) < 10$ and $10 \le \mathrm{log}(M_*/M_{\odot})$, respectively. We cannot do this directly for the GAMA sample as measurements of the gas masses are not available, however, we can make use of the relation between stellar mass and gas mass for late-type galaxies compiled by \citet{PEEPLES2014} and the $\psi_* - M_*$ relation for the \textsl{\textsc{fieldgalaxy}} sample presented in Fig.~\ref{fig_FIELDSSFRSM} to obtain a conservative estimate of the exhaustion timescales
\begin{equation}
\tau_{\mathrm{exhaust}} = M_{\mathrm{gas}}/SFR\,.
\label{eq_tauexh1}
\end{equation}
 Adopting this approach, one finds values of $2.8\,$Gyr  and $2.7\,$Gyr for the mass ranges $9.5 \le \mathrm{log}(M_*/M_{\odot}) < 10$ and $10 \le \mathrm{log}(M_*/M_{\odot})$, respectively\footnote{We have used stellar mass values corresponding to the median stellar mass in each range; $M_*=10^{9.75}M_{\odot}$ and $M_* = 10^{10.3}M_{\odot}$, respectively.}. Although quite substantial, these timescales represent the timescale on which \textit{all} gas (atomic and molecular hydrogen) of the galaxy has been consumed by star formation alone, also ignoring any potential outflows of gas from the galaxy which may considerably reduce the actual exhaustion timescale (see also \citealt{MCGEE2014}). Thus, it seems difficult to explain the lack of a large shift in the star formation activity of the majority of the satellite spiral galaxies simply in terms of the depletion of the gas reservoir (even if these were to be retained in a form comparable to the largely non-grouped galaxies in the sample of \citealt{PEEPLES2014}). Nevertheless, these exhaustion timescales of $\sim3\,$Gyr are not really decisive, as they are only comparable to the expected time spent as a satellite by the galaxies considered.\newline

To approach the question in a more quantitative manner, we therefore consider a number of simple parameterized star formation histories (SFH) for galaxies in the group environment which can be readily related to their gas-cycle (see section~\ref{GASFUELLING}) and have been chosen to bracket the range of plausible SFH for these objects. These SFH are schematically illustrated in Fig.~\ref{fig_models}. For the SFH in these models we predict the distributions of $\Delta\mathrm{log}(\psi_*)$ and compare these with our empirical results. This approach enables us to identify the SFH elements most applicable to our data, and to subsequently (see Section~\ref{GASFUELLING}) interpret our results in the physical context of the gas-cycle of galaxies, including quantitative estimates of the in- \& outflows of gas to and from the galaxy.\newline

Full details of our modelling procedure are provided in Appendix~\ref{APPEND_MODEL}. In brief, however, we proceed by creating samples of galaxies infalling into groups and becoming satellites which we evolve forwards in time to observation at $z=0.1$ following the parameterized SFH of our models. This approach requires knowledge of the time a galaxy has been a satellite, i.e the infall time, as well as of the stellar mass and SFR at the time of infall, and we have, as far as possible, adopted an empirically driven approach to determining these quantities. In creating our samples of infalling galaxies we Monte Carlo sample the $z\approx0.1$ \textsl{\textsc{FIELDGALAXY}} sample distributions of stellar mass and SFR (in bins of $M*$) and evolve the galaxy back to its infall time following the empirical parameterization of the SFMS presented by \citet{SPEAGLE2014}. An empirical determination of the infall time distribution, however, is not possible. Therefore, in determining the distribution of infall times for our model samples we have made use of the distribution of infall times found for satellite galaxies in the mock GAMA-survey light-cones produced using the Millennium dark matter simulation \citep{SPRINGEL2005} and the GALFORM semi-analytic galaxy formation model \citep{BOWER2006,MERSON2013}.\newline

As shown in Fig.~\ref{fig_SATCENTSPF}, the spiral fraction of satellite galaxies is $30$-−$40$\% lower than that of the \textsl{\textsc{fieldgalaxy}} sample, and on average we find a satellite spiral fraction of $30$\%. The observed decrease in spiral/disk fraction is often linked to the more frequent occurrence of galaxy-galaxy interactions in the group environment, which can morphologically transform disk(-dominated) galaxies to more bulge-dominated systems. However, quenching of star-formation in spiral (satellite) galaxies may also give rise to an apparent morphological transformation even without any galaxy-galaxy interaction, as a result of different degrees of fading for the largely passive bulge and the (previously) star forming disk, and may lead to disk systems no longer being identified as such \citep[e.g.][]{CAROLLO2016}. Although the selection method of \citet{GROOTES2014} is designed to allow quenched systems to enter the selection and we expect the impact of fading to be limited as previously discussed in Section~\ref{SAMPLESELECTION_SPIRALS}, we have nevertheless adopted a very conservative approach to account for this possibility, i.e. in drawing infall times from the distribution found in the mock GAMA-lightcones for our modelling purposes we assume that the spiral group member satellites correspond to the 30\% youngest group members and draw only from the corresponding fraction of the infall time distribution\footnote{This approach is conservative in the sense that it places the smallest requirements on the gas reservoirs of the satellite spiral galaxies.}. A full discussion of the modelling is provided in Appendix~\ref{APPEND_MODEL}.\newline

\subsection{One-Parameter Models}\label{SFR_SAT_1pm}
\subsubsection{The 'Infall-Quenching' Model}
\begin{figure*}
\plotone{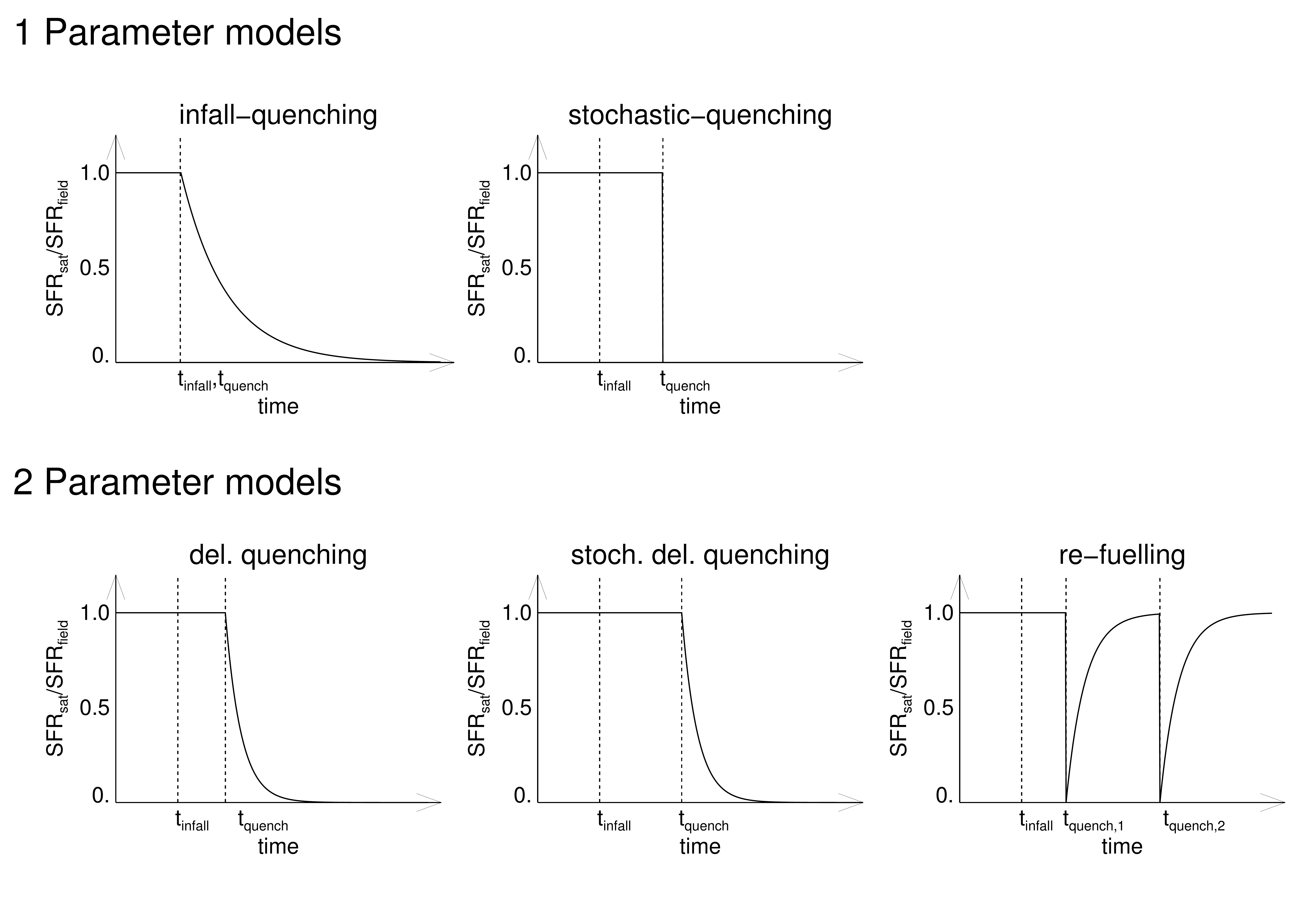}
\caption{Schematic depictions of the parameterized SFH models (relative to a comparable field galaxy) considered, split into one and two parameter families as shown. Each model name is shown on the plot above the depiction of its characteristic SFH. In all cases the dashed vertical lines shows the infall time $t_{\mathrm{infall}}$, i.e the time at which the galaxy became a satellite for the first time, as well as  the incidence of a 'quenching event', i.e either the complete shut off of star formation (stochastic quenching model, re-fuelling model) or the onset of a gradual decline (the other three models).}
\label{fig_models}  
\end{figure*}

The simplest model is the 'infall-quenching' model shown in Fig.~\ref{fig_models}. In this model,
 the star formation rate of a galaxy declines exponentially on a timescale $\tau_{\mathrm{quench}}$ upon the galaxy becoming a satellite at $t_{\mathrm{infall}}$, i.e.
 \begin{equation} 
\mathrm{SFR}(t) = \mathrm{SFR}(t_{\mathrm{infall}}) e^{-(t - t_{\mathrm{infall}})/\tau_{\mathrm{quench}}}.
\label{eq_1pexp}
\end{equation}  

The predicted distributions of the present day star formation of group satellite spiral/disk galaxies for this model in the stellar mass ranges $\mathrm{log}(M_*/M_{\odot}) \ge 10$ and $9.5 \le \mathrm{log}(M_*/M_{\odot}) < 10$ are shown in the top left panel of Figs.~\ref{fig_1pmod_hm} \& \ref{fig_1pmod_lm}, respectively, overlaid on the observed distribution in that mass range. In order to simplify the characterization of the distribution of $\Delta\mathrm{log}(\psi_*)$  we make use of three robust characteristics , the first quartile, the median, and the third quartile. For the observed distributions each of these is over-plotted. The bottom left panels of Figs.~\ref{fig_1pmod_hm} \& \ref{fig_1pmod_lm} show the locations of three characteristics of the $\Delta\mathrm{log}(\psi_*)$ distribution as a function of $\tau_{\mathrm{quench}}$ for the mass ranges $\mathrm{log}(M_*/M_{\odot}) \ge 10$ and $9.5 \le \mathrm{log}(M_*/M_{\odot}) < 10$, respectively. 
It is immediately apparent  that the infall-quenching model is incapable of simultaneously reproducing the locations of the characteristics of the distribution. Furthermore, the shape of the full distribution obtained for the value of best reproducing the median is very different from that of the data. In particular, the peak of the model distribution is shifted towards lower values of $\Delta\mathrm{log}(\psi_*)$, while displaying a smaller dispersion than the observed distribution.\newline

\subsubsection{The Stochastic Quenching Model}
In the second one parameter model, referred to as the 'stochastic-quenching' model, a galaxy becoming a satellite at time $t_{\mathrm{infall}}$ continues to form stars as if it were still a field (central) galaxy. However, with a probability per unit time $P_{\mathrm{quench}}$, the star formation of the galaxy is instantaneously completely shut off at a time $t_{\mathrm{quench}}$ with $t_{\mathrm{quench}} > t_{\mathrm{infall}}$, and the galaxy remains dormant thereafter, i.e.
\begin{equation}
\mathrm{SFR}(t) = \begin{cases} 
\mathrm{SFR}_{\mathrm{field}}(t) & \mathrm{for}\; t  < t_{\mathrm{quench}}\\
0 & t  \geq t_{\mathrm{quench}}\,. \\ 
\end{cases}
\label{eq_1pstoch}
\end{equation}\newline

\begin{figure*}
\plotone{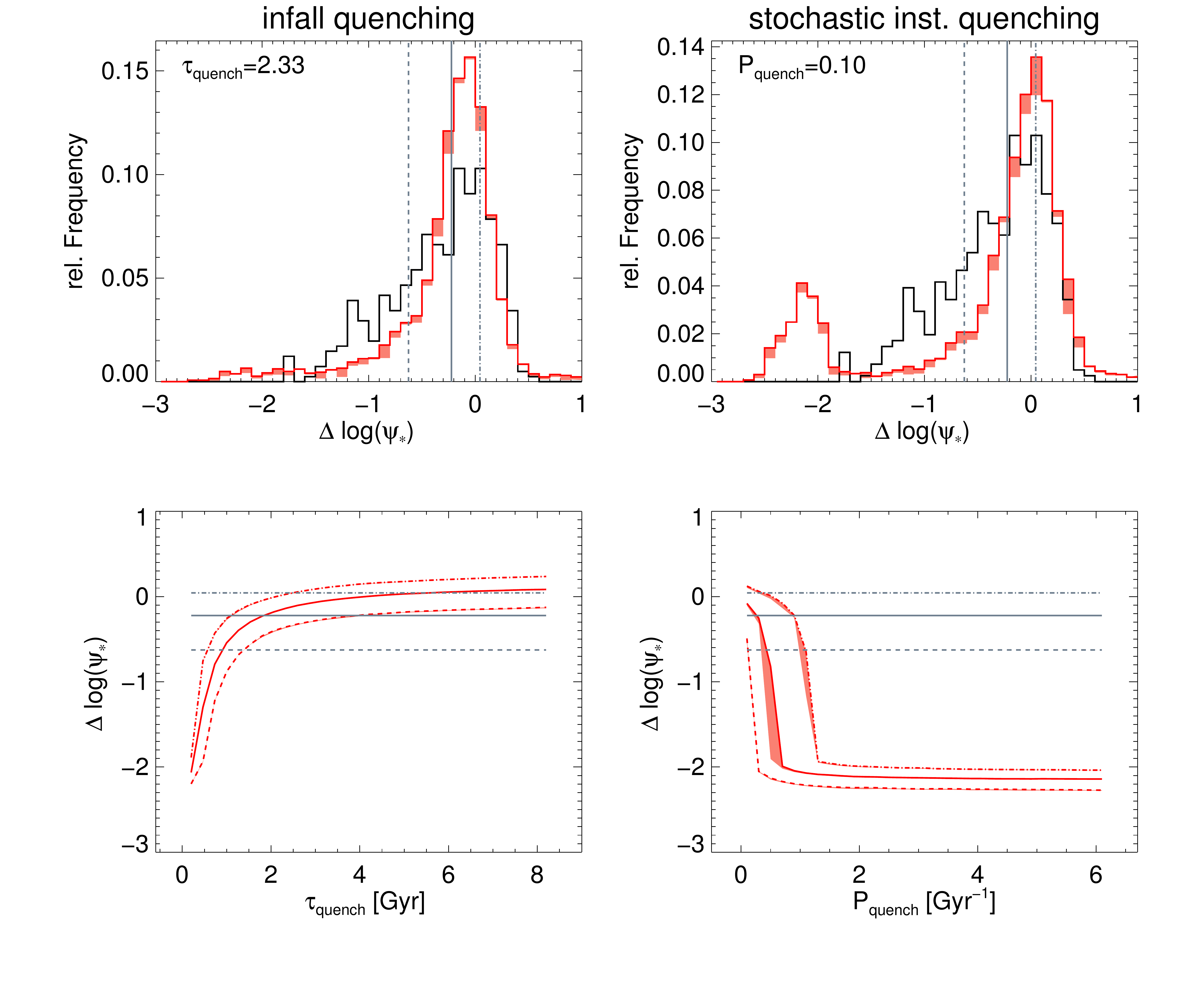}
\caption{\textbf{Top:} Observed distribution of $\Delta \mathrm{log}(\psi_*)$ for the stellar mass range $M_* \ge 10^{10} M_{\odot}$ in gray with the first quartile, median, and third quartiles the observed distribution indicated by a dashed, solid, and dashed-dotted vertical, respectively. The average $\Delta \mathrm{log}(\psi_*)$ distribution as obtained from 50 realizations of the infall-quenching model (right), respectively of the stochastic quenching model (left), is shown in red. The red shaded regions correspond to range between the 16$^{\mathrm{th}}$ and 84$^{\mathrm{th}}$ percentile in each bin in $\Delta \mathrm{log}(\psi_*)$ as found for the 50 realizations considered.  The parameter value of the model depicted (chosen to best reproduce the position of the observed median) is shown at top left. \textbf{Bottom:} Position of the first quartile, median and third quartile (from bottom to top) of the  $\Delta \mathrm{log}(\psi_*)$ distribution of the model (infall-quenching left, stochastic quenching right) as a function of the model parameter in the mass range $M_* \ge 10^{10} M_{\odot}$ in red. The red shaded regions indicate the range between the 16$^{\mathrm{th}}$ and 84$^{\mathrm{th}}$ percentile at each trial value of the model parameter as found from 50 realizations. The locations of the three characteristics of the observed distribution are over-plotted in gray, with the first quartile, median and third quartile indicated by dashed, solid, and dashed-dotted lines, respectively. }
\label{fig_1pmod_hm}  
\end{figure*}

\begin{figure*}
\plotone{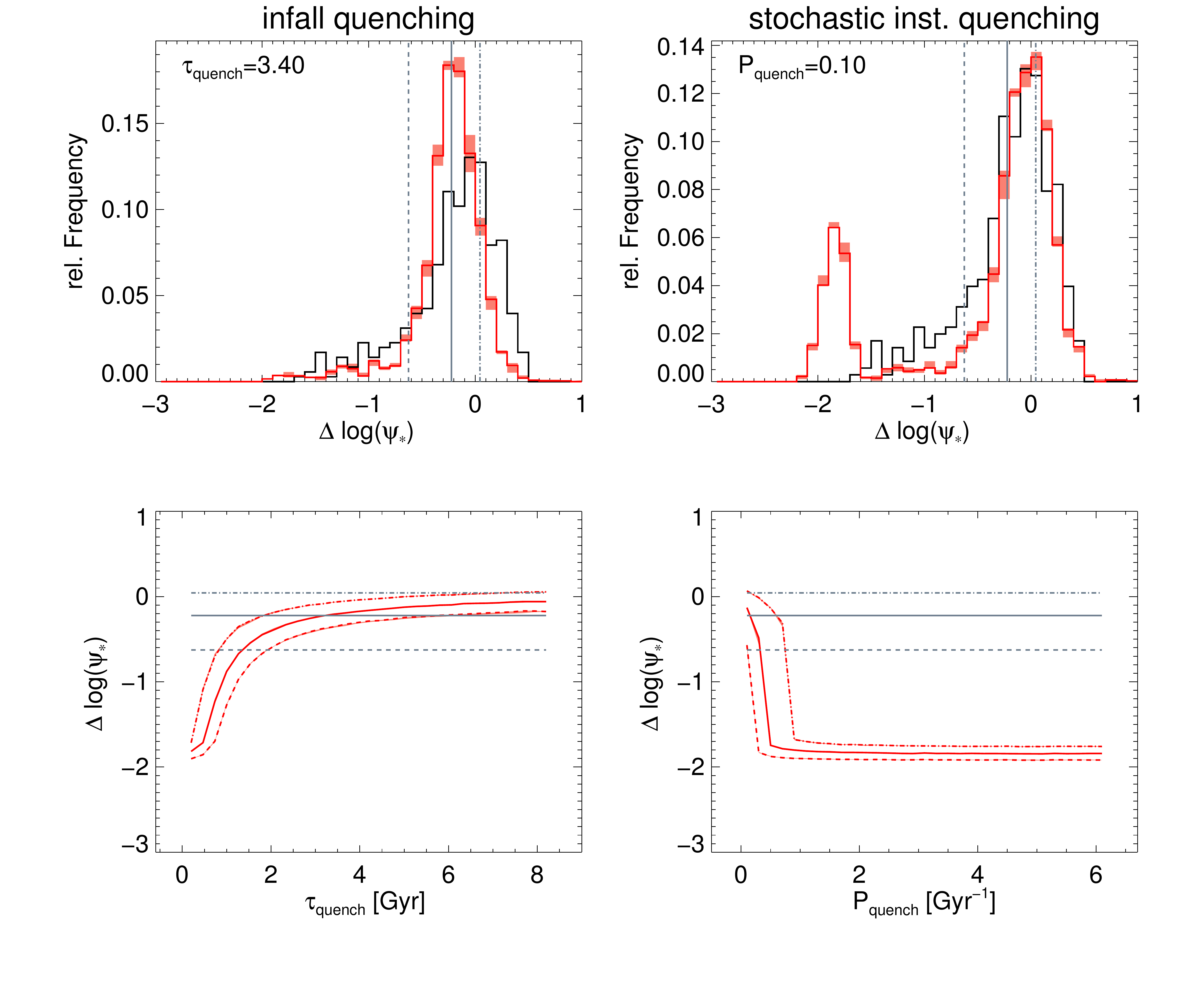}
\caption{ As Fig.~\ref{fig_1pmod_hm} but for the stellar mass range $10^{9.5} M_{\odot} \le M_* < 10^{10} M_{\odot}$.}
\label{fig_1pmod_lm}  
\end{figure*}

As for the 'infall-quenching' model, the locations of the second quartile, the median, and the third quartile of the predicted distribution of $\Delta \mathrm{log}(\psi_*)$, here as a function of  $P_{\mathrm{quench}}$, are shown in Figs.~\ref{fig_1pmod_hm} \& \ref{fig_1pmod_lm}. For very low values of $P_{\mathrm{quench}}$ the stochastic model is nearly able to reproduce the location of all three characteristics of the distribution simultaneously. However, considering the full predicted distribution, one finds that, while the location of the main peak is correct, the 'stochastic-quenching' model gives rise to a far too large population of galaxies with very strongly suppressed sSFR, and lacks the moderately suppressed galaxies found in the observational data.\newline

In summary, neither of the one parameter models considered is thus capable of satisfactorily reproducing the observed distributions of $\Delta \mathrm{log}(\psi_*)$. \newline

\subsection{Two-Parameter Models}\label{SFR_SAT_2pm}
The family of two parameters models depicted in Fig.~\ref{fig_models} encompasses three models. Unlike for the one parameter models, for two parameter models, the locations of the characteristics of the $\Delta\mathrm{log}(\psi_*)$ distribution can no longer easily be directly depicted as a function of the parameter values. Instead, we consider the topology of the expression 
\begin{equation}
\Delta q_{i} (p_1,p_2) =|q_{i,\mathrm{obs}} - q_{i,\mathrm{mod}} (p_1,p_2)|\,,
\label{eq_deltaq}
\end{equation}
where $q_{i,\mathrm{obs}}$ represents the $i^{\mathrm{th}}$ quartile of the observed distribution of $\Delta \mathrm{log}(\psi_*)$ and $q_{i,\mathrm{mod}}(p_1,p_2)$ represents the $i^{\mathrm{th}}$ quartile of the model distribution for the parameters $p_1$ and $p_2$. This is done separately for each characteristic, and a parameter combination corresponding to a good fit will simultaneously minimize the expressions for all three characteristics.\newline

In the following, to quantify the ability of a model to simultaneously reproduce the characteristics of the observed $\Delta\mathrm{log}(\psi_*)$ distribution, we  will consider the quantity
\begin{equation}
Q_{i}(p_1,p_2) = \begin{cases} 
\left[ 1 - \Delta q_{i}(p_1,p_2) \right] \cdot 0.3^{-3} & \mathrm{for}\; \Delta q_{i}(p_1,p_2) \le 0.3\\
0 & \mathrm{otherwise} \\
\end{cases}
\end{equation}
for each characteristic and formulate a composite figure of merit for the performance of the model as
\begin{equation} 
Q(p_1,p_2) = \prod\limits_{i}^{} Q_{i}(p_1,p_2)\,,
\label{eq_Qtot}
\end{equation} 
which can take values between $1$, for a perfect recovery of all characteristics, and $0$, for strong a discrepancy between model and observed distributions of $\Delta \mathrm{log}(\psi_*)$ (even in only one characteristic).\newline

Figs.~\ref{fig_2pmod_hm} \& \ref{fig_2pmod_lm} show the topologies of the three characteristics ($\Delta q_i(p_1,p_2)$), as well as the whole distribution of $\Delta \mathrm{log}(\psi_*)$ for the preferred values of $p_1$ and $p_2$ overlaid on the observed distribution, for each of the models in the high and low stellar mass ranges, respectively. Fig.~\ref{fig_Qm} shows the topology of the figure of merit $Q(p_1,p_2)$ for each model in both stellar mass ranges, while the preferred parameter values, as well as the attained values of $Q(p_1,p_2)$,  are provided in table~\ref{tab_fitpar}. In the following, we will discuss all three models individually.\newline

\begin{table*}
\centering
\begin{threeparttable}
\caption{Summary of preferred model parameter values for 2 parameter models\label{tab_fitpar}}
\begin{tabular}{ccccccccccc}
\multirow{2}{*}{$t_{\mathrm{infall}}$ dist.} & \multirow{2}{*}{$M_*$ range} & \multicolumn{3}{c}{Del. Q.} & \multicolumn{3}{c}{Stoch. Del. Q.} &\multicolumn{3}{c}{Refuel.} \\
& & $t_{\mathrm{delay}}$ [Gyr] & $\tau_{\mathrm{quench}}$ [Gyr] & $Q$ & $P_{\mathrm{quench}}$ [Gyr$^{-1}$] & $\tau_{\mathrm{quench}}$ [Gyr] & $Q$ & $P_{\mathrm{quench}}$ [Gyr$^{-1}$] & $\tau_{\mathrm{fuel}}$ [Gyr] & $Q$ \\
\hline
\multirow{2}{*}{con.} & $M_* < 10^{10}M_{\odot}$ & 2.9 & 0.5 & 0.87 & 0.3 & 1.5 & 0.79 & 0.7 & 0.58 & 0.93 \\
 & $M_* \ge 10^{10}M_{\odot}$ & 2.5 & 0.5 & 0.83 & 0.3 & 0.9 & 0.78 & 0.9 & 0.85 & 0.92 \\
\multirow{2}{*}{full} & $M_* < 10^{10}M_{\odot}$ & 4.7 & 3.7 & 0.63 & 0.1 & 3.1 & 0.71 & 0.5 & 0.45 & 0.79 \\
 & $M_* \ge 10^{10}M_{\odot}$ & 4.9 & 1.5 & 0.75 & 0.1 & 1.5 & 0.84 & 0.98 & 0.5 & 0.92 \\
 \hline
 \hline
 \end{tabular}
\begin{tablenotes}
\item{preferred parameter values and associated figure of merit $Q$ for the delayed quenching model (Del. Q.), the stochastic delayed quenching model (Stoch. Del. Q.), and the refuelling model (Refuel.) in both disjoint ranges of stellar mass considered. Values are supplied for the conservative infall time distribution (con; see Sect.~\ref{SFR_SAT}, Appendix~\ref{APPEND_MODEL} and Fig.~\ref{fig_append_tindist}), as well as for the full distribution (full; see Sect.~\ref{DEPENDINT} and Appendix~\ref{APPEND_MODEL}).}    
\end{tablenotes}
\end{threeparttable}
\end{table*}

\subsubsection{The 'Delayed-Quenching' Model}
The first of the two parameter models shown in Fig.~\ref{fig_models} is referred to as the 'delayed quenching' model. In this model a galaxy becoming a satellite at time $t_{\mathrm{infall}}$ continues to form stars as if it were still a field (central) galaxy for a fixed time $t_{\mathrm{delay}}$ until $t_{\mathrm{quench}} = t_{\mathrm{infall}} + t_{\mathrm{delay}}$ after which the SFR declines exponentially on a timescale $\tau_{\mathrm{quench}}$. Thus, the functional form of the SFR is given by 
\begin{equation} 
\mathrm{SFR}(t) = \begin{cases}
\mathrm{SFR}_{\mathrm{field}}(t) & \mathrm{for}\, t < t_{\mathrm{quench}}\\
\mathrm{SFR}_{\mathrm{field}}(t_{\mathrm{quench}})e^{\frac{-(t - t_{\mathrm{quench}})}{\tau_{\mathrm{quench}}}} & \mathrm{for}\, t> t_{\mathrm{quench}}\,.\\
\end{cases}
\label{eq_2pw}
\end{equation}\newline 

As shown in Fig.~\ref{fig_2pmod_hm} (left column) the three characteristics of the distribution of $\Delta \mathrm{log}(\psi_*)$ for the mass range $M_* \ge 10^{10} M_{\odot}$ (from top to bottom first quartile, median, and third quartile) can each be reproduced by a number of combinations of $t_{\mathrm{delay}}$ and $\tau_{\mathrm{quench}}$. However, the topology of the quantity $Q(\tau_{\mathrm{quench}},t_{\mathrm{delay}})$ shown in the top left panel of Fig.~\ref{fig_Qm} shows that only a very limited region of parameter space around $\tau_{\mathrm{quench}}=0.5  \,\mathrm{Gyr}$ and $t_{\mathrm{delay}}=2.5\,\mathrm{Gyr}$ provides a good fit to all three characteristics simultaneously ($Q=0.83$). Nevertheless, for this very narrow range of parameter space, the top panel of Fig.~\ref{fig_2pmod_hm} illustrates that these parameters provide not only a good fit of the characteristics, but also of the distribution of $\Delta \mathrm{log}(\psi_*)$ as a whole.\newline

For the stellar mass range $10^{9.5} M_{\odot} \le M_* < 10^{10} M_{\odot}$ Fig.~\ref{fig_Qm} (bottom left panel) shows the parameter space conducive to a simultaneous recovery of all three characteristics to be similarly limited as for the high stellar mass range. For the low mass range, however, the preferred parameter values are $\tau_{\mathrm{quench}}=0.5  \,\mathrm{Gyr}$ and $t_{\mathrm{delay}}=2.9\,\mathrm{Gyr}$, i.e while the preferred quenching timescale is the same, the preferred delay time before the onset of star formation quenching is slightly longer than for high mass galaxies. As for the high stellar mass range, the preferred parameters provide a good fit to the full $\Delta \mathrm{log}(\psi_*)$ distribution.\newline 

 Overall, the 'delayed quenching' model provides a good approximation of the observed distribution over the full range in stellar mass. However, the ranges in $t_{\mathrm{delay}}$ and $\tau_{\mathrm{quench}}$ for which all three characteristics can be reproduced are extremely narrow, with the solution being largely trivial, as the preferred delay time corresponds to significant fraction (or even the whole) of the satellite lifetime for a large fraction of the model group galaxies.\newline

\subsubsection{The 'Stochastic Delayed Quenching' Model}
The second two parameter model, referred to as the 'stochastic delayed quenching' model, expands on the first by replacing the fixed delay time with a probability per unit time $P_{\mathrm{quench}}$ that gradual quenching of the SFR of the galaxy begins; I.e. a galaxy becoming a satellite at time $t_{\mathrm{infall}}$ continues to form stars as if it were still a field (central) galaxy. However, with a probability $P_{\mathrm{quench}}$, the SFR of the galaxy begins an  exponential decline on a decay timescale of $\tau_{\mathrm{quench}}$. As such, the functional form of the time dependence of the SFR is identical to that given in Eq.~\ref{eq_2pw} with the difference lying in the stochastically determined delay time.\newline

As shown in the middle panels of Fig.~\ref{fig_Qm} the parameter space conducive to a simultaneous recovery of all three characteristics is very limited, with this only being possible in the vicinity of $\tau_{\mathrm{quench}} = 0.9\, \mathrm{Gyr}$ and $P_{\mathrm{quench}} = 0.3 \, \mathrm{Gyr}^{-1}$ for the stellar mass range $M_* > 10^{10} M_{\odot}$, and for $\tau_{\mathrm{quench}} = 1.5\, \mathrm{Gyr}$ and $P_{\mathrm{quench}} = 0.3 \, \mathrm{Gyr}^{-1}$ in the low stellar mass range.  As shown in the top panels of Figs.~\ref{fig_2pmod_hm} \& \ref{fig_2pmod_lm}, these parameters not only recover the three characteristics, but also the distribution of $\Delta \mathrm{log}(\psi_*)$ as a whole. However, the values of $Q$ in both stellar mass ranges are lower than those achieved by the delayed quenching model, indicative of a poorer recovery of the distributions (see table~\ref{tab_fitpar}).\newline

An important feature of the 'stochastic delayed quenching' model, is highlighted by the second, third and fourth panels from the top in Figs.~\ref{fig_2pmod_hm} \& \ref{fig_2pmod_lm}. Considering the distributions of $\Delta q_i$ for all three characteristics shown in these panels, it is apparent that the  preferred solution becomes degenerate in $P_{\mathrm{quench}}$ for $P_{\mathrm{quench}} \gtrsim 1.5\, \mathrm{Gyr}^{-1}$. This results from the fact that at and above this frequency nearly every modelled infalling galaxy will experience a quenching event. Conversely, at the preferred value of $P_{\mathrm{quench}} \approx 0.3 \,  \mathrm{Gyr}^{-1}$, a sizeable fraction ($\gtrsim 50$\%) of the infalling population does not experience a quenching event. Overall, therefore, although the stochastic delayed quenching model is capable of closely reproducing the observed distributions of $\Delta \mathrm{log}(\psi_*)$, the solution is largely trivial.\newline

\subsubsection{The 'Re-fuelling' Model}
The final two parameter model is referred to simply as the 're-fuelling' model. In this model, a galaxy becoming a satellite at time $t_{\mathrm{infall}}$ continues to form stars as if it were still a field (central) galaxy. With a probability per unit time $P_{\mathrm{quench}}$ the SFR of the galaxy is instantaneously completely shut off at a time $t_{\mathrm{quench}}$ with $t_{\mathrm{infall}} < t_{\mathrm{quench}}$ , followed by an inversely exponential recovery to the level it would have had as a field galaxy on a timescale $\tau_{\mathrm{fuel}}$. Unlike the other models with a quenching probability, where although the occurrence of the instantaneous or gradual quenching was stochastic it could only take place once, in the re-fuelling model we include the possibility of multiple such events as illustrated in Fig.~\ref{fig_models}, i.e. this model explicitly includes a resuscitation of previously quenched star formation, and the evolution of the SFR 
is given by

\begin{equation} 
\mathrm{SFR}(t) = \begin{cases}
\mathrm{SFR}_{\mathrm{field}}(t) & \mathrm{for}\, t < t_{\mathrm{quench,}1}\\
\mathrm{SFR}_{\mathrm{field}}(t)(1 - e^{\frac{-(t - t_{\mathrm{quench,i}})}{\tau_{\mathrm{quench}}}}) & \mathrm{for}\, t> t_{\mathrm{quench,i}},\,\forall i.
\label{eq_2pref}
\end{cases}
\end{equation}\newline

As shown in the right column of Fig.~\ref{fig_Qm}, the re-fuelling model can simultaneously reproduce the three characteristics of the $\Delta \mathrm{log}(\psi_*)$ distribution for a wide range of values for the parameters $\tau_{\mathrm{fuel}}$ and $P_{\mathrm{quench}}$  in both the ranges of stellar mass considered. Furthermore the re-fuelling model achieves values of $Q$ higher than the other two parameter models in both stellar mass ranges ($Q \ge 0.92$, see table~\ref{tab_fitpar}), indicative of a better simultaneous recovery of the observed distribution of $\Delta \mathrm{log}(\psi_*)$.\newline

\begin{figure*}
\plotone{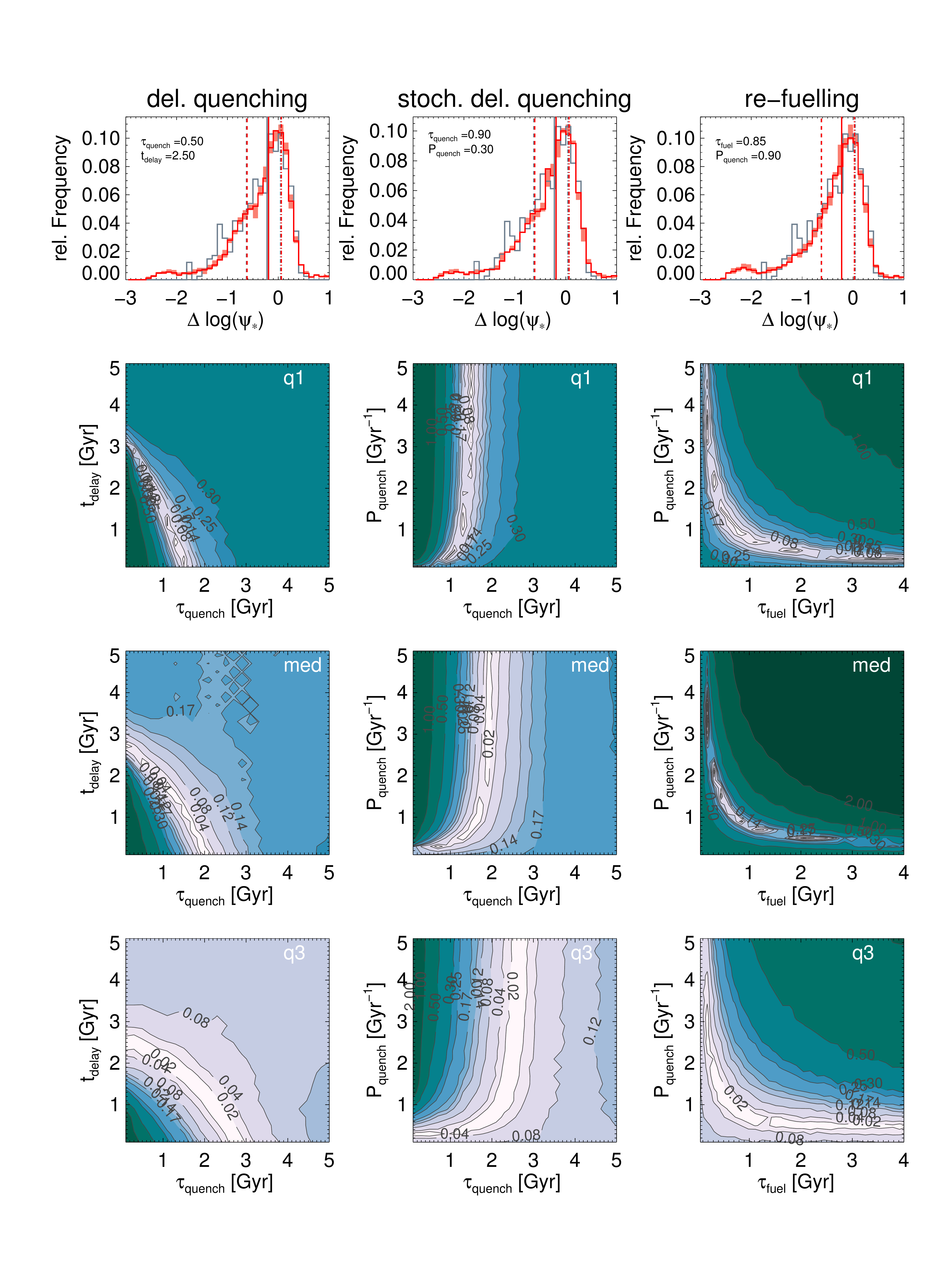}
\caption{This figure depicts the performance of the family of two parameter models (the delayed quenching model, the stochastic delayed quenching model, and the re-fuelling model, from left to right) in the stellar mass range $M_* \le 10^{10} M_{\odot}$. \textbf{Top:} This panel shows the observed distribution of $\Delta \mathrm{log}(\psi_*)$ in gray, with the three characteristics of the observed distribution (the second quartile, the median, and the third quartile) depicted as vertical gray dashed, solid , and dash-dotted lines, respectively. The average of 50 realization of the model with the parameter values listed at top left (chosen to provide the best simultaneous fit to the three characteristics as determined using $Q_{\mathrm{m}}$ in Fig.~\ref{fig_Qm}) is over-plotted in red, with the three characteristics of the model distribution indicated by vertical red lines (of the same line style). The red shaded region shows the  range between the 16$^{\mathrm{th}}$ and 84$^{\mathrm{th}}$ percentile in each bin of $\Delta \mathrm{log}(\psi_*)$ as found from 50 realizations of the model. \textbf{Second from top}: This panel shows the topology $\Delta q_1$.
The color coding of the contours goes from dark green via blue to white for decreasing values of $\Delta q_1$. \textbf{Third from top:} As for the panel above but for $\Delta q_2$. \textbf{Bottom}: As for the two panels above, but for $\Delta q_3$.} 
\label{fig_2pmod_hm}  
\end{figure*}

For the high stellar mass range, the preferred parameter values are $\tau_{\mathrm{fuel}} = 0.85 \, \mathrm{Gyr}$ and $P_{\mathrm{quench}} = 0.9 \, \mathrm{Gyr}^{-1}$, with these values also providing a good fit to the distribution of  $\Delta \mathrm{log}(\psi_*)$ as a whole, as shown in the top right panel of Fig.~\ref{fig_2pmod_hm}. However, unlike for the other two parameter models, there is a pronounced degeneracy between the model parameters, with the models with 
$\tau_{\mathrm{fuel}} = 1.9 \, \mathrm{Gyr}$ and $P_{\mathrm{quench}} = 0.5 \, \mathrm{Gyr}^{-1}$, respectively $\tau_{\mathrm{fuel}} = 0.3 \, \mathrm{Gyr}$ and $P_{\mathrm{quench}} = 2.1 \, \mathrm{Gyr}^{-1}$, also performing similarly well.\newline

\begin{figure*}
\plotone{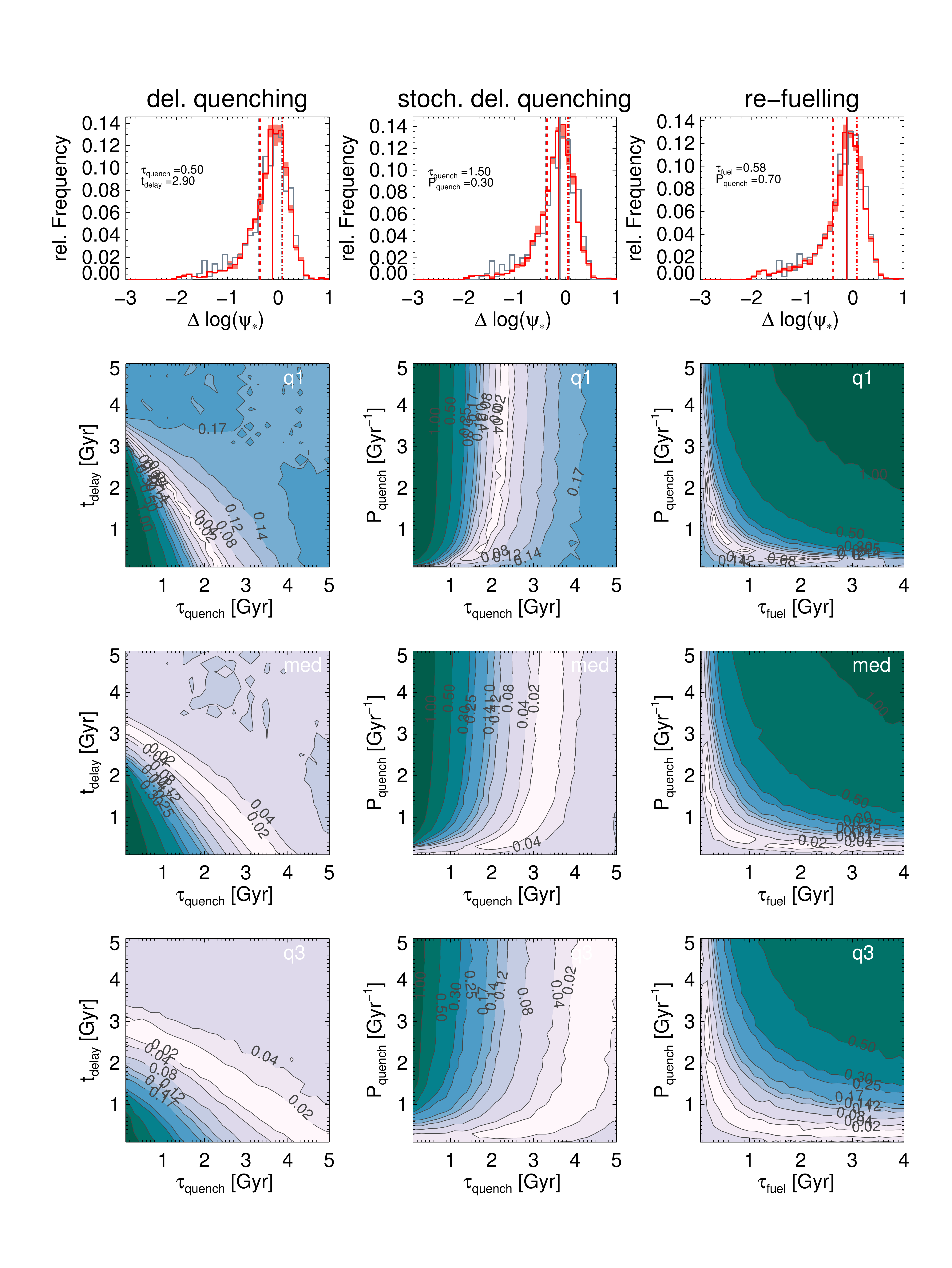}
\caption{As Fig.~\ref{fig_2pmod_hm}, but for the mass range $10^{9.5}M_{\odot} \le M_* < 10^{10} M_{\odot}$. }
\label{fig_2pmod_lm}  
\end{figure*}

\begin{figure*}
\plotone{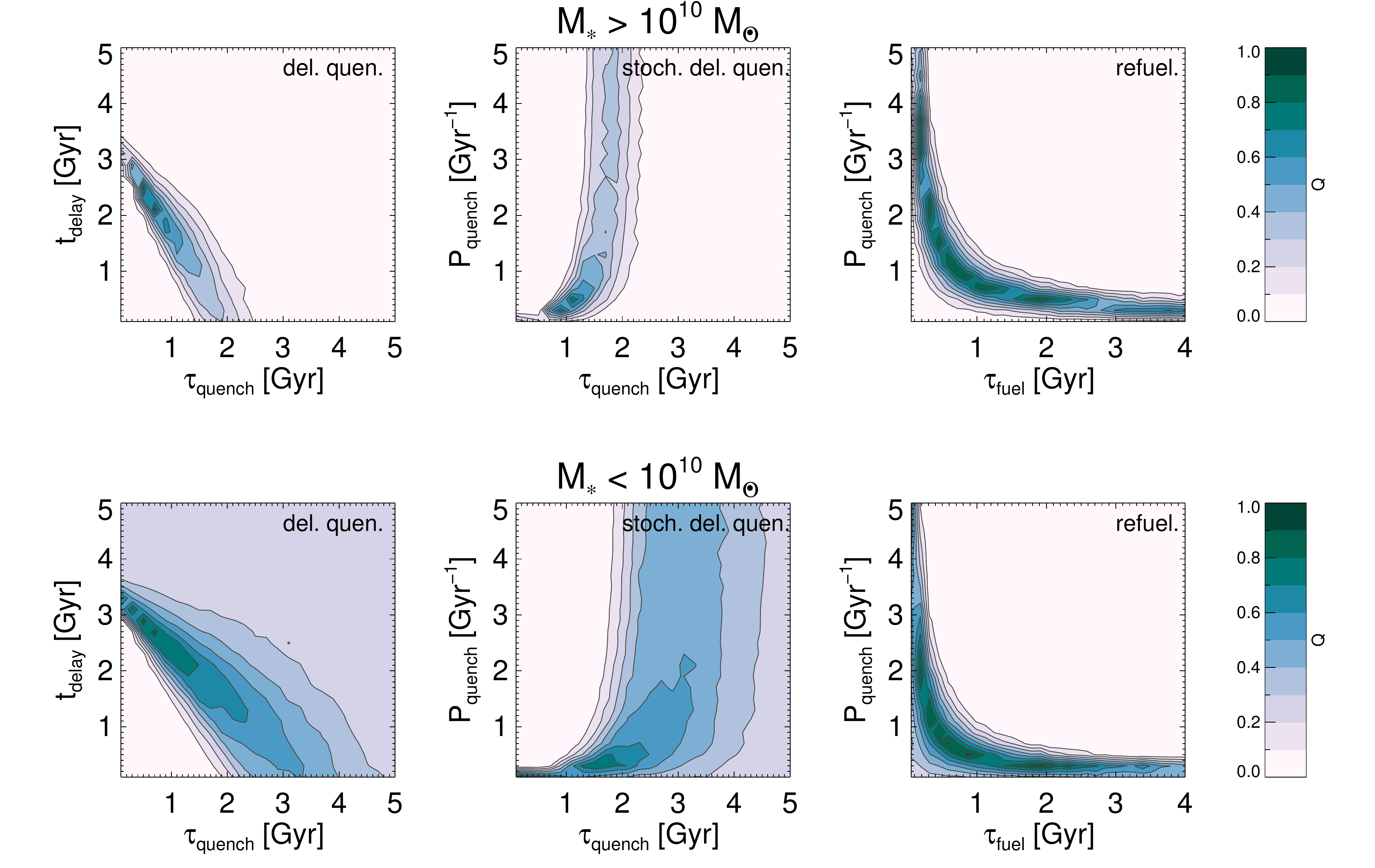}
\caption{Figure of merit $Q(p_1,p_2)$ indicating the ability of the two parameter model to simultaneously recover all three characteristics (first quartile, median, third quartile) as a function of both parameters. $0 \le Q \le 1$, with $1$ corresponding to a perfect simultaneous recovery. The top row shows the results in the high stellar mass range $M_* > 10^{10} M_{\odot}$, while the bottom row shows those for the low stellar mass range $10^{9.5} M_{\odot} \le M_* \le 10^{10} M_{\odot}$. From left to right the columns shown the results for the delayed quenching model, the stochastic delayed quenching model, and the refuelling model.}
\label{fig_Qm}
\end{figure*}

For the low stellar mass range the results are qualitatively similar (as shown in Figs.~\ref{fig_Qm} \& \ref{fig_2pmod_lm}), albeit with preferred values of $\tau_{\mathrm{fuel}} = 0.58 \, \mathrm{Gyr}$ and $P_{\mathrm{quench}} = 0.7 \, \mathrm{Gyr}^{-1}$. Overall, the re-fuelling model is capable of closely reproducing the observed distributions of $\Delta \mathrm{log}(\psi_*)$ for a wide range of parameter values, common to both ranges of stellar mass. Finally, it also remains to be noted that the majority of these solutions are non-trivial, as for the higher values of $P_{\mathrm{quench}}$ the majority of the model satellites experiences at least one quenching and re-fuelling cycle.\newline 

In summary, we thus find that all three two parameter models considered can reproduce the observed distributions of $\Delta \mathrm{log}(\psi_*)$, albeit that the re-fuelling model performs best. Independent of the chosen SFH model however, we find a prolonged (or indefinite) period of star formation at the level of a comparable field galaxy while already a satellite to be required in order to recover the observed distributions.\newline

\section{Implications for the Gas-fuelling of Satellite Spiral Galaxies}\label{GASFUELLING}
In the previous sections we have presented a detailed empirical analysis of the star formation in satellite spiral/disk galaxies considering the distributions of sSFR and $\Delta\mathrm{log}(\psi_*)$. Making use of simple models for the star-formation histories (SFH) of satellite spiral galaxies we have shown that the empirical distributions of $\Delta\mathrm{log}(\psi_*)$ favour SFH for satellite spirals with extended periods of star formation at the level of comparable field spiral galaxies and rapid quenching of star formation, respectively a rapid recovery of star formation. In the following we will consider the implications of these findings in the context of the gas-cycle of spiral satellite galaxies and use them to broadly constrain the gas-fuelling of these objects.We begin by outlining our methodology, followed by a derivation of estimates for the in- an outflows in the context of our model SFH. Finaly, we discuss the implications of our results with respect to the reservoirs form which gas-fuelling can potentially be sourced. \newline

\subsection{Constraining the Gas-Cycle}\label{GASFUELLING_ISMEVOL}
To obtain broad quantitative constraints on the flows of gas into and out of the ISM we will make use of the equations describing the gas-cycle of galaxies. Here we present an overview of our methodology, while a full detailed derivation and discussion is supplied in Appendix~\ref{APPEND_GASCYCLE}.\newline

In a general form, the ISM gas content of a galaxy, respectively its time-dependent evolution, can be expressed as 
\begin{equation}
\dot{M}_{\mathrm{ISM}}  =  \dot{M}_{\mathrm{in}} - \dot{M}_{\mathrm{out}} - (1-\alpha) \Phi_*\,, 
\label{eq_ISMevol1}
\end{equation}
where $\dot{M}_{\mathrm{in}}$ is the inflow rate of gas into the ISM of the galaxy, $\dot{M}_{\mathrm{out}}$ is the outflow rate of gas from the ISM of the galaxy, $\Phi_*$ is the current SFR, and $\alpha$ is a positive constant less than unity which accounts for the recycling of gas from high mass stars back into the ISM  (as detailed in appendix~\ref{APPEND_MODEL}, $\alpha=0.3$ throughout). Assuming a volumetric star formation law
\begin{equation}
\Phi_* = \tilde{\kappa} M_{\mathrm{ISM}}
\label{eq_volSFlaw}
\end{equation}
following \citet{KRUMHOLZ2012}, and that the outflow rate can be re-expressed in terms of the ISM content and a typical residence time $\tau_{\mathrm{res}}$ of a unit mass of gas in the ISM\footnote{we note that this formulation is equivalent to the widely used mass-loading approximation for parameterizing outflows.}, Eq.~\ref{eq_ISMevol1} can be written as
\begin{eqnarray}
\dot{M}_{\mathrm{ISM}}  & = &\dot{M}_{\mathrm{in}} - \frac{M_{\mathrm{ISM}}}{\tau_{\mathrm{res}}} - (1 - \alpha ) \tilde{ \kappa} M_{\mathrm{ISM}} \nonumber \\
 & = & \dot{M}_{\mathrm{in}} - \frac{M_{\mathrm{ISM}}}{\tau_{\mathrm{res}}} - \kappa M_{\mathrm{ISM}}\,. \label{eq_ISMevol2}
\end{eqnarray}
As $\kappa = 1/\tau_{\mathrm{exhaust}}$, where $\tau_{\mathrm{exhaust}}$ corresponds to the exhaustion timescale of the ISM in a closed box model, Eq.~\ref{eq_ISMevol2} can be reformulated and simplified as 
\begin{equation}
\dot{M}_{\mathrm{ISM}}  = \dot{M}_{\mathrm{in}} - \frac{M_{\mathrm{ISM}}}{\tilde{\tau}}
\label{eq_ISMevol4}
\end{equation}
using an effective timescale $\tilde{\tau} = \tau_{\mathrm{res}} \tau_{\mathrm{exhaust}} / (\tau_{\mathrm{res}} + \tau_{\mathrm{exhaust}})$. For our analysis, we assume that $\tilde{\kappa}$, respectively $\tau_{\mathrm{exhaust}}$ and $\tau_{\mathrm{res}}$, is set by galaxy specific processes alone, i.e. is independent of environment. As detailed in Appendix~\ref{APPEND_GASCYCLE}, this quantity has been individually calibrated for both stellar mass ranges considered in our analysis using the model of \citet{POPPING2014} and our median $\psi_*$--$M_*$ relation for the \textsl{\textsc{fieldgalaxy}} sample.\newline

For spiral galaxies in the field, the SFR is found to evolve only very slowly with redshift and is thought to be determined by a very gradually evolving self-regulated balance between inflow, outflow, and consumption of ISM via star formation \citep[e.g.][]{KERES2005,DAVE2011,LILLY2013,SAINTONGE2013} such that at any given time their SFR can be considered quasi-constant. As we show in detail in Appendix~\ref{APPEND_GASCYCLE}, this quasi-steady state allows an estimate of the inflow rate to be derived as 
\begin{equation}
\dot{M}_{\mathrm{in}} \approx \frac{M_{\mathrm{ISM}}}{\tilde{\tau}} = \frac{\Phi_*}{\tilde{\kappa} \tilde{\tau}}\,,
\label{eq_MinSS}
\end{equation} 
given knowledge of the effective timescale $\tilde{\tau}$ and the SFR\footnote{We note that the estimate of the inflow rate given by Eq.~\ref{eq_MinSS} represents a conservative estimate in our derived framework.} . As shown in Appendix~\ref{APPEND_GASCYCLE}, the corresponding outflow rate in the steady state can then be estimated as 
\begin{equation}
\frac{\dot{M}_{\mathrm{out}}}{\Phi_*} = \frac{1}{\tilde{\kappa} \tilde{\tau}} - (1- \alpha) = \frac{\dot{M}_{\mathrm{in}}}{\Phi_*} - (1 - \alpha)\,,
\label{eq_MoutSS}
\end{equation}
where $\dot{M}_{\mathrm{in}}$ is determined using Eq.~\ref{eq_MinSS}. This approximation will hold as long as the rate at which the inflow changes is small compared to the timescale $\tilde{\tau}$. As discussed in Appendix\ref{APPEND_GASCYCLE} the deduced values for $\tilde{\tau}$ are $\lesssim1\,$Gyr, retroactively justifying our use of this approximation.\newline

As the volumetric star formation law - Eq.~\ref{eq_volSFlaw} - linearly couples the SFR and ISM mass (at a given stellar mass), in the following, we will proceed by identifying solutions of Eq.~\ref{eq_ISMevol4} which correspond to the parameterized SFH of the models and directly interpret the preferred values of the model parameters identified in section~\ref{SFR_SAT_2pm} in terms of mass flows into and out of the ISM. Specifically, we make use of the evolution of the SFR during the quenching, respectively the refuelling, phases of the two parameter models to constrain $\tilde{\tau}$, enabling us to use Eq.~\ref{eq_MinSS} to estimate the value of $\dot{M}_{\mathrm{in}}$ required during the prolonged periods of star-formation activity comparable to that of field galaxies observed for all these models.\newline

At this point, we note that we will take the term $\dot{M}_{\mathrm{in}}$ to represent a pure inflow term, i.e. mass coming into the ISM from outside of the volume of the galaxy occupied by its stellar component. In reality, the ISM of a galaxy will also be fuelled by the mass loss from evolved intermediate and low mass stars (e.g. TP-AGB stars) not included in the definition of $\kappa$.  For the Milky Way bulge, where the mass return is dominated by these evolved stars, studies find stellar mass normalized mass return rates of $\mathcal{O}(10^{11})\,$yr$^{-1}$ \citep[e.g.][]{OJHA2007}. Comparing this to the sSFR of even the higher stellar mass galaxies (i.e. those with a larger old stellar component), one finds that the mass return rate which must be considered is likely $\lesssim10$\% of the observed sSFR. In the following, we have therefore chosen to ignore this contribution, but will return to and justify this choice later.\newline

\subsection{Estimates of In- and Outflows}
In the following we present the results of applying the approach to constraining the gas-cycle outlined above to the two parameter model family.
The estimated in- and outflow rates for each model are listed in table~\ref{tab_INFLOW}.\newline 

\subsubsection{The Delayed Quenching Model}\label{IMP_DQM}
Inserting the SFH of the delayed quenching model into Eq.~\ref{eq_volSFlaw} we find
\begin{equation} 
M_{\mathrm{ISM}}(t) = \begin{cases}
M_{\mathrm{ISM,field}}(t) & \mathrm{for}\, t < t_{\mathrm{quench}}\\
M_{\mathrm{ISM,field}}(t_{\mathrm{quench}})e^{\frac{-(t - t_{\mathrm{quench}})}{\tau_{\mathrm{quench}}}} & \mathrm{for}\, t> t_{\mathrm{quench}}\,,\\
\end{cases}
\label{eq_MISM2pw}
\end{equation}\newline 
corresponding to the balance between inflow, outflow, and consumption of the ISM via star formation - assumed to be in place for field galaxies - being maintained upon infall of a galaxy for the time $t_{\mathrm{delay}}$ until $t=t_{\mathrm{quench}}$, followed by an exponential decline of the ISM mass. Solving Eq.~\ref{eq_ISMevol4} for this scenario, i.e. with the inflow being cut off for $t >t_{\mathrm{quench}}$ ($\dot{M}_{\mathrm{in}} (t >t_{\mathrm{quench}}) \equiv 0$), we find a solution of the form 
\begin{equation}
M_{\mathrm{ISM}}(t > t_{\mathrm{quench}}) = M_{\mathrm{ISM}}(t=t_{\mathrm{quench}}) e^{-\left(\frac{t - t_{\mathrm{quench}}}{\tilde{\tau}}\right)}\,.
\label{eq_delsol1}
\end{equation} 
With the balance between inflow, outflow, and consumption of ISM being maintained as for a corresponding field galaxy while $t_{\mathrm{infall}} < t \le t_{\mathrm{infall}}$, we can identify $M_{\mathrm{ISM}}(t) = M_{\mathrm{ISM,field}}(t) \,\forall t \,\in (t_{\mathrm{infall}}, t_{\mathrm{quench}}]$. Comparing Eqs.~\ref{eq_delsol1} \& \ref{eq_MISM2pw} we can then immediately identify
\begin{equation}
\tilde{\tau} = \tau_{\mathrm{quench}}\,.
\label{eq_idtauquench}
\end{equation}
\newline

Inserting Eq.~\ref{eq_idtauquench} into Eq.~\ref{eq_MinSS} we can estimate the rate of inflow from our observations. As previously, we consider the stellar mass ranges $10^{9.5} M_{\odot} \le M_* < 10^{10} M_{\odot}$ and $M_* \ge 10^{10} M_{\odot}$ separately. In this fashion, we thus estimate an inflow rate of $4.46$ ($4.56$) times the SFR for the high (low) stellar mass range, and find a corresponding outflow of $3.76$ ($3.86$) times the SFR. The estimated in- and outflows are listed in table~\ref{tab_INFLOW}.\newline

\subsubsection{The Stochastic Delayed Quenching Model}\label{IMP_SDQM}
The basic functional form of the stochastic delayed quenching model corresponds to that of the delayed quenching model as discussed above, with the only difference being that the fixed delay time $t_{\mathrm{delay}}$ is replaced by a probability per unit time that quenching occurs $P_{\mathrm{quench}}$. Accordingly we estimate the in and outflows based on the preferred parameter values as for the delayed quenching model, finding a required inflow of $2.38$ ($1.46$) times the SFR in high (low) stellar mass range, and a corresponding outflow of $1.68$ ($0.76$) times the SFR as listed in table~\ref{tab_INFLOW}.\newline

As previously discussed, for the stochastic delayed quenching model we find the preferred solution for each characteristic to become degenerate in $P_{\mathrm{quench}}$ for $P_{\mathrm{quench}} \gtrsim 1.5\, \mathrm{Gyr}^{-1}$. As argued above, for the preferred value of $P_{\mathrm{quench}}=0.3\,$Gyr$^{-1}$ (for both stellar mass ranges) this implies that $\gtrsim 50$\% of the infalling population does not experience a quenching event, and hence must continue to maintain the balance between inflow, outflow and star formation at the level of a comparable field galaxy over their entire satellite lifetime. In turn this enables us to interpret the preferred value of $P_{\mathrm{quench}}$ in terms of an effective requirement on the duration of the extended period of star formation for satellite galaxies in the group environment. Based on the value of $P_{\mathrm{quench}}=0.3\,$Gyr$^{-1}$ and using the extremely conservative distribution of infall times adopted in our modelling, we find that $30$\% ( $20$\%, $10$\%) of the satellite galaxies have resided in the group environment without quenching for $\gtrsim2\,$Gyr ($\gtrsim2.5\,$Gyr, $\gtrsim3\,$Gyr).\newline

\subsubsection{The Re-fuelling Model}\label{IMP_RFM}
As for the delayed quenching model, we insert the SFH embodied by the re-fuelling model - Eq.~\ref{eq_2pref} - in Eq.~\ref{eq_volSFlaw} finding the re-fuelling model to correspond to a case in which the assumed balance between inflow, outflow, and star formation is initially maintained by galaxies upon becoming satellites, albeit with a probability per unit time $P_{\mathrm{quench}}$ that at least a large fraction of the ISM of the galaxy is quasi-instantaneously removed. However, even with the occurrence of a quenching event the inflow continues as for a comparable field galaxy, so that the self regulated balance eventually reinstates itself. Solving Eq.~\ref{eq_ISMevol4} for these boundary conditions, i.e. $\dot{M}_{\mathrm{in}} = \mathrm{const.}$ \footnote{In this derivation we at all times assume the rate of change of the inflow to be small compared to the other timescales (quenching and replenishment) involved, and treat $\dot{M}_{\mathrm{in}}$ as (quasi-)constant.}, and $M_{\mathrm{ISM}}(t_{\mathrm{quench,i}}) = 0 \forall i$, we find a solution of the form
\begin{equation}
M_{\mathrm{ISM}} (t > t_{\mathrm{quench,i}}) = \dot{M}_{\mathrm{in}}\tilde{\tau}(1 - e^{\frac{-\left(t - t_{\mathrm{quench,i}}\right)}{\tilde{\tau}}})\,,
\label{eq_refsol}
\end{equation}
where we have identified  $M_{\mathrm{ISM,field}} = \dot{M}_{\mathrm{in}}\tilde{\tau}$ by making use of the special case of the occurrence of only a single quenching event and taking the limit $t >> t_{\mathrm{quench,1}}$.\newline

This enables us to identify $\tilde{\tau} = \tau_{\mathrm{fuel}}$, and we make use of Eqs.~\ref{eq_MinSS} \& \ref{eq_MoutSS} to estimate the in- and outflow rates for  both mass ranges, finding an estimated inflow of $2.52$ ($3.78$) times the SFR and an associated outflow of $1.82$ ($3.08$) times the SFR for the high (low) stellar mass range, as listed in table~\ref{tab_INFLOW}.\newline

In summary, for all our disparate models, we find a requirement of a rapid cycle of gas into and out of the ISM, with inflow rates well in excess of the SFR (ranging up to $\gtrsim 3$ times the SFR), accompanied by similarly high outflow rates. Comparing these inflow rates with the estimate of the mass return rate from evolved intermediate and low mass stars of $\lesssim 10$\% of the SFR, we see that ignoring their contribution to the fuelling is retroactively justified for all models considered\footnote{Even when considering the total gas return rate from stellar populations, i.e including the contribution from short-lived stars absorbed in the value of $\kappa$, this rate is generally found to be at most several tens of percent of the SFR, both for young lower mass galaxies such as the LMC and SMC \citep{MATSUURA2009,MATSUURA2013}, as well as for more massive \& mature spiral galaxies \citep{TIELENS2005}, and thus well shy of the required fuelling rates.}. \newline

\begin{table*}
\centering
\begin{threeparttable}
\caption{Summary of Estimated In- \& Outflow Rates\label{tab_INFLOW}}
\begin{tabular}{cccccc}
\multicolumn{2}{c}{} & \multicolumn{2}{c}{Inflow ($\dot{M}_{\mathrm{in}}$/SFR)} & \multicolumn{2}{c}{Outflow ($\dot{M}_{\mathrm{out}}$/SFR)} \\
Model & $t_{\mathrm{infall}}$ dist. & $M_* < 10^{10} M_{\odot}$ & $M_* > 10^{10} M_{\odot}$ & $M_* < 10^{10} M_{\odot}$ & $M_* > 10^{10} M_{\odot}$ \\
\hline
\multirow{2}{*}{Del. Q.} &  \multicolumn{1}{c}{con.} & $4.56$& $4.46$&$3.86$ &$3.76$  \\
& \multicolumn{1}{c}{full} &  -$^{\alpha}$&$1.43$ & - $^{\alpha}$&$0.73$  \\
\multirow{2}{*}{Stoch. Del. Q.} &  \multicolumn{1}{c}{con.} & $1.46$& $2.38$&$0.76$ &$1.68$  \\
& \multicolumn{1}{c}{full} & -$^{\alpha}$&$1.43$ & -$^{\alpha}$ &$0.73$ \\
\multirow{2}{*}{Refuel.} &  \multicolumn{1}{c}{con.} & $3.78$& $2.52$&$3.08$ &$1.82$  \\
& \multicolumn{1}{c}{full} & $4.87$&$2.18$ &$4.17$ &$1.48$  \\
\hline
\hline
\end{tabular}
\begin{tablenotes}
\item{Inflow and corresponding outflow rates as multiples of the star formation rate SFR for the delayed quenching model (Del. Q.), the stochastic delayed quenching model (Stoch. Del. Q.), and the refuelling model (Refuel), as determined under a quasi-steady state assumption following Eqs.~\ref{eq_MinSS} \& \ref{eq_MoutSS}, as detailed in Sections~\ref{IMP_DQM},\ref{IMP_SDQM}, and \ref{IMP_RFM}. Both the conservative distribution of infall times (con; see Sect.~\ref{SFR_SAT}, Appendix~\ref{APPEND_MODEL} and Fig.~\ref{fig_append_tindist}), as well as the full distribution (full; see Appendix~\ref{APPEND_MODEL}) are considered. In- and outflows have been estimated separately for the stellar mass ranges $10^{9.5} M_{\odot} \le M_* < 10^{10} M_{\odot}$ and $M_* \ge 10^{10}M_{\odot}$.}
\item[$\alpha$]{Assumption that quenching timescale is short w.r.t rate of change of inflow violated. Estimate not possible.}
\end{tablenotes}
\end{threeparttable}
\end{table*}

\subsubsection{Dependence on the choice of infall time distribution}\label{DEPENDINT}
In considering the implications of the preferred parameters of our models in terms of the gas flows in satellite spiral/disk-dominated galaxies, we have adopted an extremely conservative assumption concerning the distribution of infall times, i.e. that \textit{only} the youngest satellites retain a spiral morphology. The other possible extreme assumption on the distribution of infall-times is to sample the full distribution, assuming that the time spent in the group environment does not influence the probability of the morphological transformation of a galaxy. This is almost certainly not the case. Instead, the true distribution of infall times will fall somewhere between these two extremes.\newline

The results of adopting this latter extreme distribution of infall times are shown in Figs.~\ref{fig_append_2pmod_FD_hm} \& \ref{fig_append_2pmod_FD_lm} in Appendix~\ref{APPEND_FD}, with the composite figure of merit $Q$ shown in Fig.~\ref{fig_append_Qm} (the preferred parameter values and associated figure of merit for each model are listed in table~\ref{tab_fitpar}). In the high stellar mass range, all three models are formally capable of reproducing the observed distribution of $\Delta \mathrm{log}(\psi_*)$. However, while the re-fuelling model (unsurprisingly) favors parameters comparable to those previously found ($\tau_{\mathrm{fuel}} = 0.98\,$Gyr, $P_{\mathrm{quench}} = 0.5\,$Gyr$^{-1}$) and achieves $Q=0.92$, comparable to that previously obtained, the delayed quenching model and the stochastic delayed quenching model favor a longer quenching timescale ($\tau_{\mathrm{quench}} =1.5\,$Gyr in both models) as well as longer delay times ($4.9\,$Gyr)/lower quenching probabilities ($0.1\,$Gyr$^{-1}$). While the stochastic delayed quenching model performs as previously in terms of recovering the distribution and characteristics, even achieving a higher value of $Q$, the performance of the delayed quenching model is considerably worse, only attaining $Q=0.75$. In terms of required inflow rates, the preferred parameter values imply a rate of $1.43$ times the SFR for the delayed and stochastic delayed quenching models and a rate of $2.18$ times the SFR for the refuelling model, as listed in table~\ref{tab_INFLOW}. Again, it is immediately apparent that these inflow rates cannot be supported by mass return from evolved intermediate and low mass stars, hence justifying the treatment of $M_{\mathrm{in}}$ as a pure inflow.\newline 

In the low stellar mass range, the models struggle to reproduce the observed distribution of $\Delta \mathrm{log}(\psi_*)$ (see fig.~\ref{fig_append_Qm}). Of the three models, only the re-fuelling model reasonably recovers the observed distribution, achieving a value of $Q =0.79$, compared to $Q=0.63$ and $Q=0.71$ for the delayed and stochastic delayed quenching models, respectively. The preferred parameter values for the refuelling model are $\tau_{\mathrm{fuel}} = 0.45\,$Gyr, $P_{\mathrm{quench}} = 0.5\,$Gyr$^{-1}$, again comparable to those previously obtained. Both the delayed quenching model and the stochastic delayed quenching model, on the other hand, over-predict the relative number of largely unquenched galaxies, and in the case of the delayed quenching model, markedly under-predict the number of strongly quenched galaxies, i.e. both fail to recover the observed distribution. This is a result of the long preferred quenching timescales of $\tau_{\mathrm{quench}} = 3.1\,$Gyr and $\tau_{\mathrm{quench}}=3.7\,$Gyr (for the stochastic delayed and delayed quenching models, respectively), as well as of the low quenching probability $P_{\mathrm{quench}}=0.1\,$Gyr$^{-1}$ and the long delay time $t_{\mathrm{delay}}=4.3\,$Gyr - driven by the large peak of unquenched galaxies - which result in the satellite galaxy population largely mimicking the evolution of a comparable field galaxy. Furthermore, these delay timescales are longer than the previously discussed gas exhaustion timescales.\newline

Converting the preferred model parameters into in- and outflows rates for the  re-fuelling model, one obtains a required inflow rate of $4.87$ times the SFR with a corresponding outflow of $4.17$ times the SFR. For the  delayed quenching models, however, the basic requirement, that the timescale on which the inflow rate changes be large compared to $\tau_{\mathrm{res}}$ and $\tau_{\mathrm{SF}}$ is violated, making a estimate of the inflow rate using Eq.~\ref{eq_MinSS} unreliable. \newline 

Overall, we thus find our result of a rapid cycle of gas into and out of the ISM with inflow rates in excess of the SFR to be upheld even under the assumption of the opposite extreme infall time distribution, lending confidence that this finding is robust w.r.t the actual infall time distribution.\newline

\subsection{Sources for Replenishment}\label{SOURCESREP}
In general, we find the observed distributions of $\Delta \mathrm{log}(\psi_*)$ for satellite spiral/disk-dominated galaxies to imply that, upon becoming satellites, these objects must experience star formation at the level of comparable field galaxies for prolonged periods (several Gyr) if not continuously. In turn, this requires a replenishment of the ISM consumed by star formation, naturally raising the question as to the nature of the gas reservoir from which the replenishment is fuelled. In particular we wish to establish whether 
the reservoir can be entirely comprised of gas associated with the galaxy upon infall, i.e the ISM and the more loosely bound circum-galactic medium (CGM), or
must instead/also be sourced from the IHM of the galaxy group.
As mechanisms to replenish the HI reservoirs of galaxies from ionized hydrogen have been put forward \citep[e.g.][]{HOPKINS2008}, we include the dominant ionized component of the CGM in our considerations. For the ISM, where the mass fraction of ionized gas is generally  found to be $\lesssim10$\% in spiral galaxies such as those in our sample, we consider the cold/neutral gas mass as representative of the total ISM mass.\newline 

\subsubsection{Gas associated with the galaxy upon infall}\label{gasatinfall}
We begin by considering the gas associated with the galaxy at the time it first became a satellite as a possible reservoir from which to the observed on-going star formation might be fuelled. This reservoir consists of the ISM of the galaxy, distributed on scales of ca. $10\,$kpc, as well as of the more loosely bound CGM. Recent work on the CGM of isolated typical $L^*$-galaxies out to $z=0.35$ has found that it may contain a gas mass comparable to the stellar mass of the galaxy within a physical radius of $150\,$kpc \citep[e.g.][]{TUMLINSON2011,TUMLINSON2013}, with $1 - 10$ \% of this gas being cold neutral and/or molecular hydrogen.\newline

Based on our results, we rule out the ISM of the galaxy upon infall as the sole reservoir of fuel for star formation. All disparate models which recover the observed distributions of sSFR require strong flows of gas both into and out of the ISM (see Table~\ref{tab_INFLOW}). These findings are in conflict with the ISM being the only source of fuel, because the outflows reduce the residence timescale
of the ISM to around a Gyr, and because in any case the models
require inflows originating exterior to the ISM. Furthermore, if the ISM were the only reservoir of fuel for star formation,
the Schmidt-Kennicutt relation would lead one to expect a gradual decline
of the SFR, beginning upon a galaxy becoming a satellite. Such a SFH, however, would correspond to the infall-quenching model rejected in Sect.~\ref{SFR_SAT_1pm}, and not to the preferred two parameter models.\newline 

Although our empirical results favour significant flows of gas into and out of the ISM it is possible that these outflows remain bound to the galaxy, i.e have their end-point in the CGM, and can be recycled into the ISM at a later time. Accordingly, we consider the ability of the combined ISM and CGM to support the required star-formation, initially assuming all gas to remain bound, and the complete CGM and ISM to be retained upon infall. To this end, we begin by comparing the stellar mass growth of the satellite galaxies during the period of on-going star formation to the expected total ISM and CGM mass at infall, again distinguishing between the low and high stellar mass ranges. To estimate the stellar mass growth, we consider our fiducial galaxies with $M_* = 10^{9.75} M_{\odot}$ and $M_* = 10^{10.3} M_{\odot}$ and assume a star formation rate for these galaxies based on the $\psi_* - M_*$ relation defined using the \textsl{\textsc{fieldgalaxy}} sample. We can then estimate their stellar mass growth over the period of a characteristic delay time by evolving the galaxies backwards in time as detailed in Appendix~\ref{APPEND_MODEL}. For the purposes of this estimate we conservatively adopt a delay time of $2.5\,$Gyr as a fiducial time for sustained star formation at the level of a field galaxy. This time corresponds to the shortest fixed delay time preferred by the models considered.\newline

Estimating the stellar mass growth in this fashion, we find the lower stellar mass galaxy to have increased its stellar mass by $2.2\times10^{9} M_{\odot}$ ($80$\%; $M_{*,\mathrm{infall}} = 10^{9.45}M_{\odot}$) over the $2.5\,$Gyr prior to observation, and the stellar mass of the higher stellar mass galaxy to have increased by $5.4\times10^{9}M_{\odot}$ ($37$\%; $M_{*,\mathrm{infall}} = 10^{10.17}M_{\odot}$).\newline

As current satellite galaxies first became satellites at an earlier time (corresponding to a higher redshift) when galaxies were relatively more gas rich, the initial mass of the ISM upon infall may have been larger than that still present at the redshift of observation. In fact, recent work modelling the evolution of the total gas fraction of star-forming disk galaxies indicates that the total (cold) gas fraction $f_{gas} = (M_{\mathrm{H I}} + M_{\mathrm{H}_2} )/ (M_{\mathrm{H I}} + M_{\mathrm{H}_2} + M_* )$ was $\sim1.5\,$times greater at $z\gtrsim0.5$ then at $z=0$\citep{LAGOS2011,POPPING2014} for a given stellar mass\footnote{While the evolution of the total cold gas fraction is mild, \citet{POPPING2014} predict the molecular gas fraction to evolve strongly with redshift, in agreement with observations \citep{TACCONI2010,GEACH2011}.}. Following \citet{POPPING2014}\footnote{We make use of the total cold gas fraction as a function of stellar mass at different redshifts provided in \citet{POPPING2014}, interpolating between these to a redshift of $z=0.35$.} we estimate an ISM mass of $2.1\times 10^{9} M_{\odot}$ ($0.75\times M_{*,\mathrm{infall}}$) for the low stellar mass galaxy at infall and a mass of $5.3 \times 10^{9} M_{\odot}$ ($0.36\times M_{*,\mathrm{infall}}$) for the ISM of the high stellar mass galaxy. Combined with an estimate of the CGM mass at infall being equal to the stellar mass at that epoch, we thus estimate a total reservoir mass at infall of $5\times10^9 M_{\odot}$ and $2\times10^{10}M_{\odot}$ in the low and high stellar mass case, respectively. Contrasting these masses with the stellar mass growth, we find that $41$\%, respectively $27$\%, of the gas associated with the galaxy at infall would be required to fuel the stellar mass growth, making fuelling of the required star formation from the joint ISM and CGM at infall a seemingly feasible proposition.\newline

In this estimation we have assumed that the outflows from the ISM remain bound to the galaxy in their entirety. However, at least of order $10$\% of outflows from the ISM of the galaxy are likely to be unbound and escape \citep[e.g.][]{LOEB2006}. Taking this additional  mass loss into consideration, i.e. assuming that $10$\% of the outflows from the galaxy are lost and making use of the ratios of outflow to star formation rate as determined from our models and listed in Table~\ref{tab_INFLOW}, we find that in the low stellar mass range $49 - 69$\% of the combined CGM and ISM must be retained in order to replenish the mass loss due to star formation and outflows, while for the high stellar mass range the figure is $33 - 41$\%. Nevertheless, it thus appears potentially feasible that the mass of gas associated with the galaxy at infall is sufficient to support the required inflows, outflows and star formation, provided it can be retained and can cool efficiently.\newline

\subsubsection{Stripping and Dependence on Galaxy mass/Sub-halo mass\label{stripping}}
Our previous estimate that the gas initially bound to the galaxy at infall in the form of the CGM, as well as the ISM cycled into the CGM as a result of outflows, suffices to support the observed on-going star formation is predicated on this diffuse component of gas, distributed over scales of $\sim100\,$kpc, remaining bound to the galaxy upon its becoming a satellite, i.e. falling into the more massive DMH of an other galaxy/galaxy group. Thus, the question arises to what extent this extended diffuse reservoir of gas can be retained when the galaxy to which it is initially bound is moving relative to the pressurized diffuse IHM of the satellite spiral's host galaxy group, i.e. to what extent ram pressure and/or tidal stripping will unbind and remove this gas reservoir.\newline 

We begin by addressing this question empirically, making use of the observed distributions of $\Delta \mathrm{log}(\psi_*)$. As we have shown in Section~\ref{SFR_SAT}, these are only weakly dependent on the mass of the galaxy, with, if anything stronger effects observed in higher mass galaxies. As the ability of a galaxy to retain gas against the effects of ram-pressure stripping is expected to increase with the depth of its potential well, i.e with its mass, one would expect higher mass satellite spiral galaxies to be better able to retain their CGM than lower mass galaxies.  Furthermore, the probability for a low mass system ($M_* \le 10^{10} M_{\odot}$) to experience a tidal stripping encounter with a higher mass system which can remove gas from the lower mass partner is greater than that for a higher mass system ($M_* > 10^{10} M_{\odot}$). As a result, if retained and recycled CGM and ISM were to constitute the main source of replenishment of the ISM, one would expect shorter delay times $t_{\mathrm{delay}}$ and higher quenching probabilities $P_{\mathrm{quench}}$ in lower mass galaxies/sub-halos than in higher mass systems, as well as a higher rate of the occurrence of strongly quenched galaxies at lower galaxy stellar mass. In other words, one would expect the distribution of $\Delta \mathrm{log}(\psi_*)$ to be more strongly skewed towards low values for low stellar
mass galaxies than for high stellar mass galaxies. Our empirical results, however, are completely
contrary to this expectation. Comparison of the distribution of $\Delta \mathrm{log}(\psi_*)$ for satellite spiral galaxies and field galaxies in Fig.~\ref{fig_SATCENT_2} shows that the group environment actually has a stronger
effect on the sSFR of higher mass galaxies: the distributions of $\Delta \mathrm{log}(\psi_*)$ for low stellar
mass galaxies are also less skewed towards low value of $\Delta \mathrm{log}(\psi_*)$ than those of higher mass galaxies.
In terms of our modelling in Sect.~\ref{SFR_SAT} this translates into delay times for low mass galaxies
which are equal to, or longer than those for higher mass galaxies,
and quenching probabilities which are smaller for low mass galaxies than for high mass galaxies. The delay timescales (direct and implied by the quenching probabilities), moreover are longer than the expected time to group pericentric passage ($1.5 - 2\,$Gyr; \citealt{HESTER2006}), which is the point in the orbit at which stripping effects will be strongest. Thus, we find our empirical results to disfavour retained CGM as the dominant source for the replenishment of the ISM. At this point we reiterate that we have previously shown that our sample of spiral galaxies is mass complete and volume-limited, even in the low stellar mass range  (Sects.~\ref{SAMPLESELECTION}, \ref{sSFRFIELD}  and recovers the parent sSFR distribution well  (Section~\ref{SAMPLESELECTION} and Appendix~\ref{APPEND_MORPHSEL}) . Therefore, the empirical basis for these findings should be considered robust and physical in nature.\newline

Further support for the findings disfavouring the CGM as the dominant reservoir is also provided by a range of theoretical work considering the ram-pressure stripping of satellite galaxies in galaxy groups and clusters \citep[e.g.][]{HESTER2006,MCCARTHY2008,BAHE2015}. While the temperature and pressure of the diffuse IHM of the galaxy group, responsible for the ram-pressure stripping, is expected to increase with group halo mass, thus enhancing the stripping, the ability of a satellite galaxy to retain its gas reservoirs is expected to increase with increasing depth of its potential well, i.e. with increasing stellar mass, respectively DM sub-halo mass. As shown by \citet[e.g][]{HESTER2006} the ram-pressure stripping process is not scale-free, so that the degree of stripping depends both on the ratio of satellite to group/cluster DMH mass, as well as on the absolute DMH mass of the group/cluster.\newline 

\citet{HESTER2006} has considered the case of satellite spiral galaxies subjected to ram-pressure stripping in groups and clusters of galaxies using a multi-component semi-empirical model, finding that ram-pressure stripping is expected to completely remove any extended gas halo by the first pericentric passage of the satellite (i.e within $1.5 - 2\,$Gyr), even in the galaxy group environment with DMH masses of $\sim10^{13} M_{\odot}$ (but see also \citealt{MCCARTHY2008}). In fact, \citet{HESTER2006} shows that depending on the mass ratio of the satellite galaxy DM sub-halo and the group DMH even the extended ISM disk of the galaxy may be partially stripped in both group and cluster environments. These predictions are also supported by recent detailed simulations presented by \citet{BAHE2015}.\newline

Splitting our sample of satellite spirals into two mass ranges ($10^{9.5} M_{\odot} \le M_* < 10^{10} M_{\odot}$ and $M_* \le 10^{10} M_{\odot}$) we find the median dynamical masses of the galaxy groups hosting the satellite spirals to be $10^{13.4} M_{\odot}$ for the low stellar mass galaxies, and $10^{13.5} M_{\odot}$ for the high stellar mass galaxies. As our fiducial galaxies with $M_* = 10^{9.75} M_{\odot}$ and $M_* = 10^{10.3}$ correspond to the median stellar masses for systems in the low and high stellar mass range respectively, we can use these stellar masses to estimate the median mass of the DM sub-halos by making use of the average stellar mass to halo mass relation presented by \citet{MOSTER2010}. This results in estimated DM sub-halo masses of $10^{11.5} M_{\odot}$ and $10^{11.9}$ for the low and high stellar mass range, respectively, corresponding to logarithmic DM sub-halo to group halo mass ratios of $-1.9$ and $-1.6$. For these ratios of DM sub-halo to group halo mass \citet{HESTER2006} predicts that by pericentric passage, even in galaxy groups and low mass clusters, the extended ISM disk of the galaxy will be affected by ram-pressure stripping in approximately its outer third to half, with $\sim50$\% of its total mass being stripped, in addition to the complete removal of the diffuse gaseous halo. For the low and high stellar mass ranges of our sample, the removal of the CGM alone would reduce the fraction of gas retained to to $43$\% and $27$\%, respectively, i.e. below the required levels as previously estimated. Moreover, the expected outflows predicted in our models would easily suffice to cycle $>50$\% of ISM into the CGM within the delay time of $2.5\,$Gyr. \newline

Thus, we find that the results of recent work considering the ram-pressure stripping of satellite spiral galaxies in the group environment disfavour the retention of a large fraction of the diffuse CGM and recycled ISM, in line with our empirical finding. Accordingly, although the mass \textit{at infall} of the CGM and ISM of a spiral galaxy becoming a satellite may, in principle, be sufficient to sustain the required inflow and star-formation over prolonged periods of the galaxy's satellite lifetime, it seems likely that only an insufficient fraction of this reservoir can actually be retained in the group environment and contribute to the fuelling of the galaxy. Accordingly, it is probable that at most part of the fuel required for the observed on-going star formation of satellite spiral galaxies can be sourced from gas which was associated with the galaxy upon infall, with at least a significant fraction being sourced from gas not initially associated with the galaxy.

\subsubsection{Replenishment from the IHM}\label{repIHM}
With our empirical results disfavouring the combined CGM and ISM as the source for the inflow of gas required, instead favouring a further reservoir not associated with the galaxy, by process of elimination we conclude that our empirical analysis implies an inflow of gas from the IHM of the group into the ISM of the satellite spiral galaxy as a mechanism to meet the demands of the extended period of star formation at the level of a comparable field galaxy implied for satellite spiral galaxies. Although the exact rate of inflow from the IHM required depends on the degree to which the CGM can be retained, as well as the fraction of ISM mass lost to winds or while being cycled through the CGM, it nevertheless appears that an inflow from the IHM of order the SFR is required, contrary to the standard paradigm.\newline

Before discussing the physical implications we first consider whether a  possible factor to at least partially
ameliorate this conclusion may lie in our simplifying assumption in our modelling that
the residence time $ \tau_{\mathrm{res}}$ (i.e. the the time a unit mass of
gas spends in the disk of a spiral galaxy before being
expelled, introduced in Sect.~\ref{GASFUELLING} and Eq.~\ref{eq_taures} of Appendix~\ref{APPEND_GASCYCLE}) is
independent of environment, depending only on galaxy specific properties, in particular galaxy mass.
For satellite spiral galaxies in galaxy groups, the surrounding medium may, in fact, be more pressurized than for a similar stellar mass galaxy in the field, decreasing the outflows from the satellite galaxy and increasing $\tau_{\mathrm{res}}$. 
However, if the medium surrounding the galaxy were sufficiently pressurized to fully suppress wind-driven outflows from the ISM, this medium would also act to enhance the efficiency of the removal of the CGM via ram-pressure stripping \citep{HESTER2006}. As discussed in Sect.~\ref{stripping}, ram pressure stripping is likely sufficient to remove not only the CGM of spiral satellite galaxies, but also part of the ISM of these systems, even in groups of the mass scale considered. The resulting requirement of fuelling from sources external to the ISM, most notably the IHM is further compounded by the fact that the estimated total ISM mass associated with the galaxy upon infall is less than the increase in stellar mass over the fiducial $2.5\,$Gyr delay timescale for both fiducial galaxies considered as detailed in Sect.~\ref{gasatinfall}.
Nevertheless, the suppression of outflows is potentially amenable to testing by considering the metallicity of satellite spiral galaxies (\citealt[e.g.][]{PASQUALI2012}), and will be pursued for this sample in future work. We note, however, that e.g. \citet{PENG2014} have investigated the gas-phase metallicity of star-forming satellite and field galaxies, interpreting their results in the sense of a metal-enriched inflow onto star forming group satellite galaxies, in agreement with our presented results.\newline

Having identified the IHM as a plausible source of fuel to support the inflows and star formation of spiral satellite galaxies, we investigate the viability of this option, considering the IHM as the sole reservoir for the sake of argument. Making use of our group dynamical mass estimates and assuming a universal baryon mass fraction of $\Omega_{b} / \Omega_M = 1/6$, we can estimate the IHM mass for the galaxy groups in our sample\footnote{We equate the dynamical mass of the system to the halo mass and estimate the total baryonic mass using the universal baryon fraction. We then subtract the total stellar mass of all group members to obtain an estimate of the IHM mass.}. Comparing this with the total (current) star formation rate associated with the member satellite spiral galaxies of each  group contained in the \textsl{\textsc{groupgalaxy}} sample, we find that the timescale on which the IHM would be consumed by this star formation activity is $\gtrsim500\,$times the Hubble time. As, however, not only satellite spiral galaxies in the \textsl{\textsc{groupgalaxy}} sample will be forming stars, as a hypothetical limiting case, we also consider the total star formation of \textit{all} member galaxies of a each group, assuming star formation rates based on their stellar mass and the $\psi_*$ - $M_*$ relation for the \textsl{\textsc{fieldgalaxy}} sample. Nevertheless, even in this case, the median exhaustion timescale of the IHM is $\gtrsim150\,$times the Hubble time. \newline

Furthermore, numerical simulations indicate that ambient dark and baryonic matter is being accreted onto the dark matter halos of galaxy groups \citep[e.g.][]{MCBRIDE2009,VANDEVOORT2011,WETZEL2015}. As a result, the IHM of galaxy groups is constantly being replenished.
For the galaxy groups in our sample, we estimate the inflow rate of baryons using Eq.~9 of \citet{MCBRIDE2009}, equating the group dynamical mass estimate to the group halo mass and applying our universal baryon mass fraction. Contrasting this inflow rate with the estimate of the total star formation rate of the \textsl{\textsc{groupgalaxy}} satellite spirals in the group, we find the star formation to equate, on average, to $\sim1$\% of the baryon inflow. Even considering the hypothetical limiting case for the total star formation of group member galaxies, we find that the star formation equates to $\sim10$\% of the baryon inflow.\newline

In summary, the rate of replenishment of the IHM as well as the size of the reservoir imply that, if even only a small fraction of the IHM can cool and be accreted, this reservoir is easily sufficient to support the inflows and prolonged star formation in satellite spiral galaxies required by our empirical results, making the required fuelling from the IHM a viable option.\newline

\subsubsection{Variability of SFR of satellite spiral galaxies}
Finally, in the context of the replenishment of the ISM of satellite spiral galaxies it is interesting to consider the variability of the SFR of these objects. If the required fuelling of satellite galaxies is indeed sourced largely from the IHM of the group, it may be expected to take place indefinitely, rather than only occurring for a limited time. In such a scenario both the quenching and the fuelling of star formation, and as a result the SFR, might be expected to vary on timescales comparable to the orbital timescale of the galaxy as it transits regions in which gas-stripping of the ISM/CGM and the accretion of gas from the IHM into the ISM, respectively, are more/less efficient, e.g. via a dependence on the density and temperature profile of the IHM. Indeed, for the refuelling model we find preferred quenching probabilities of $\sim 0.8 - 1.2\,$Gyr$^{-1}$, corresponding to the inverse of the typical dynamical timescales and which might be related to pericentric passage of the satellite. This would introduce an additional intrinsic scatter in the $\psi_*$--$M_*$ relation at fixed $M_*$ which would potentially offer an explanation
to our empirical result that the intrinsic scatter of this relation is higher (at $0.59\,$dex; $0.44\,$dex $1-\sigma$ equivalent) for grouped
spirals than for field spirals (at $0.36\,$dex; $0.27\,$dex $1-\sigma$ equivalent).\newline

Overall, in a statistical sample, one thus might expect to find satellite galaxies with increasing and declining star formation rates, rather than only such with declining SFR as would be expected if fuelling were sourced from a gas reservoir tightly associated with the satellite galaxy and being slowly depleted. We have shown that a refuelling model is indeed consistent
with the distribution of $\Delta\mathrm{log}(\psi_*)$ for satellites
and indeed provides the best fit to the data of the
models considered. A ready means to investigate
the refuelling hypothesis further in future work would be to
consider the distribution of short wavelength
colors of the galaxy sample, e.g. $\mathrm{FUV} - \mathrm{NUV}$ or $\mathrm{NUV}$ -- $\mathrm{u}$. Provided the timescale the color is sensitive to is short enough, a population of galaxies with increasing star formation should have different colors than one with gradually declining SFR, potentially enabling a distinction between the scenarios. \newline

\section{Discussion}\label{DISCUSSION}
Overall, our investigation of the star formation and gas-fuelling of spiral/disk-dominated galaxies in the local universe has found that these objects are characterized by a rapid cycle of gas into and out of the ISM replenishing the gas consumed by star formation. Based on our consideration of satellite spiral galaxies and regardless of the details of the gas-fuelling model considered, we conclude that the fuelling of spiral galaxies is largely independent of environment, with substantial flows of gas into and out of the ISM of the satellite galaxies on timescales of several Gyr \textit{while} the galaxy is a satellite. Furthermore, consideration of the reservoirs of gas available to these satellite galaxies on infall and the dependence of SFH on galaxy mass favor
scenarios in which this sustained accretion
is fuelled from the IHM of the DM halo of their host galaxy group, rather
than from gas associated with the galaxy before it became a satellite.\newline

\subsection{Implications for the IHM}
The fundamental question posed by our findings, is that of the nature of the mechanism which enables the accretion of gas from the IHM into galaxies in general and into satellite galaxies in galaxy groups in particular. In the mass range of our groups, the virial temperature is generally $\gtrsim10^{6}\,$K, yet IHM gas must be at least as cool as $T_{\mathrm{gas}} \lesssim 10^5\,$K in
order to accrete directly onto the ISM of massive spiral galaxies in our sample
and cooler still to if the accretion occurs first onto the CGM. Our results
also require IHM fuelling of the low mass galaxies in our sample, for which
even lower IHM temperatures are required.
Thus our findings fundamentally require the IHM of galaxy groups to be a two phase medium, encompassing a cold phase, as well as a warm/hot phase, with fuelling of the galaxies occurring more or less continuously in small increments from this IHM, largely independent of the galaxies' environment. This is unlike the standard picture of cold mode accretion with its associated dominant streams and halo mass dependence (\citealt{KERES2005,DEKEL2006,TORREY2012}, but see also \citep{KERES2009a}). Further indirect observational support for a multiphase IHM in galaxy groups may also be supplied by the recent findings that the distribution of Mg II absorption around 'isolated' central galaxies, indicative of a cold clumpy component, is self-similar as a function of dark matter halo mass, extending to halos of mass $M_{\mathrm{halo}} \approx 10^{14}M_{\odot}$ \citep{CHURCHILL2013},
impling that a multiphase medium can exist in halos of the mass scale of the
galaxy groups studied here.\newline

This cold phase of the IHM, by necessity, must have a relatively small volume filling factor and pervade the volume sampled by the orbits of the satellite galaxies, rather than be associated with the CGM of the individual galaxies, as we have previously argued. Future work will have to focus on understanding the origin of this two phase medium, and in particular that of the cold phase. One possible solution is that the hot gas
atmosphere does not extend to the group virial radius, but
rather ends at much lower group-centric distances due
to cooling processes operating in the IHM of
galaxy groups spanning the mass range
from ca. $10^{11.75} to 10^{14.4} M_{\odot}$
of the groups considered in this work.
This in turn would imply
that the tipping point for the free fall timescale
of the groups to exceed the cooling timescale of the
IHM in the groups occurs at higher group masses than
expected, due to some cooling mechanism operating
in the IHM that has not previously been considered.\newline

One possibility is cooling of the IHM due to inelastic collisions
of ions and electrons in the plasma with dust particles. This process
is the most efficient coolant for hot gas with $T \gtrsim 10^{5.5}\,$K,
\citep[see e.g.][]{DWEK1981}.  Simulations
by \citet{MONTIER2004} show that dust cooling
exceeds gas phase cooling processes if the
for dust-to-gas ratio in the IHM exceeds ca. $10^{-4}$ by mass,
(ca. one percent of the value in the ISM).
There is some observational evidence that there
is sufficient dust in the IHM for this mechanism to be operating
in Stefan's Quintet compact galaxy group
\citep{NATALE2010}. The mechanism requires a continuous
injection of dust into the IHM to balance losses of grains
through sputtering in the hot plasma. Possible sources
for dust can be the injection via stars released from
galaxies into the IHM during galaxy-galaxy interactions \citep{NATALE2010}, or through winds 
driven out of satellite galaxies depositing
dust along their orbits. Support for the latter is provided
through observations of individual edge-on spiral galaxies
in the field revealing copious amounts of dust in their CGM
scattering non-ionising UV light from massive stars in the disk
(\citealt{HODGES-KLUCK2014} and \citealt{SEON2014}). If,
as we have argued in Sects.~\ref{stripping} \& \ref{repIHM}, the CGM will be stripped
from the host galaxy on entry into a group, the dust in the CGM 
would thereby be injected into the IHM \citep[see also][]{POPESCU2000b}.
Future analysis of diffuse FIR emission of groups on scales of $0.1-1\,$Mpc
could in principle determine the total cooling rate of the hot
component of the IHM due to dust.\newline

\subsection{Implications for the color--density relation}\label{implications_CD}
Having controlled both for morphology and environment in our analysis, we can also leverage our results to shed light on the mechanisms underlying the color density relation 
\citep[e.g][]{PIMBBLET2002,LEWIS2002,GOMEZ2003,KAUFFMANN2004,BALOGH2004,BALDRY2004,BLANTON2007,BAMFORD2009,CUCCIATI2010,ZEHAVI2011}
. This relation simply states that the colors of galaxies are redder (indicative of less star formation activity) in denser environments. Since the pioneering work of \citet{HUBBLEHUMASON1931,DRESSLER1980}, it is also known, that early type galaxies predominantly reside in denser regions, and it has also been shown, that color/SFR and morphological type of a galaxy are correlated \citep[][]{JAMES2008}. Finally, although local density (as measured in fixed apertures or out to a specified $n^{th}$ neighbor) and host dark matter halo mass are correlated, mapping from one to the other is non-trivial, in particular in the regime of galaxy groups, due to the considerable scatter \citep{HAAS2012}. \newline

The question thus arises whether the color
density relation is driven by galaxy-galaxy interactions or some other
process changing the morphological mix of galaxies,
or whether it is driven by a changed thermodynamic state
of the IHM leading to a decrease in
availability of gas sufficiently cold
to be accreted onto galaxies and fuel star formation.
Since we have controlled for morphology, we can differentiate
between these scenarios. In particular, we have presented
evidence in favour of the on-going gas-fuelling of
a highly pure morphologically selected sample of
disk-dominated galaxies in the group environment,
with this fuelling being sourced from reservoirs
extraneous to the galaxy - in particular the IHM.
We therefore conclude that the color-density relation
is not predominantly due to the gas-fuelling
rate as determined by the host DMH mass, but
rather, to a large part, is due to a change in
the morphological type of galaxies towards more
bulge-dominated systems (a comparison of spiral
fraction for field and group galaxies is shown in
the upper panel of Fig.~\ref{fig_GFCP_SPFRAC})
Most particularly for satellite galaxies in groups,
there seems to be no obvious reason why the rate
of inflow of gas on $> 100\,$kpc scales from the
IHM onto a galaxy should be influenced by whether the
galaxy is disk- or spheroid-dominated. As such, the underlying
physical mechanism of the colour-density relation
likely links the ability of a galaxy to retain gas
and convert it into stars to its morphology, i.e. to
the relative importance of the bulge. This is in line with
findings that galaxies with prominent bulges (which we have
deliberately excluded from our analysis) are
driving the downturn in slope of the SFMS at
higher stellar mass. In this picture, the rate at which star formation is
decreased over time within the group environment is
controlled by the rate at which the galaxy morphology is transformed,
in which case one might expect the bulge-to-disk ratio
to increase with decreasing group centric distance (since the latter is
a proxy for lookback time since a galaxy first entered the group).
Indeed, \citet{GEORGE2013} find exactly this, even amongst the
quenched population (but see \citealt{CAROLLO2016} for a contrasting
view\footnote{\citet{CAROLLO2016} find the morphological mix amongst quenched galaxies in the group environment is constant as a function of group centric distance. Based on this finding they argue in favour of a process linked to the large-scale group DMH halo driving the quenching of satellite galaxies, with the efficiency in terms of number of affected satellites increasing towards the group center, and identifying secular differential fading of the disk component in quenched galaxies as being responsible for the differences in the morphological mix between star-forming and quenched satellites. Considering the implications of the scenario suggested by \citet{CAROLLO2016} in the context of our empirical analysis, we find that, as our morphological selection should be robust to disk fading on the timescale of several Gyr (see Appendix~\ref{APPEND_MORPHSEL}) and the differential fading mechanism proposed by \citet{CAROLLO2016} is predicated on a rapid quenching of star-formation, under this alternative scenario we would expect a substantial population of disk galaxies with very highly suppressed SFR in the group environment, in excess of that found in our analysis. Furthermore, we note that the steep radial age gradients in the stellar population required in terms of the differential fading model appear to be in conflict with the shallow gradients observed in local spiral galaxies \citep[e.g.][]{MACARTHUR2009,SANCHEZ-BLAZQUEZ2011,SANCHEZ-BLAZQUEZ2014}.}).\newline

If morphological transformation is indeed mainly driven by
galaxy--galaxy interactions, the interpretation of the colour-density
relation presented is entirely consistent with the idea that
galaxy--galaxy interactions are the main factor driving the evolution of galaxies in the group environment, as often argued in recent works \citep[e.g.][]{ROBOTHAM2013,ROBOTHAM2014,DAVIES2015,ALATALO2015,BITSAKIS2016}. It seems to be the morphological transformations (in the sense of an increase of the bulge component) triggered by these events, rather than the increasing dominance of hot gas in collapsing structures, which is the main factor
causing the star formation of the
galaxy population to switch off in the present epoch on the Mpc scale of
the composite haloes of galaxy groups.\newline

Nevertheless, we have also shown that there are also (rare) events by which the star formation of a spiral/disk can be quenched without it undergoing morphological transformation, as evidenced by our population of quenched spiral/disks. These findings are in line with \citet{MASTERS2010}, who also identify a population of red spirals with intrinsically low sSFR. However, these will be primarily of interest in what they can tell us about the process of gas-fuelling, rather than in their direct effect on the observed properties of the galaxy population in groups, which as we have shown, is relatively small.\newline

Finally, we may also note that the more complex shorter term variations in SFR
exhibited by interacting galaxy pairs, which are observed to die down
as a function of separation \citep[e.g.][]{DAVIES2015}, are also consistent
with, and may require, a recovery of gas fuelling to its
pre-interaction level
relatively soon after an interaction event has changed the amount and/or
distribution of gas in the ISM of a galaxy.
In this sense, it would appear that galaxy-galaxy interactions
may only be able to manifest as the
dominant process influencing SFR due to the surprising constancy of
the gas fuelling process in all non-interacting systems, independent
of the larger scale environment,
as evidenced by the present analysis. \newline

\subsection{Implications for the morphological transformation of satellite galaxies}
Although the morphological transformation of galaxies from disk- to bulge-dominated systems is generally ascribed to galaxy-galaxy interactions and the secular fading of the disk component after quenching of star formation, the prolonged, substantial accretion (see table~\ref{tab_INFLOW}) of group IHM onto satellite galaxies will impact the morphology of these galaxies and may provide an additional pathway for morphological transformation.\newline

Unlike for central galaxies, where angular momentum is added coherently from the angular momentum of the group, for satellite galaxies accreting gas from the IHM, the accreted gas will have no preferred angular momentum vector with respect to that of the galaxy, resulting in a net angular momentum of zero for the accreted IHM. Considering our fiducial high and low stellar mass galaxies over the characteristic delay time of $2.5\,$Gyr the on-going accretion will result in between $3.4$ and $6.5$ ($2.2$ and $6.9$) times  the ISM mass upon infall being accreted into the ISM of the galaxy in the high (low) stellar mass ranges, and between $2.4$ and $5.4$ ($1.2$ and $5.9$) times the ISM mass being expelled. This will clearly suffice to obliterate the original angular momentum of the gaseous ISM disk. Thus, the fact that star-forming disks are observed in satellite spirals
would seem to imply that angular momentum  from some reservoir can be transferred to the accreted gas. For satellite galaxies being fuelled from the IHM the available reservoirs of angular momentum are (i) the stellar disk, and (ii) the dark-matter sub-halo.\newline

If the angular momentum of the ISM is (partly) replenished from that of the stellar component, then the specific angular momentum of the stellar and gas disks will decrease with continuing accretion. In addition, the formation of additional stars, potentially in part from gas accreted with zero net angular momentum, will further reduce the specific angular momentum of the stellar disk. One consequence of such a decrease in specific angular momentum is that the gas of the galaxy will settle more towards the center of the galaxy. Thus, the more centrally concentrated distribution of gas in satellite galaxies \citep{CAYATTE1994,KOOPMANN2004,CORTESE2010,BRETHERTON2013,DENES2016} may, at least in part, be the result of the continued inflow of gas rather than of the stripping of gas due to environmental processes. Furthermore, this will potentially result in less star formation and redder colors in the outer disks of satellite spiral galaxies than in comparable field galaxies, i.e in different color gradients for these two categories of spiral galaxies.\newline 

In addition, a decrease in the specific angular momentum of the stellar component (primarily built up when the galaxy was a central and subsequently
diluted by stars formed from accreted gas with low angular momentum), will cause the stellar disk of the galaxy to shrink and the old stellar population to compactify (see also \citealt{ELMEGREEN2014}). As a result, at given stellar mass, satellites with on-going accretion of IHM material would be
predicted to have a more dominant bulge component and smaller disks than comparable field galaxies, driving them towards more lenticular/early-type morphologies. As such, on-going accretion onto satellite galaxies may represent a further secular pathway for the morphological evolution of a galaxy from late- to early-type.\newline

To obtain a simple conservative order of magnitude estimate of the reduction in specific angular momentum of the composite gas + stellar system we consider the effect of mass growth as a result of accretion with zero net angular momentum, disregarding, in first instance, the effects of mass outflows and the probable stochastic nature of gas fuelling from the IHM. We consider two cases making use of our fiducial galaxies by (i) adding the total cycled gas mass\footnote{We make use of the inflow rates derived for the re-fuelling model.} to the gas + stellar mass at infall and (ii) by adding (only) the mass of newly formed stars. Assuming the size of the disk of a galaxy is proportional to the specific angular momentum \citep[e.g.][]{BULLOCK2001} one would expect a decrease in disk size by (i) $\sim 0.29\,$dex ($\sim0.54\,$dex) in the high (low) stellar mass range, respectively by (ii) of $0.1\,$dex ($0.16\,$dex) for the high (low) stellar mass range, over a period of $2.5\,$Gyr.\newline

In the light
of the ongoing accretion of gas from the IHM implied by our study,
this would seem to imply that
the specific angular momentum of the ISM and stellar
component would need to be replenished from the DM sub-halo of the
satellite galaxy. As shown in Fig.~\ref{fig_append_atten_1}, which displays the distributions of galaxy size as a function of stellar mass for the \textsl{\textsc{groupgalaxy}} and \textsl{\textsc{fieldgalaxy}} samples, the median size of the \textsl{\textsc{groupgalaxy}} sample is smaller, though only by  $\sim0.03\,$dex (at all stellar masses). In light of the result of ongoing accretion this may imply that the specific angular momentum of the ISM and stellar component is replenished from the DM sub-halo of the satellite galaxy. However, theoretical studies indicate that strong stellar feedback and outflows from galaxies may enhance the effective retention of specific angular momentum in galaxies by preferentially removing low angular momentum gas \citep[e.g.][]{SOMMER-LARSEN1999,GOVERNATO2007,AGERTZ2011,DALLAVECCHIA2012,VOGELSBERGER2013,UEBLER2014,GENEL2015}, in which case our simple consideration of case (ii) might be more appropriate, mitigating the need for angular momentum transfer from the DM sub-halo.\newline 
A detailed consideration of these processes for satellite
galaxies is beyond the scope of this paper, and will require
future detailed theoretical and empirical consideration.
Here, we limit ourselves to drawing attention to the possibility that
that the inflow of IHM gas onto satellite galaxies at the rate
implied by our measurements of SFR represents a secular process
through which disk-dominated galaxies can evolve into spheroid-dominated
ones. Moreover, as discussed in Sect.\ref{implications_CD}, our analysis of the
gas flows required to reproduced the observed distributions of SFR in
our very pure sample of disk-dominated satellite galaxies show that
it is most likely a change in the mix of galaxian morphologies
in the group environment compared to the field, rather than a
reduced propensity of the IHM to cool and fuel star formation, that is
primarily responsible for the reduction in star formation
activity of the galaxy population as a whole in groups
compared to the field. It therefore follows,
somewhat paradoxically, that the ongoing gas fuelling of disk galaxies
in the group environment may itself lead to a secular quenching of the
star formation of galaxies after falling into groups. This route for
the quenching of star formation in disk galaxies would be tantamount
to death by gluttony, in marked contrast to a death by starvation
which most previous studies have invoked.
Future work will place further constraints on the efficacy and timescales
for this secular quenching mechanism and the relative importance to
mechanisms for morphological transformation and quenching related to
galaxy-galaxy interactions by considering the
star-formation rates and star-formation histories of group galaxies
divided into to finer morphological classifications, most particularly
in the range S0 to Sa.\newline

\section{Summary \& Conclusions}\label{CONCLUSION}
Making use of morphologically selected samples of disk-dominated/spiral galaxies we have conducted a detailed investigation of the impact of the group environment on the star formation activity of central and satellite group spiral galaxies, as well as of a sample of largely isolated field (central) spiral galaxies, isolating the effects of galaxy--IHM interactions from those of galaxy--galaxy interactions. We have described the samples in detail and present the results of our analysis as an empirical reference for current and future theoretical work aimed at understanding the importance and impact of galaxy--IHM interactions, including gas-fuelling, for the evolution of galaxies in the group environment.\newline

This analysis has made use of the NUV emission of a galaxy as a tracer of its SFR, rather than H$\alpha$. In addition to reliably sampling the total star formation activity of the galaxy (which may be inhomogeneously distributed), this choice also renders our analysis largely robust against uncertainties of the IMF as well as against stochastic variations in the SFR, while providing enough time resolution to resolve (in time) processes linked to the environment and its characteristic timescale of $\sim1\,$Gyr. Furthermore, we have employed newly developed radiation transfer based techniques to account for the effect of dust on the ratio between observed and intrinsic NUV emission, enabling the intrinsic SFR of the galaxy sample to be determined with great precision, including the SFR of a full set of quenched spirals.\newline

Having made the isolation of the effects of galaxy--IHM interactions from those of galaxy--galaxy interactions a main objective of our sample construction, we have been able to interpret our empirical results on the SFR distribution of our galaxy sample in terms of the gas cycle of these galaxies via implementation of the Schmidt-Kennicutt relation\footnote{Although our inferences are thus, by necessity indirect, they will be testable by SKA pathfinders - e.g. the DINGO survey of the ASKAP will provide HI data of sufficient depth covering the GAMA regions to test if and how the gas content of spiral galaxies reflects the NUV based sSFR.}, with particular focus on their gas-fuelling, i.e the accretion of gas from the IHM onto the galaxy. This has led us to a number of new results, some of which force us to question our knowledge of the process and regulatory agents of gas accretion by galaxies in the group environment. In the following, we briefly summarize our main results and conclusions as presented and discussed in the preceding sections.\newline

\begin{itemize}

\item{\textbf{Central spiral galaxies}\newline
\noindent In our analysis we have considered the isolated field central spiral galaxies and group central spiral galaxies separately. In doing so, we have found that:

\begin{itemize}
	\item{the $\psi_*$--$M_*$ relation for largely isolated central spiral galaxies is well characterized by a single power law $\psi_* \propto M_*^{\gamma}$ with $\gamma = -0.45 \pm 0.01$ and a very low scatter of $0.36\,$dex interquartile (1-$\sigma$ $0.27\,$dex) around the relation (see Fig.~\ref{fig_FIELDSSFRSM} and table~\ref{tab_PLfits}). This also implies that the turnover in the main sequence of star-forming galaxies reported by other authors \citep[e.g.][]{SCHREIBER2015,LEE2015,ERFANIANFAR2016} is due to an increase of (more) bulge-dominated galaxies at higher mass in their samples of star-forming galaxies. }
	
	\item{the existence of a remaining population of strongly quenched field spiral galaxies, predominantly at higher stellar mass, in spite of the reduction in scatter, implies the existence of a possibly mass dependent secular quenching mechanism for field spiral galaxies.} 

	\item{the normalization of the  $\psi_*$--$M_*$ relation for largely isolated central spiral galaxies evolves gradually but noticeably over the short redshift range of $z=0.05 - 0.1$ (see Fig.~\ref{fig_SSFRSMFIELD_LOCDIST}). This evolution is shown to be in agreement with that predicted by the empirical fit to the evolution of the main sequence of star-forming galaxies presented by \citet{SPEAGLE2014}.}

	\item{the $\psi_*$--$M_*$ for group central spiral galaxies is very close to and largely coincides with that for field central spiral galaxies over the full mutual range of stellar mass, implying a lack of environmental dependence of the gas-fuelling of central spiral galaxies (see Fig.~\ref{fig_SATCENT_1}.}

	\end{itemize}
		
	A further discussion of these findings is deferred to a future paper in this series.\newline	}

\item{\textbf{Satellite Spiral Galaxies}\newline
\noindent Considering the $\psi_*$--$M_*$ relation for satellite spiral galaxies we find the median relation to be offset from that of the field and group central spiral galaxies by $\sim0.1 -0.2$ dex at all $M_*$ (see FIg.~\ref{fig_SATCENT_1} and table~\ref{tab_PLfits}). Making use of the full distribution of the offset of $\psi_*$ around the median for a given stellar mass $M_*$, we find the offset in the median to arise from a minority population of galaxies with strongly suppressed sSFR with respect to that of comparable field spiral galaxies, while the majority of satellite spiral galaxies at all $M_*$ displays sSFR akin to those of comparable field spiral galaxies (see Fig~\ref{fig_SATCENT_2}).\newline 

\noindent Contrasting the observed distributions of sSFR for group spiral galaxies with those obtained from a number of empirically informed models of the star formation history of spiral galaxies in the group environment designed to bracket the range of plausible star formation histories (see Figs.~\ref{fig_2pmod_hm}, \ref{fig_2pmod_lm}, \& \ref{fig_Qm}), we find that:

	\begin{itemize}
		\item{The gas-cycle of spiral/disk galaxies is characterized by a rapid cycle of gas into and out of the ISM, with rates of inflow and outflow comparable to or larger than the star formation rate. Furthermore, this rapid cycle is largely independent of the galaxies environment, being inferred for field, group central, and group satellite spirals.}
		
		\item{In order to reproduce the observed distributions, contrary to the standard paradigm of satellite galaxy evolution, we require the on-going replenishment of the ISM of spiral satellite galaxies \textit{while} they are satellites in a galaxy group. This replenishment must take place over Gyr timescales and be comparable to that which is generally assumed to support the quasi-constant star formation in field spiral galaxies.}
		
		\item{Furthermore, simple conservative considerations of the depletion timescales and gas reservoirs for group satellite spiral galaxies favour the IHM of the host group being accreted into the ISM of satellite galaxies, rather than material associated with the galaxy at the time it became a satellite, as source of fuel for this replenishment, also contrary to the standard paradigm of satellite galaxy evolution.}

	\item{The on-going fuelling of spiral satellite galaxies implies that the color-density relation is the result of an increase in the fraction of morphologically late-type galaxies in denser environments rather than to an environmental effect on the gas-fuelling of galaxies. The dichotomy in sSFR at given stellar mass between early- and late-type galaxies, accordingly, is driven by galaxy-specific processes likely linked to their morphology.} 

	\item{The implied on-going substantial accretion of gas with zero net angular momentum by satellite spiral galaxies represents an additional efficient mechanism capable of facilitating the morphological transformation of late-type galaxies to more bulge-dominated earlier types in the (denser) environment of galaxy groups. Potentially, therefore, this continued gas accretion and star-formation will lead to a gradual build up of spheroidal components in satellite disk galaxies which, in turn, will lead to a secular quenching of the star-formation, representing a 'death by gluttony', in sharp contrast to the 'death by starvation' scenario previously invoked for such mechanisms.}
	
	\end{itemize}
  }
\end{itemize}

Overall, our analysis has returned a number of surprising results which are difficult to reconcile with the standard picture of galaxy evolution in the group environment. The emerging picture is that of an on-going process of gas-fuelling for both central and satellite spiral galaxies, largely independent of environment, supporting spiral galaxies as systems characterized by a rapid in- and out-flow of gas, cycling the fuel needed to support star formation in and out of the ISM of the disk and replenishing it as required. Nevertheless, a small minority of satellite spiral galaxies with strongly quenched star formation is observed, whose provenance remains unclear. Overall, we are left to conclude that our current understanding of galaxy evolution in the group environment remains incomplete.\newline 

\section*{Acknowledgements}
We would like to thank the anonymous referee for insightful comments which helped us improve the manuscript. MWG and RJT thank Jay Gallagher for helpful discussions during the formative phases of this project.\newline
MWG gratefully acknowledges the support provided by the International Max-Planck Research School on Astronomy and Astrophysics Heidelberg (IMPRS-HD) and the Heidelberg Graduate School for Fundamental Physics (HGSFP) during the performance of some of the work presented here. CCP acknowledges support from the Leverhulme Trust Research Project Grant RPG-2013-418 and from a previous grant from the UK Science and Technology Facilities Council (STFC; grant ST/J001341/1).\newline
GAMA is a joint European-Australasian project based around a spectroscopic campaign using the Anglo-Australian Telescope. The GAMA input catalogue is based on data taken from the Sloan Digital Sky Survey and the UKIRT Infrared Deep Sky Survey. Complementary imaging of the GAMA regions is being obtained by a number of independent survey programs including GALEX MIS, VST KiDS, VISTA VIKING, WISE, Herschel-ATLAS, GMRT, and ASKAP providing UV to radio coverage. GAMA is funded by the STFC (UK) , the ARC (Australia), the AAO, and the participating institutions. The GAMA website is: http://www.gama-survey.org. \newline
GALEX (Galaxy Evolution Explorer) is a NASA Small Explorer, launched in April 2003. We gratefully acknowledge NASA's support for construction, operation, and science analysis for the GALEX mission, developed in cooperation with the Centre National d'Etudes Spatiales (CNES) of France and the Korean Ministry of Science and Technology. We are grateful to the institutions in the
GAMA/H-Atlas/Dingo Consortia which made possible
further GALEX observations of GAMA fields as part of the
GALEX Sky Completion Project.\newline

\appendix

\section{Appendix A: Selection of Disk/Spiral Galaxies}\label{APPEND_MORPHSEL}
A main requirement of our analysis is that the sample of spirals used be selected purely based on morphology, and provide an unbiased representation of the SFR distribution of spiral galaxies both in the field and in groups. Furthermore, the sample used must be as pure and simultaneously complete as possible. Recently, \citep{GROOTES2014} have presented a method of morphologically identifying spiral galaxies, capable of meeting these requirements. Their method identifies the morphology of a galaxy based on its position in a three dimensional parameter space spanned by a range of optical wavelength galaxy properties. In our analysis we have chosen to use the parameters S\'ersic index $n$, $r$-band effective radius $r_e$ as determined from the single S\'ersic fit, and $i$-band absolute magnitude $M_i$, i.e. the combination (log($n$),log($r_e$),$M_i$), which has been shown to recover highly complete, pure, and unbiased samples of spiral galaxies.\newline

\begin{figure}
\plotone{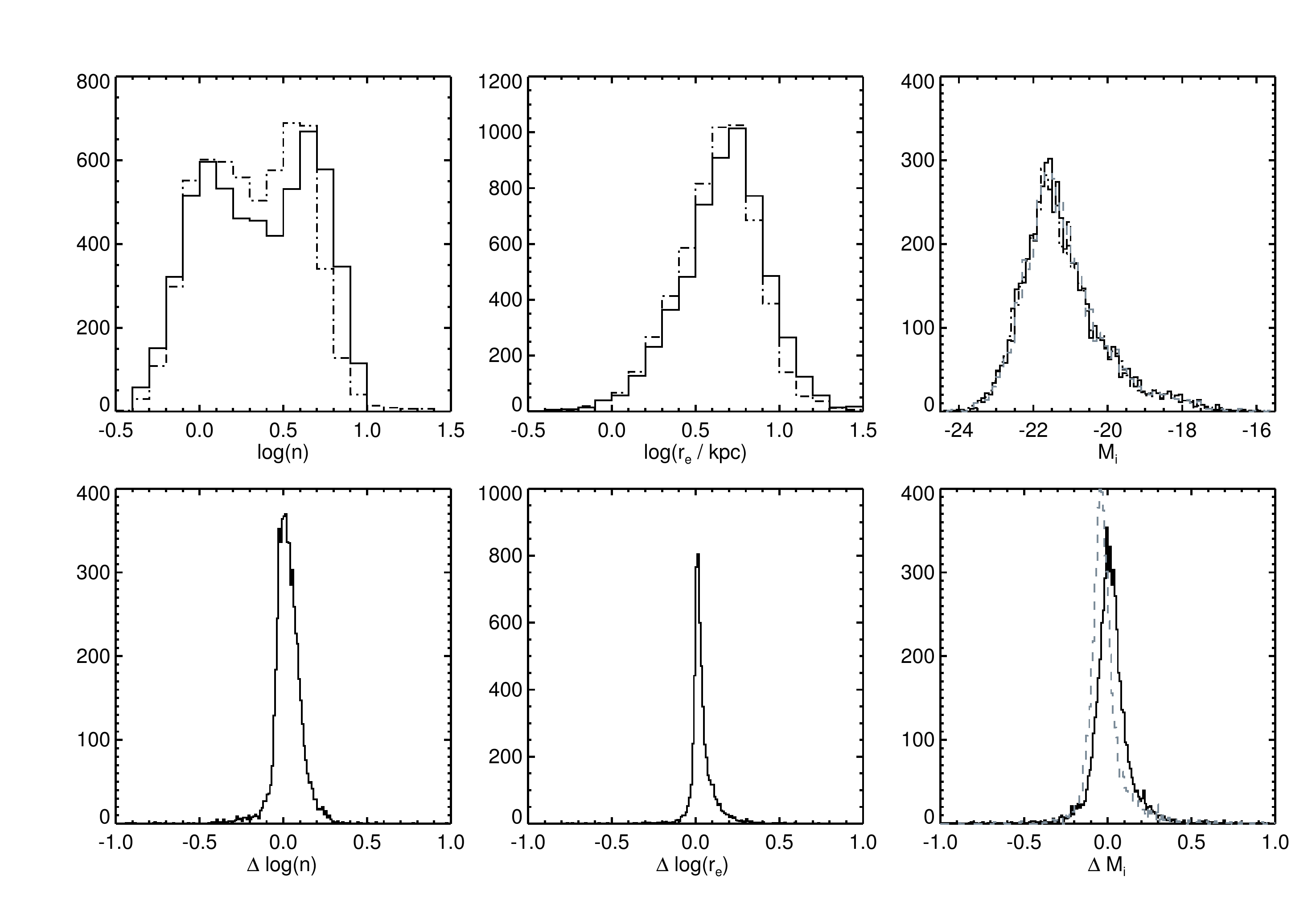}
\caption{\textbf{Top row:} Distributions of the parameters log($n$), log($r_e$), and $M_i$ for the 5747 galaxies common to GAMA and the data set used by \citet{GROOTES2014}. The values from \citet{GROOTES2014} are shown as solid lines and those determined by GAMA are shown as dash-dotted lines. For the parameter $M_i$ the GAMA single S\'ersic magnitudes are shown as a black dash-dotted line, and the fixed aperture magnitudes are shown as a gray dashed line.
\textbf{Bottom row:} Distributions of the differences of the parameter values (value \citet{GROOTES2014} - value GAMA) for common galaxies. For $M_i$ the difference to the GAMA single S\'ersic profiles is shown as a black solid line, while the difference to the fixed aperture photometry is shown as a gray dashed line.}
\label{fig_append_morph_1}  
\end{figure}

As emphasized in \citet{GROOTES2014}, their classification tables have been calibrated using SDSS DR7 photometry, and single S\'ersic fits performed by \citet{SIMARD2011} using \texttt{GIM2D}, and researchers using the classifications are cautioned to check whether their data are compatible, ideally by using a common sub-sample. Fig.~\ref{fig_append_morph_1} shows the distributions of log($n$),log($r_e$), and $M_i$
for 5747 galaxies common to GAMA and the data set used in \citet{GROOTES2014}. The agreement in the parameter values is very good for all parameters, so that we find the selection scheme to be applicable as calibrated.
While \citet{GROOTES2014} extrapolate the SDSS DR 7 Petrosian photometry to total S\'ersic magnitudes using the prescription of \citet{GRAHAM2008}, GAMA provides both single S\'ersic profile and fixed aperture $i$-band photometry. For the purpose of identifying galaxies using the method of \citet{GROOTES2014} we have made use of the single S\'ersic (total) magnitudes. However, with the GAMA data used in our analysis reaching a depth of $r \le 19.4$ and extending to a redshift of $z=0.13$, a non-negligible fraction of the fainter sources is only marginally resolved. For these, S\'ersic profile fitting may not always provide the most accurate or reliable estimate of the total galaxy flux. In selecting our sample, we have therefore independently classified our sample using the GAMA $i$-band single S\'ersic profile and fixed aperture photometry. Sources with differing classifications have been visually inspected and manually classified to obtain our final sample of spiral galaxies. In selecting spiral galaxies we have chosen to use the calibration of the combination (log($n$),log($r_e$),$M_i$) using threshold values of $\mathcal{F}_{\mathrm{sp}} \ge 0.4$ and  $\Delta \mathcal{F}_{\mathrm{sp,rel}} \le 1$ as specified in \citet{GROOTES2014}. Finally we emphasize that, although largely complete, as demonstrated in \citet{GROOTES2014}, this selection places slightly more emphasis on the purity of the spiral samples.\newline

While \citet{GROOTES2014} have demonstrated the performance of the selection method on samples with no distinction of galaxy environment, for the purposes of defining the \textsl{\textsc{groupgalaxy}} sample, we must consider an additional possible difficulty. If a spiral galaxy consists of an old stellar bulge component with no (or little) star formation, as well as a disk in which the bulk of the star formation takes place, a cessation of star formation will cause the disk to fade relative to the bulge. In turn, this might affect the values of $n$ and $r_e$, causing a spiral galaxy to no longer be classified as such without any actual change in morphology. Importantly, this would lead to a preferential loss of quenched spiral galaxies.\newline

To gain a simple insight into the potential impact of this scenario we consider a fiducial spiral galaxy which consists of a bulge component with luminosity $B_0$ and a disk component with luminosity $D_0$ such that the total luminosity $T_0 = B_0 + D_0$. We further assume the age of the stellar population of the bulge to be such that any fading over a timescale of several Gyr prior to observation is negligible, i.e $B_{\mathrm{fade}} = B_0$, and assume all recent and mid-term star formation ($\lesssim 8\,$Gyr prior to observation) to have taken place in the disk. For a given quenching/fading scenario we can express the luminosity of the faded disk component as $D_{\mathrm{fade}} = \eta D_0$, which enables us to estimate the faded bulge-to-total ratio as 
\begin{equation}
\left(B/T\right)_{\mathrm{fade}} = \frac{\left(B/T\right)_{0}}{\eta + (1- \eta)\left(B/T\right)_{0}}\,
\label{eq_fade}
\end{equation}
following \citet{CAROLLO2016}. \newline

Combining the visual morphological classifications of GAMA sources with $z\le 0.06$ presented by \citet{KELVIN2014} and the bulge-disk decompositions of this sample presented by \citet{LANGE2016} and cross-matching with our samples using GAMA's uniqe source identifier,we find
an average $r$-band $B/T$ value of $B/T = 0.3$ ($0.18$) in the high (low) stellar mass range of the \textsl{\textsc{fieldgalaxy}} sample and a comparable value of  $B/T = 0.32$ ($0.19$) for those galaxies in the \textsl{\textsc{fieldgalaxy}} parent sample visually classified as S0/Sa -- Sd/Irr. As fading will only act to increase the $B/T$ ratio, we restrict ourselves to considering the high stellar mass range as our fiducial galaxy. As such, for the star formation history of our disk component we adopt an exponentially declining SFH over the last $8\,$Gyr prior to observation with a timescale of $3.5\,$Gyr, which provides a good fit to the SFH of our fiducial high stellar mass galaxy evolving according to the SFMS of \citet{SPEAGLE2014} over that time. We use this SFH as our baseline and modify it in accordance with a range of quenching scenarios. Subsequently, we make use of \texttt{STARBURST99} \citep{LEITHERER1999,LEITHERER2014} to model the (intrinsic) spectral energy distribution of the disk component and determine the degree of fading for each scenario.\newline

\begin{figure}
\plotone{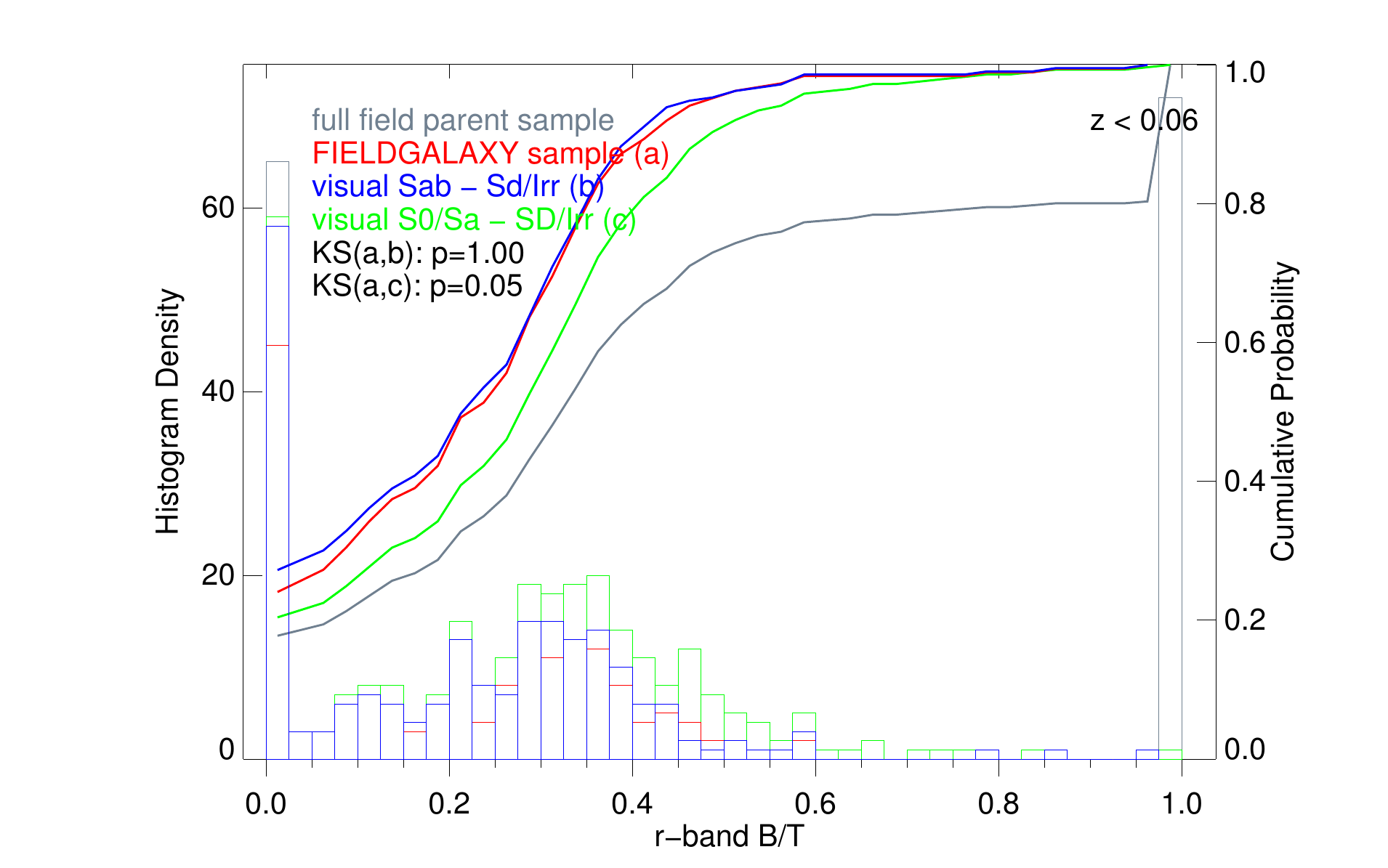}
\caption{Distribution of $r$-band $B/T$ values for GAMA sources in the \textsl{\textsc{Fieldgalaxy}} (red) and \textsl{\textsc{Fieldgalaxy}} parent sample (gray) with $z \le 0.06$ as determined by \citet{LANGE2016}. Those members of the \textsl{\textsc{Fieldgalaxy}} parent sample visually classified \citep{KELVIN2014} as Sab -- Sd/Irr galaxies are shown in blue, while the superset additionally including S0/Sa galaxies is shown in green. Cumulative distribution functions are overploted as colored lines. The p-values of Kolmogorov-Smirnoff tests comparing the distributions are shown in the figure.}   
\label{fig_append_BD}  
\end{figure}

Assuming an exponentially declining star formation rate over the $2.5\,$Gyr ($1\,$Gyr, $2\,$Gyr, $3\,$Gyr) prior to observation with a time constant of $0.5\,$Gyr, i.e a rapid quenching \citep[e.g.][]{WETZEL2013} , we find $\eta = 0.64$ ($0.82$,$0.69$,$0.59$) and  $\left(B/T\right)_{\mathrm{fade}} =0.40$ ($0.34$, $0.38$, $0.42$). Considering slower declines in the star formation activity, i.e. longer time constants as favoured by more gradual quenching models and some quenching models in our analysis, the change in $B/T$ ratio is even smaller. For a time constant of $1\,$Gyr we find a $\left(B/T\right)_{\mathrm{fade}}$ of $0.37$ ($0.32$, $0.35$, $0.39$) for $2.5\,$Gy ($1\,$Gy, $2\,$Gyr, $3\,$Gy) evolution, while for a time constant of $1.5\,$Gyr the $B/T$ ratio decreases to $0.35$ ($0.31$,$0.34$,$0.36$) for an onset of quenching $2.5\,$Gyr ($1\,$Gyr, $2\,$Gyr, $3\,$Gyr) prior to observation. As such, we do not expect disk fading due to quenching over timescales of several Gyr to shift a significant fraction of sources out of the range of $B/T$ values recovered by the method of \citet{GROOTES2014} (see Fig.~\ref{fig_append_BD}). Accordingly, we conclude that , while the secular fading of stellar disks may lead to the loss of some quenched galaxies from the \textsl{\textsc{groupgalaxy}} sample, this will affect only a minority of potential sample members and won't strongly bias the samples. Furthermore, as the adopted selection method recovers the underlying SFR distribution of disk galaxies (see \citealt{GROOTES2014} and Section~\ref{SAMPLESELECTION_SPIRALS}), our conservative treatment of the infall time distribution can adequately limit any potential remaining bias.\newline

\section{Appendix B: Deriving Attenuation Corrections}\label{APPEND_ATTCOR}
Dust in the inter-stellar medium of galaxies can strongly affect the ratio of observed to intrinsic emission from these objects, typically attenuating the emission of late-type galaxies by a factor of 2-10 in the NUV. Thus, in order to make use of the NUV emission of a galaxy as a tracer of its SFR, it is essential to correct for this attenuation and make use of the \textit{intrinsic} emission. In principle, an accurate correction is only possible by modelling the full FUV-FIR SED using radiation transfer techniques in conjunction with independent knowledge of the geometry and orientation of the galaxy. However, the majority of galaxies in our sample are not detected in the FIR. We therefore make use of the method of obtaining accurate radiation-transfer-based attenuation corrections for samples of spiral galaxies without available FIR data presented by \citet{GROOTES2013}. These authors have shown that the critical parameter determining the attenuation of emission from a galaxy, i.e the opacity due to dust, can be accurately estimated using the stellar mass surface density. As demonstrated by \citet{GROOTES2013} using the GAMA survey, this technique, which makes use of the radiation transfer model of \citet{POPESCU2011}, enables the derivation of highly accurate attenuation corrections for large samples of spiral galaxies.\newline

In determining attenuation corrections we have proceeded as follows:\newline
The GAMA measurements of galaxy stellar mass and size have been used to determine the effective stellar mass surface density $\mu_*$ as
\begin{equation}
\mu_* = \frac{M_*}{2 \pi D^{2}_{A}(z)\theta_{e,ss,r}^{2}}\;,
\end{equation}
where $D_{A}(z)$ is the angular diameter distance corresponding to the redshift $z$, $M_*$ is the stellar mass, and $\theta_{e,ss,r}$ is the angular size corresponding to the effective radius of the $r$-band single S\'ersic profile. Using this estimate of $\mu_*$, we have used Eq.~5 of \citet{GROOTES2013} to determine the central face-on opacity in the $B$-band $\tau^f_B$\footnote{$\tau^f_B$ constitutes a reference value for the radiation transfer model of \citealt{POPESCU2011}. The reader is referred to \citet{GROOTES2013} and \citet{POPESCU2011} for details of the parameters.}. \newline

Next the $r$-band axis ratio of each galaxy, as measured by the single S\'ersic profile fit, is used to estimate its inclination. These inclinations are then corrected for the effects of finite disk thickness as detailed in \citet{GROOTES2013} and in Section 3 of \citet{DRIVER2007}, with an assumed intrinsic ratio of scale-height to semi-major axis of 0.12.\newline
Combining the estimates of the galaxy's $\tau^f_B$ and inclination, we then determine the attenuation of the (NUV) emission using the model of \citet{POPESCU2011}.\newline

The relation between $\tau^f_B$ and $\mu_*$ presented in \citet{GROOTES2013} has been calibrated on a subsample of spiral galaxies and accounts for the effect of dust on observed disk size \citep[e.g.][]{MOELLENHOFF2006,PASTRAV2013a} using the corrections of \citet{PASTRAV2013b}. As this relation was calibrated on a sample of galaxies chosen with no regard to their
environment, it is likely to have been dominated by galaxies which would belong to our \textsl{\textsc{fieldgalaxy}} sample.\newline

\begin{figure}
\plotone{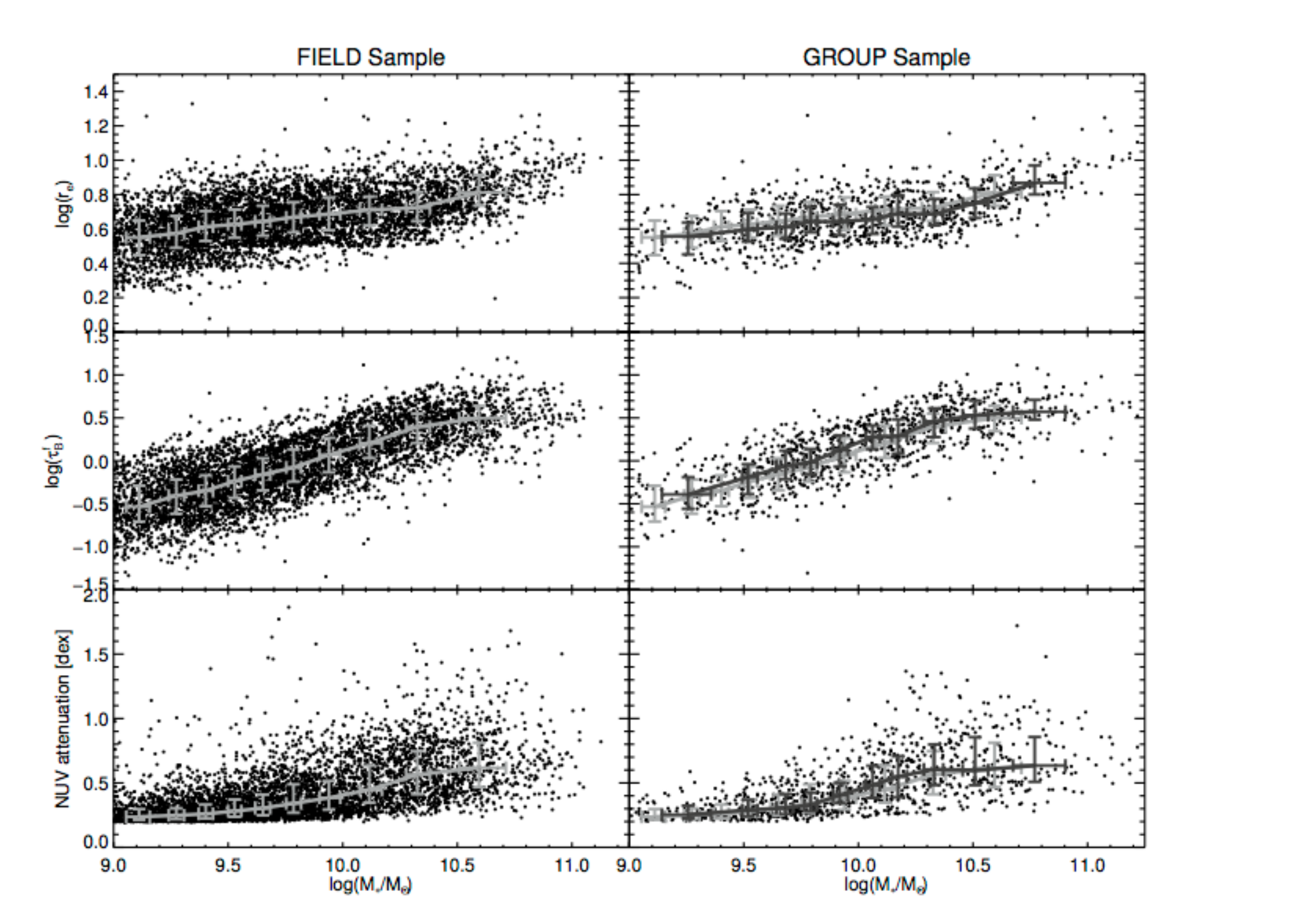}
\caption{Distribution of $r$-band effective radius (top) as a function of $M_*$ for the \textsl{\textsc{fieldgalaxy}} (left) and \textsl{\textsc{groupgalaxy}} (right) samples. The resulting distributions $\tau^f_B$ and of the attenuation corrections applied in the NUV are shown in the middle and bottom panels, respectively. The median distributions for the \textsl{\textsc{fieldgalaxy}} sample are shown as solid gray lines, while that of the \textsl{\textsc{groupgalaxy}} sample is shown as a solid black line in the right-hand panels. The errorbars indicate the interquartile ranges in bins containing equal numbers of galaxies (10\% of the respective parent samples). }
\label{fig_append_atten_1}  
\end{figure}

If the size of a galaxy at given mass varies with environment, this will affect the attenuation corrections applied in a systematic manner. Fig.~\ref{fig_append_atten_1} shows the distribution of galaxy size $r_e$ as a function of $M_*$ for the \textsl{\textsc{groupgalaxy}} and \textsl{\textsc{fieldgalaxy}} samples, as well as the distributions of the derived parameter $\tau^f_B$ and the attenuation corrections as a function of $M_*$. The corrections distributions are very similar for both samples, although group galaxies appear to be slightly smaller at a given stellar mass than field galaxies ($\lesssim0.03\,$dex). However, the resulting shift in attenuation correction is negligible, as shown in the right bottom panel of Fig.~\ref{fig_append_atten_1}. Thus, under the assumption that the $\tau^f_B - \mu_*$ relation is independent of environment, the method of \citet{GROOTES2013} should supply accurate attenuation corrections.\newline 
However, environment driven shifts in the spatial distribution of gas and dust with respect to the stellar component, as observed, e.g. in galaxies in the Virgo cluster \citep{PAPPALARDO2012,CORTESE2012b}, as well as systematic differences in the dust content of galaxies of a given mass as a function of environment, can be envisaged, and represent a major source of uncertainty in the attenuation corrections applied and by extension in our analysis.\newline

As discussed in \citet{GROOTES2013}, the underlying driver of the $\tau^f_B - \mu_*$ appears to be the near linear relation between $M_*$ and $M_{\mathrm{dust}}$. If the dust content of galaxies is systematically different in the group and field environment, this will affect the attenuation corrections applied. However, with the stripping of material from galaxies by various mechanisms known to affect galaxies in groups, as seen in the Virgo cluster \citep[e.g.][]{CHUNG2009,PAPPALARDO2012}, it appears likely that any systematic difference will tend towards the ratio of gas to stars being smaller in groups. This would lead to over-corrections of the observed emission, making any observed suppression of star formation activity a lower limit on the actual suppression. However, it is also likely that this effect may be balanced by an increase in metallicity of the ISM of galaxies in the cluster environment, leading to higher dust-to-gas ratios. This might account for the empirical result that the dust content of spiral galaxies in the Virgo cluster shows a lack of strong variation as a function of cluster-centric radius \citep{TUFFS2002,POPESCU2002}. In addition, the observed radial gradients in the dust-to-gas ratio indicate that gas is much more efficiently removed than dust \citep{CORTESE2012a,PAPPALARDO2012}, especially within the optical stellar disk. As the dust in the outer regions of the disk has a smaller effect on the observed $NUV$ flux than that in the inner regions, this will mitigate the effect of stripping on the attenuation corrections.\newline

Unfortunately, a detailed analysis of the $\tau^f_B - \mu_*$ relation for galaxies in different group environments has not yet been performed, due largely to the lack of FIR data for these objects. Therefore, in this analysis, we \textit{adopt} the assumption that the $\tau^f_B - \mu_*$ relation is (largely) independent of environment. The systematic uncertainties due to environmental effects in the attenuation corrections are probably the largest systematic uncertainty in the study. In this context, we note that to explain a shift of $\sim0.1\,$dex in $NUV$ flux by a shift via the calibration of the $\tau^f_B - \mu_*$ relation alone, would typically require a systematic shift in $\tau^f_B$ by $\sim25\,$\%.

\section{Appendix C: Modelling the Star Formation History of Group Satellite Spiral Galaxies}\label{APPEND_MODEL}
In order to gain a quantitative understanding of the potential requirement of on-going gas-fuelling in satellite spiral galaxies we have constructed a number of models with parameterized star formation histories (see Section~\ref{SFR_SAT}) from which the distributions of $\Delta \mathrm{log}(\psi_*)$ can be predicted, and which we can contrast with those observed for the \textsl{\textsc{groupgalaxy}} sample in the mass ranges $10^{9.5} M_{\odot} \le M_* < 10^{10} M_{\odot}$ and $10^{10} M_{\odot} \le M_*$. Here we describe how the model populations are constructed.\newline

The construction of a model realization of the group galaxy population for a given star formation history as satellite galaxies in the group environment requires the knowledge of the time since a galaxy first became a satellite $t_{\mathrm{infall}}$, the SFR of the galaxy at the time of infall $\Phi_{*,\mathrm{in}}$, and the stellar mass at time of infall $M_{*,\mathrm{in}}$. Provided this information we can simply evolve the galaxy forward in time according to the chosen parameterized SFH.\newline

Our goal in this modelling is to make as few a priori assumptions about the underlying physical drivers of the SFH as possible, and rather to identify the best fitting parameterized SFH and then in a second step interpret their physical implications. Therefore, we have chosen to rely on structure-growth/galaxy evolution simulations only to determine the distribution of infall times (which is predominantly linked to the underlying DM halo merger history) , and to obtain the values of $\Phi_{*,\mathrm{in}}$, and $M_{*,\mathrm{in}}$, using empirical relations.\newline

\begin{figure*}
\plotone{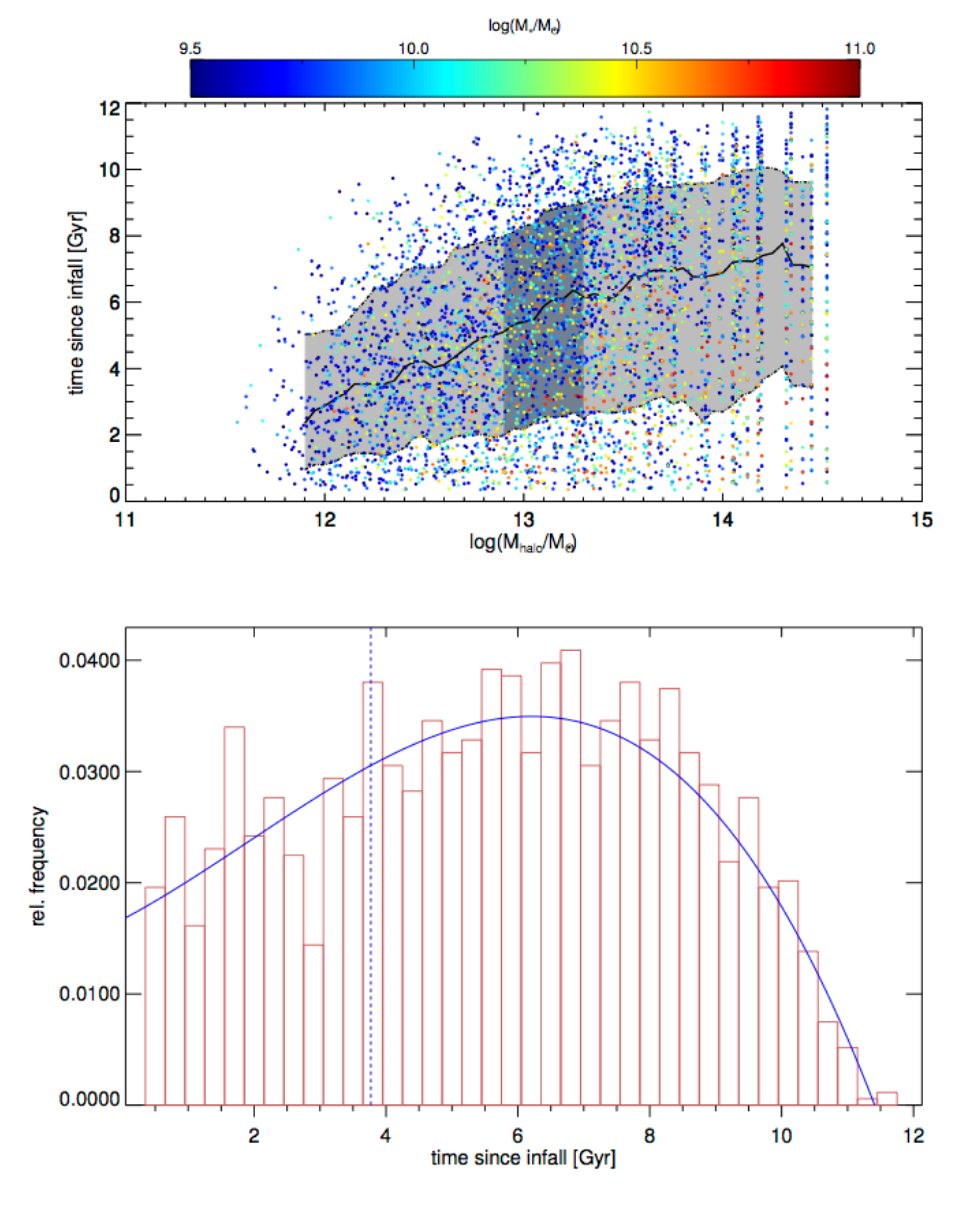}
\caption{\textbf{Top:} Time since infall as a function of host dark matter halo mass for all satellite galaxies with $M_* \ge 10^{9.5}$ from the mock GAMA survey light-cones with $z\le 0.13$. The stellar mass of the galaxy is encoded by the color of the symbol as indicated in the figure. The strong quantization at high halo masses is a result of the scarcity of such objects in the limited volume of the mock survey, however, this does not impinge on the mass range relevant to this analysis. The light gray shaded are indicates the range between the 16$^{the}$ and 84$^{the}$ percentiles (both shown as black dash-dotted lines), in a moving bin of $0.2\,$dex width in $M_{\mathrm{halo}}$ with a step size of $0.05\,$dex. The median in these bins is shown as a solid black line. Superimposed on this is the dark gray shaded area showing the range of dark matter halo mass from which the distribution of infall times has been constructed. \textbf{Bottom:} Distribution of time since infall for satellite galaxies residing in host dark matter halos with $12.9 \le \mathrm{log}(M_{\mathrm{halo}}/M_{\odot}) \le 13.3$ in bins of $0.3\,$Gyr width. The blue solid lines shows the result of a third order polynomial fit ($a_0=0.01668$,$a_1=0.00314$,$a_2=0.00041$,$a_3=-7.1\times10^{-5}$) to the distribution (assuming Poisson errors on each bin). The vertical blue line at $3.77\,$Gyr corresponds to the $30$\% shortest times since infall, the extremely conservative limit used in our modelling.}
\label{fig_append_tindist}  
\end{figure*}

In order to obtain a distribution of infall times for satellite galaxies we make use of the mock GAMA light-cones produced using the Millennium dark matter structure formation simulation \citep{SPRINGEL2005} and the GALFORM semi-analytic galaxy evolution model \citep{BOWER2006,MERSON2013}. Fig.~\ref{fig_append_tindist} shows the distribution of time since becoming a satellite galaxy for all satellite galaxies with $M_* \ge 10^{9.5} M_{\odot}$ in the mock GAMA survey out to a redshift of $z=0.13$, as a function of the mass of the halo in which they reside. We find the distribution of time since infall to be very broad, with the median time since becoming a satellite increasing towards more massive host dark matter halos. Considering the satellite spiral galaxies of the \textsl{\textsc{groupgalaxy}} sample, we find that $68$\% of these reside in dark matter halos with masses between $10^{12.9} M_{\odot}$ and $10^{14.03} M_{\odot}$, with a median DMH of $10^{13.5} M_{\odot}$ . Therefore, in order to obtain a conservative estimate of the infall time distribution for our modelling purposes, we use mock satellite galaxies residing in halos in the mass range $10^{12.9} M_{\odot} \le M_{\mathrm{halo}} \le 10^{13.3} M_{\odot}$, and fit the resulting distribution of infall times with a third order polynomial, as shown in Fig.~\ref{fig_append_tindist}.\newline
 
As is immediately apparent from Fig.~\ref{fig_SATCENTSPF} showing the spiral fractions of the \textsl{\textsc{fieldgalaxy}} sample and of the satellite galaxies, not all galaxies which fall into a group as a spiral retain this morphology. Instead, the spiral fraction of group satellite galaxies decreases by $\sim30 - 40$\% with respect to the group environment over the stellar mass range $M_* \ge 10^{9.5} M_{\odot}$. Performing a weighted average over the spiral fraction of satellite galaxies in stellar mass bins of $0.2\,$dex covering the range $M_* \ge 10^{9.4} M_{\odot}$, we find an average spiral fraction of $30$\%. Here, we make the conservative assumption that \textit{only} the youngest 30\% of all satellite galaxies have a spiral morphology. Accordingly, in assigning infall times, we only consider the age range corresponding to the youngest 30\% of the distribution, resulting in a maximum time since infall of $3.77\,$Gyr.\newline

Having obtained a distribution of infall times we proceed in constructing our model group satellite spiral populations as follows:
\begin{enumerate}
\item{Create a Monte-Carlo realization of the observed \textsl{\textsc{fieldgalaxy}} sample in terms of stellar mass by sampling the observed distribution of $M_*$.}
\item{Assign each galaxy an infall time by sampling from the infall time distribution.}   

\item{Assign each galaxy a SFR by sampling from the observed SFR distribution of the \textsl{\textsc{fieldgalaxy}} sample in bins of $0.2\,$ dex in $M_*$, using the bin which contains the assigned stellar mass of the galaxy}.

\item{Assign each galaxy an  NUV background value and an NUV effective exposure time by sampling the observed distributions of both quantities.}

\item{Evolve the galaxy backwards to $t_{\mathrm{infall}}$ using the empirical relation describing the evolution of the SFMS presented by \citet{SPEAGLE2014}. Here, we use a timestep of $10^7$ yr and assume that a fraction of $\alpha=0.3$ of the total ISM mass converted to stars is immediately returned \citep[e.g.][]{CALURA2014}. We calculate the stellar mass at $t=t_{i-1}$ as 
\begin{equation}
M_{*}(t_{i-1}) = M_*(t_i) - (1-\alpha)\Phi_*(t_i)*\Delta t
\end{equation}
and the SFR as 
\begin{equation}
\Phi_*(t_{i-1}) = \Phi_*(t_{i})\frac{S(t_{i-1},M_*(t_{i-1}))}{S(t_{i},M_*(t_{i}))}
\end{equation}
where $S(t,M_*)$ is the empirical relation describing the evolution of the SFMS provided in Eq.~28 of \citet{SPEAGLE2014}\footnote{\citeauthor{SPEAGLE2014} use the age of the universe in their equation. For our purposes, we have assumed an age of the universe at time of observation $t_{\mathrm{obs}} = 12.161\,$Gyr corresponding to a redshift of $z=0.1$ - the median redshift of the \textsl{\textsc{groupgalaxy}} sample for $M_* \ge 10^{9.5} M_{\odot}$ - in our adopted cosmology. The input time to $S(t,M_*)$is then correctly expressed as $t_i = t_{\mathrm{obs}} - i*\Delta t$.}.}

\item{Evolve the galaxy forward in time to the assumed observation redshift of $z=0.1$ according to the desired parameterized SFH, again assuming a return fraction of $\alpha=0.3$. For each galaxy we convert its SFR into a UV flux by inverting Eq.~\ref{eq_SFR} and assuming it is a pointsource at $z=0.1$. We then compare this with our S/N requirement of $\mathrm{S}/\mathrm{N} \ge 2.5$ using the assigned background and exposure time values. Where the predicted flux would lie below the $\mathrm{S}/\mathrm{N} \ge 2.5$ limit, we have instead adopted an NUV flux at this level and have converted this back to a SFR, replacing the original estimate. This treatment is entirely analogous to our treatment of upper limits in the observed data.}

\end{enumerate} 

For each of our one and two parameter models we create $50$ realizations of $\approx 1000$ galaxies (corresponding to the size of the \textsl{\textsc{groupgalaxy}}) sample at $31$, respectively $31\times31$ sampling positions in the 1d and 2d parameter space. These model populations are then treated completely analogously to the observed satellite spiral galaxy sample in terms of the construction of the distributions of $\Delta\mathrm{log}(\psi_*)$.We subsequently average over the models to obtain an estimate of the distribution as well as of the variability.\newline

We note that our forward modelling assumes that the evolution of the gas and stellar content of satellite spiral galaxies as well as their SFR is not significantly affected by galaxy interactions and/or merger events. Clearly, this assumption will not hold for non-spiral satellite galaxies, but is reasonable for morphologically classified spiral galaxies. Indeed, \citet{ROBOTHAM2014} show that, on average, the growth of stellar mass in galaxies is dominated by continuous accretion rather than (minor) mergers for galaxies with $M_*\le 10^{10.7} M_{\odot}$. This will only be exacerbated for our sample of spiral galaxies \textit{not} in close pairs, as the average merger rate for these objects will be lower than that determined for the close pairs of galaxies on which the analysis of \citet{ROBOTHAM2014} is based.
In essence, we thus assume that the mass accretion history of the galaxies in our sample is dominated by smooth continuous accretion of gas and that any galaxy - galaxy interaction which significantly affects the stellar mass, gas mass, and star formation of a galaxy alters its morphology sufficiently to remove it from our sample. 
The second assumption, is that the we have assumed the SFR and $M_*$ distributions of field galaxies currently on the SFMS and of present day satellites were the same at the epoch when the present day satellites first became satellites. Given the only very weak dependence of infall time on stellar mass, this seems to be sufficiently fulfilled for the purposes of our simple models.\newline

\section{Appendix D: Constraining the Gas-Cycle of Satellite Spiral Galaxies}\label{APPEND_GASCYCLE} 
In a general form the ISM content of a galaxy and its time-dependent evolution can be expressed as a balance between an inflow of gas into the ISM with a rate $\dot{M}_{\mathrm{in}}$, an outflow of gas from the ISM with a rate $\dot{M}_{\mathrm{out}}$ and the consumption of ISM by star formation as
\begin{equation}
\dot{M}_{\mathrm{ISM}} = \dot{M}_{\mathrm{in}} - \dot{M}_{\mathrm{out}} - (1 - \alpha)\Phi_*\,,
\label{eq_Append_ISMevol1}
\end{equation}
where the factor $(1-\alpha)$ accounts for the (instantaneous) recycling of gas from high mass stars back into the ISM.\newline

Departing from Eq.~\ref{eq_Append_ISMevol1} it is reasonable to assume that the outflow rate $\dot{M}_{\mathrm{out}}$ is proportional to the ISM mass of the galaxy, i.e. can be re-expressed as
\begin{equation}
\dot{M}_{\mathrm{out}} = \frac{1}{\tau_{\mathrm{res}}}M_{\mathrm{ISM}}\,,
\label{eq_taures}
\end{equation}
where the constant of proportionality is cast in terms of a typical residence time $\tau_{\mathrm{res}}$ of a unit mass of gas in the ISM.\footnote{We note that for a volumetric star formation law, as we will motivate in the following, this formulation is equivalent to the widely used mass-loading parameterization $\dot{M}_{\mathrm{out}} = \lambda \Phi_*$.} In the following we assume $\tau_{\mathrm{res}}$ to be determined by galaxy-specific processes, i.e. to be constant for galaxies of a given stellar mass. $\tau_{\mathrm{res}}$ may, however, be expected to vary as a function of stellar mass, e.g. as result of variations in the ratio of the feedback energy per unit mass to the depth of the potential well. Making use of $\tau_{\mathrm{res}}$ as defined above, we can reformulate Eq.~\ref{eq_Append_ISMevol1} as  
\begin{equation}
\dot{M}_{\mathrm{ISM}}  = \dot{M}_{\mathrm{in}} - \frac{M_{\mathrm{ISM}}}{\tau_{\mathrm{res}}} - (1-\alpha) \Phi_*\,.
\label{eq_Append_ISMevol2}
\end{equation}

Empirically, it is well known that star-formation and galaxy gas-content are connected, with star-forming galaxies, and in particular star forming spiral galaxies, being found to follow the Schmidt-Kennicutt relation \citep{SCHMIDT1959,KENNICUTT1998a}, i.e. $\dot{\Sigma}_* \propto \Sigma^{n}$ with $n\approx 1.5$. Although this empirical relation connects the surface densities of gas and star formation, recent theoretical work \citep[e.g.][and references therein]{KRUMHOLZ2012} suggests that this results from an underlying volumetric star formation law. Specifically, \citet{KRUMHOLZ2012} argue that the underlying physical relation has the form $\dot{\rho}_* \propto \rho/\tau_{\mathrm{col}}$, where $\tau_{\mathrm{col}} = \sqrt{3\pi/32G\rho}$ is the timescale for the \textit{star forming cloud} to collapse under its self-gravity. In terms of extragalactic observations, where only surface densities are available, this becomes $\dot{\Sigma}_* \propto \Sigma /\tau_{\mathrm{col}}$. In fact, \citet{KRUMHOLZ2012} show that in this formulation objects from giant molecular clouds to high-redshift star-burst galaxies all fall on the same relation. In the redshift range considered in this analysis the typical locus of star formation may be assumed to be molecular clouds, for which, to first order, the density, and hence $\tau_{\mathrm{col}}$ can be assumed to be constant \citep{KRUMHOLZ2012}. Thus, in a spatially integrated form, we can assume 
\begin{equation}
\Phi_* = \tilde{\kappa}M_{\mathrm{ISM}}\,,
\label{eq_PHIkappaISM}
\end{equation}
and indirectly obtain information on a galaxy's gas content by measuring its SFR.\newline

A caveat to this, however, is the fact that \citet{KRUMHOLZ2012} considered only H$_2$, while our analysis considers the total HI + H$_2$ in the ISM. The ratio of molecular to total neutral hydrogen will vary with, e.g. galaxy stellar mass, and even if that were not the case, the numerical value needed to link $M_{\mathrm{ISM}}$ to $\Phi_*$ would differ from that provided by \citet{KRUMHOLZ2012}. For the purpose of our analysis, we have therefore chosen to recalibrate $\tilde{\kappa}$ for the two mass ranges considered, using the fiducial stellar mass for the relevant range in each case ($10^{9.75} M_{\odot}$, resp. $10^{10.3}M_{\odot}$). To determine the total HI + H$_2$ gas mass for both values of $M_*$ we make use of the model of \citet{POPPING2014} which shows good agreement with the measurements of \citet{LEROY2008},\citet{SAINTONGE2011},\citet{CATINELLA2013}, and \citet{BOSELLI2014}. We then use the average SFR derived from our \textsl{\textsc{fieldgalaxy}} sample at the appropriate masses to empirically determine the constant of proportionality, finding $\tilde{\kappa} = 0.46 \mathrm{Gyr}^{-1}$ and  $\tilde{\kappa} = 0.47 \mathrm{Gyr}^{-1}$ for the low and high stellar mass ranges, respectively.\newline

Inserting Eq.~\ref{eq_PHIkappaISM} into Eq.~\ref{eq_Append_ISMevol2} we obtain
\begin{eqnarray}
\dot{M}_{\mathrm{ISM}}  & = &\dot{M}_{\mathrm{in}} - \frac{M_{\mathrm{ISM}}}{\tau_{\mathrm{res}}} - \kappa M_{\mathrm{ISM}}\label{eq_Append_MdotMISM}\\
& = & \dot{M}_{\mathrm{in}} - \lambda \Phi_* - (1 - \alpha)\Phi_* \label{eq_Append_MdotPhi}
\end{eqnarray}
where we have defined $ \kappa = (1 - \alpha) \tilde{ \kappa }$ in Eq.~\ref{eq_Append_MdotMISM} and $\lambda = 1/( \tau_{\mathrm{res}}  \tilde{ \kappa})$ in Eq.~\ref{eq_Append_MdotPhi};  A volumetric star formation law enables the general time-dependent evolution of the ISM content of a galaxy to be equivalently formulated in terms of ISM mass and star formation rate.\newline

Finally, considering the last term in Eq.~\ref{eq_Append_MdotMISM}, we can express the constant of proportionality $\kappa$ in terms of a timescale $\tau_{\mathrm{exhaust}}$, where
\begin{equation}
\tau_{\mathrm{exhaust}} = \frac{1}{\kappa} =  \frac{1}{(1-\alpha)\tilde{\kappa}}
\label{eq_tauexhaust}
\end{equation}
corresponds to the exhaustion timescale of the ISM in a closed box model, i.e due to star formation alone. Inserting Eq.~\ref{eq_tauexhaust} in Eq.~\ref{eq_Append_MdotMISM}, and introducing the effective timescale
\begin{equation}
\tilde{\tau} = \frac{\tau_{\mathrm{res}}\tau_{\mathrm{exhaust}}}{\tau_{\mathrm{res}} + \tau_{\mathrm{exhuast}}}\,
\label{tautilde}
\end{equation}
we obtain 
\begin{equation}
\dot{M}_{\mathrm{ISM}}  = \dot{M}_{\mathrm{in}} - \frac{M_{\mathrm{ISM}}}{\tilde{\tau}} \,,
\label{eq_Append_ISMevol4}
\end{equation}
i.e. Eq.~\ref{eq_ISMevol4} of section~\ref{GASFUELLING_ISMEVOL}.\newline

As stated in Section~\ref{GASFUELLING_ISMEVOL}, for spiral galaxies in the field, the SFR is found to evolve only very slowly with redshift, and is thought to be determined by a self-regulated balance between inflow, outflow, and consumption of the ISM via star formation which only evolves very gradually \citep{KERES2005,DAVE2011,LILLY2013,SAINTONGE2013} such that at any given time their SFR can be considered quasi-constant. Via the volumetric star formation law this implies a constraint on $\dot{M}_{\mathrm{ISM}}$ which can be used to enable estimates of $\dot{M}_{\mathrm{in}}$ from Eqs.~\ref{eq_Append_MdotMISM} \& \ref{eq_Append_MdotPhi}.

As suggested by \citet{LILLY2013} a good description of the quasi steady state is given by the requirement 
\begin{equation}
\mu = \frac{M_{\mathrm{ISM}}}{M_*} = \mathrm{const.}\,,
\label{eq_Append_mu}
\end{equation}
i.e. that the ISM mass per unit stellar mass, and accordingly the sSFR be constant. As these authors show, this formulation allows for the dependence of gas metallicity on SFR in line with observations.  However, as the stellar mass increases, the galaxy will also change its position along the $ \psi_* $--$ M_* $ relation, shifting towards a higher value of $M_*$ and an accordingly lower expected value of the sSFR. Considering a period of time over which the accumulated stellar mass is negligible, an alternative implementation of the quasi steady-state is that the ISM mass remains constant, i.e. $\dot{M}_{\mathrm{ISM}} = 0$. Clearly, these scenarios bracket behaviour of $M_{\mathrm{ISM}}$ which can be expected in the quasi steady-state. In the following we will derive an estimate of the inflow rate using the requirement given by Eq.~\ref{eq_Append_mu} following \citet{LILLY2013}, and subsequently compare it to the result obtained using $\dot{M}_{\mathrm{ISM}} = 0$.\newline

We begin by considering the total time derivative of $\mu$ given by
\begin{eqnarray}
\frac{d}{dt}\mu & = & \frac{1}{M_*}\frac{\partial M_{\mathrm{ISM}}}{\partial t} + \frac{M_{\mathrm{ISM}}}{M^{2}_{*}}\frac{\partial M_{*}}{\partial t} \\
& = & \frac{1}{M_*}\frac{\partial M_{\mathrm{ISM}}}{\partial t} + \frac{1}{M_*} \mu \left( 1-\alpha \right) \phi \label{dmudt}\,,
\end{eqnarray} 
from which we obtain
\begin{equation}
\dot{M}_{\mathrm{ISM}} = \frac{\partial M_{\mathrm{ISM}}}{\partial t} = \mu  \left( 1-\alpha \right) \phi + M_*\frac{d}{dt}\mu\,. 
\label{dmismdt}
\end{equation}
Inserting Eq.~\ref{dmismdt} into Eq.~\ref{eq_Append_MdotPhi} and isolating $M_{\mathrm{in}}$ we obtain
\begin{equation}
M_{\mathrm{in}} = \left[ \left(1-\alpha \right) \left(1 + \mu \right) + \lambda \right] \phi + M_* \frac{d}{dt}\mu \,. 
\label{Min}
\end{equation}  
As we assume $\mu$ to be quasi constant in the (quasi) steady state, the term $\frac{d}{dt}\mu$ will be negligible compared to the other terms in Eq.~\ref{Min}. Thus, for the quasi steady state, the inflow of gas to the galaxy will be given by
\begin{equation}
M_{\mathrm{in}} \approx \left[ \lambda + (1 - \alpha) + \mu (1 - \alpha) \right] \phi  \,.
\label{Minss}
\end{equation}\newline

Comparing the estimate for the inflow given in Eq.~\ref{Minss} with that which can immediately be obtained from Eq.~\ref{eq_Append_MdotPhi} for $\dot{M}_{\mathrm{ISM}} = 0$, i.e.
\begin{equation}
\dot{M}_{\mathrm{in}} = \left[ \lambda + (1- \alpha) \right] \Phi_* = \frac{\Phi_*}{\tilde{\kappa} \tilde{\tau}},
\label{Minssalt}
\end{equation}
(where we have made use of Eqs.~\ref{eq_Append_ISMevol4} \& \ref{eq_volSFlaw} for the last equality). It is clear that the inflow given by Eq.~\ref{Minss} is larger by a factor of $\mu (1- \alpha)$, with the end-point of this fraction of the flow being the (growing) ISM of the galaxy. As the actual inflow will lie somewhere between the bracketing cases considered here, we have chosen to adopt Eq.~\ref{Minssalt}, as a conservative estimate of the inflow to the galaxy in the quasi steady-state in the context of this work.\newline

Although we have derived an estimate if the inflow of gas into the ISM making use of the quasi steady state, this is still predicated on our knowledge of the outflow, i.e of the parameters $\lambda$, respectively $\tau_{\mathrm{res}}$. To this end, we consider the effective timescale
\begin{equation}
\tilde{\tau} = \frac{\tau_{\mathrm{res}}\tau_{\mathrm{exhaust}}}{\tau_{\mathrm{res}} + \tau_{\mathrm{exhaust}}} = \frac{1}{\tilde{\kappa}\left( 1 - \alpha + \lambda \right)} \,,
\end{equation} 
from which we obtain
\begin{equation} 
\lambda = \frac{1}{\tilde{\kappa} \tilde{\tau}} + \alpha - 1  \,.
\label{lambdadq}
\end{equation}
As discussed in Section~\ref{GASFUELLING_ISMEVOL}, independent of any assumptions with regard to the quasi steady state, $\tilde{\tau}$ can determined from the quenching, respectively, refuelling phase(s) of the fitted parameterized star formation histories. Accordingly, the outflow estimate given by Eq.~\ref{lambdadq} holds regardless of steady state description adopted, and if we adopt Eq.~\ref{Minssalt} as an inflow estimate and insert it into Eq.~\ref{lambdadq} we obtain the expression given by Eq.~\ref{eq_MoutSS} in Section~\ref{GASFUELLING_ISMEVOL}.\newline 

Finally, we reiterate that the derivation presented above with the corresponding approximations will hold as long as the rate at which the inflow changes is small compared to the timescale $\tilde{\tau}$. For the majority of our considered cases we find the adopted approximations to be retroactively justified as we find $\tilde{\tau} \lesssim 1\,$Gyr following Eq.~\ref{tautilde}. Where this is not the case we can no longer derive meaningful constraints using the equations derived above.\newline

\section{Appendix E: Sampling the Full Distribution of $t_{\mathrm{infall}}$ - Results for Two Parameter Models}\label{APPEND_FD}     
\begin{figure*}
\plotone{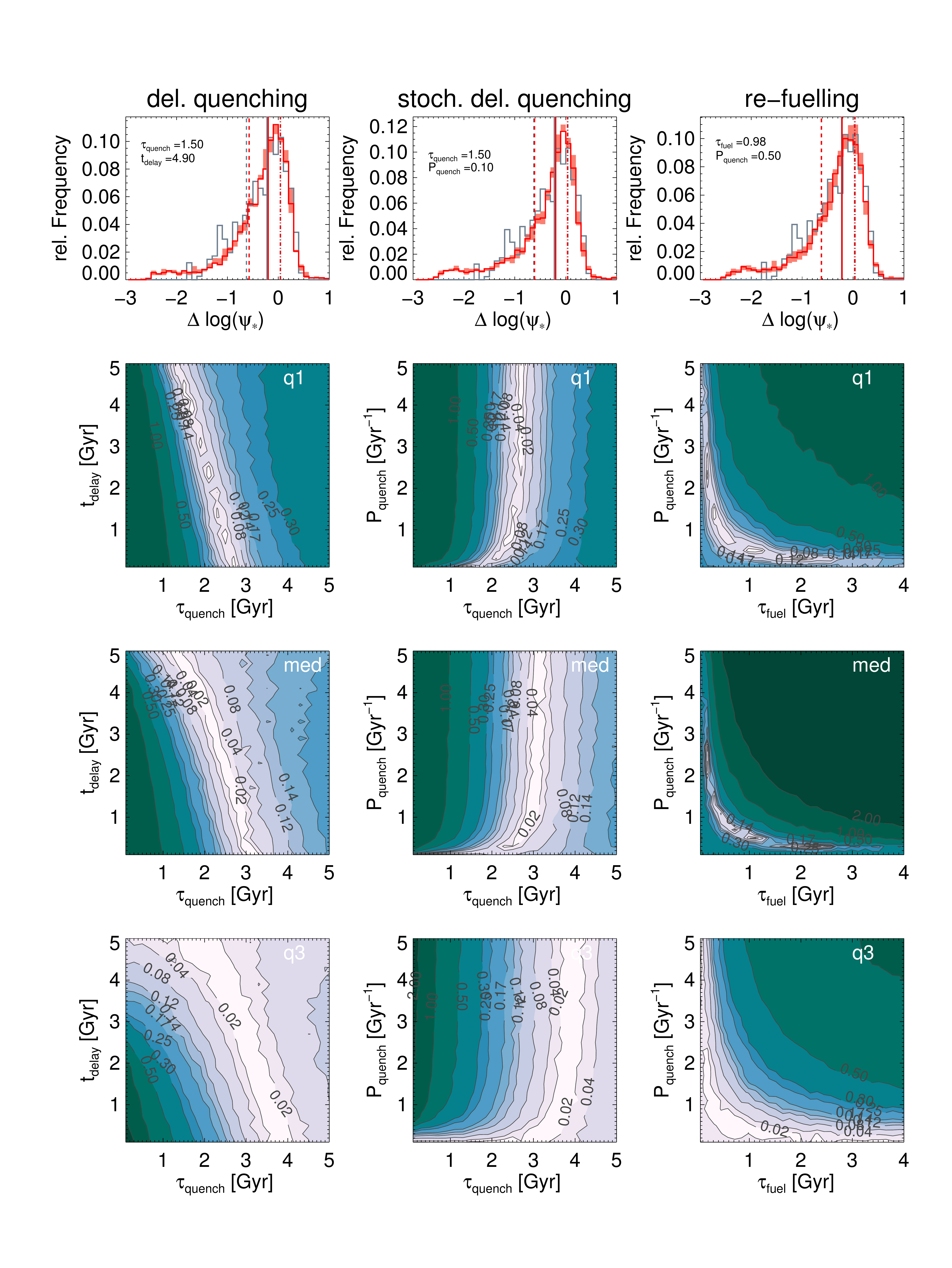}
\caption{As Fig.~\ref{fig_2pmod_hm}, i.e. for the mass range $M_* \ge 10^{10} M_{\odot}$, but for models sampling the full distribution of infall times as shown in Fig.~\ref{fig_append_tindist}.}
\label{fig_append_2pmod_FD_hm}  
\end{figure*}

\begin{figure*}
\plotone{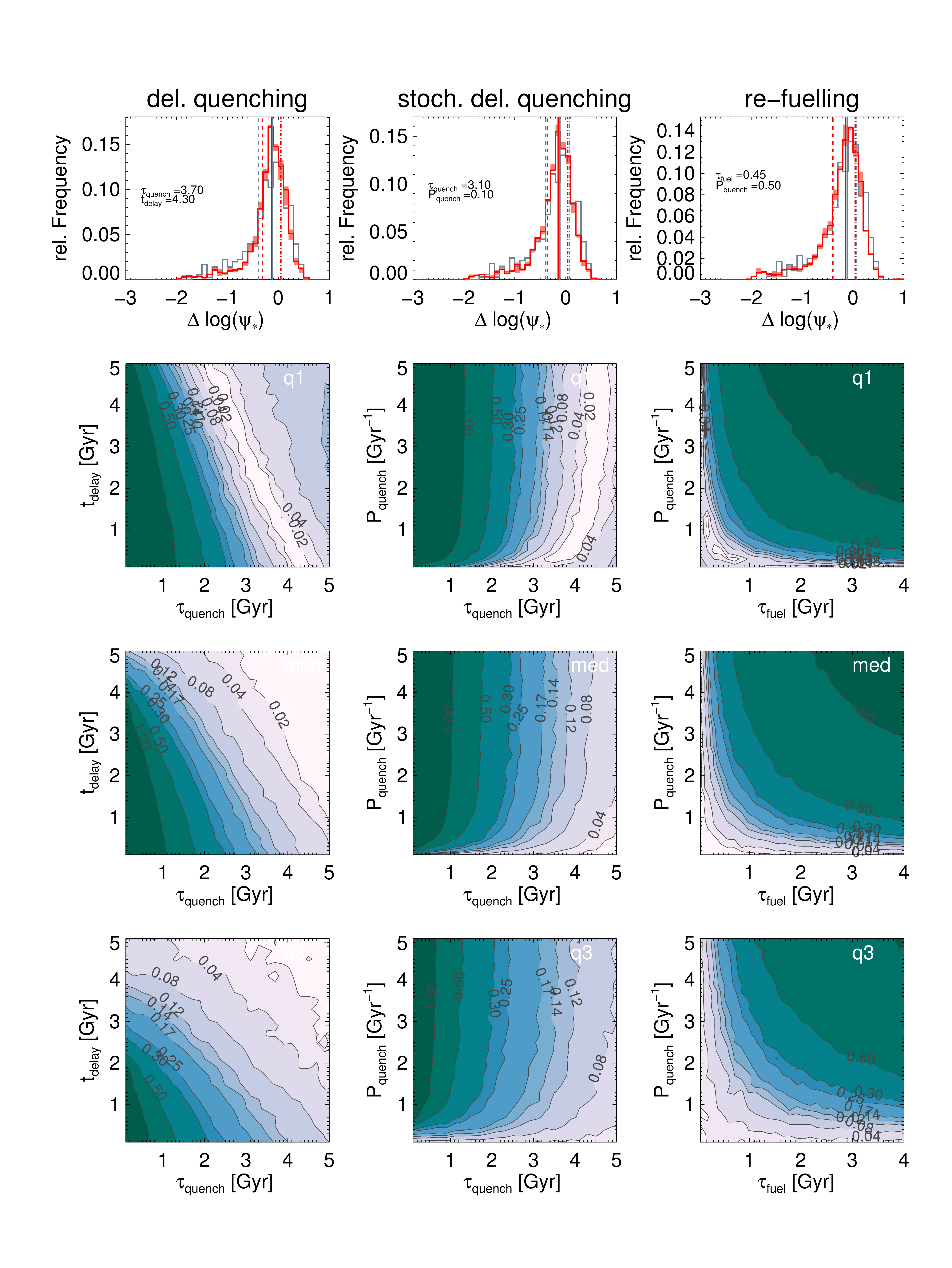}
\caption{As Fig.~\ref{fig_append_2pmod_FD_hm}, but for the mass range $10^{9.5}M_{\odot} \le M_* \ge 10^{10} M_{\odot}$. }
\label{fig_append_2pmod_FD_lm}  
\end{figure*}

\begin{figure*}
\plotone{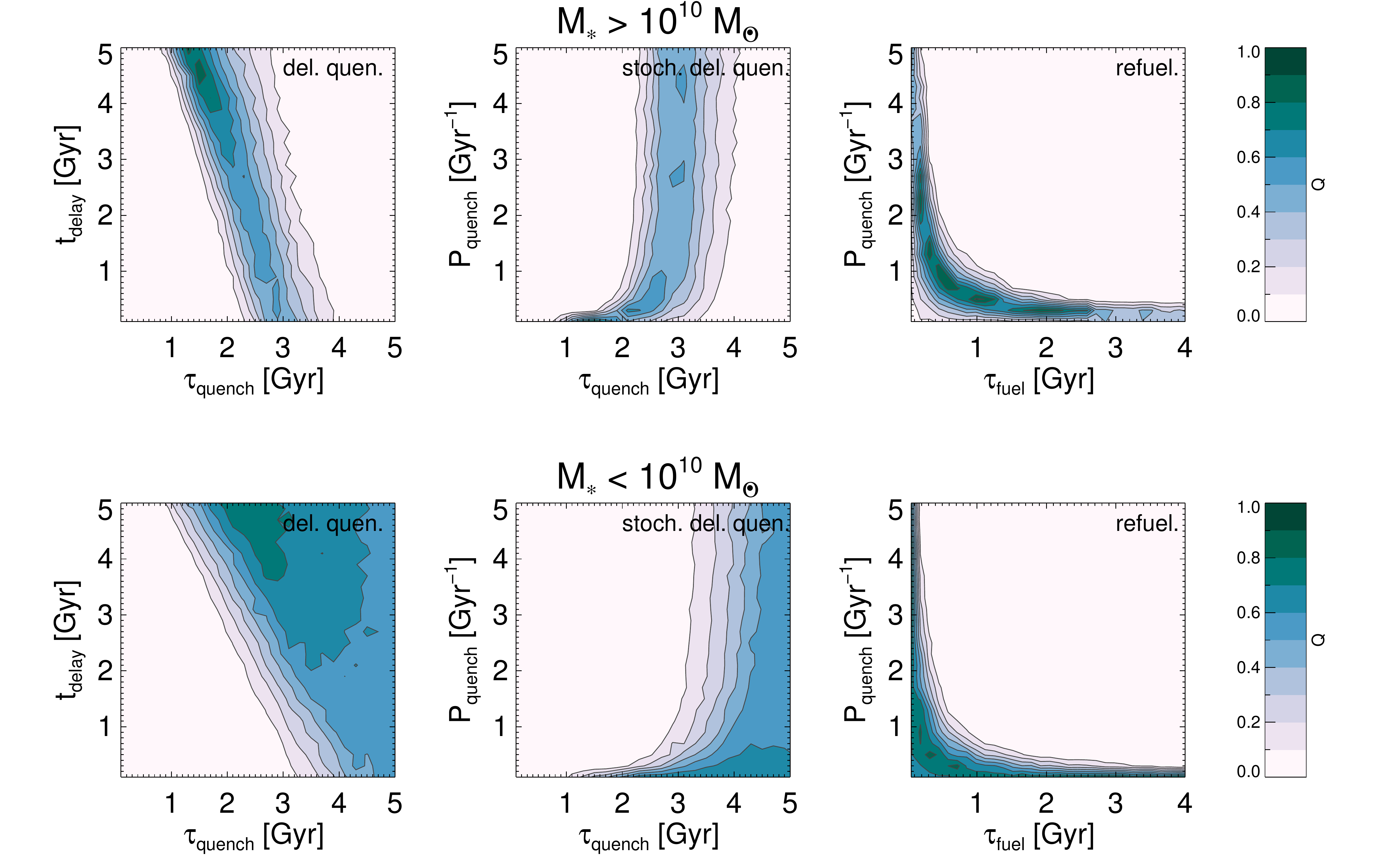}
\caption{As Fig.~\ref{fig_Qm}, but for models sampling the full distribution of infall times as shown in Fig.~\ref{fig_append_tindist}.}
\label{fig_append_Qm}
\end{figure*}

For the main purpose of our analysis we have adopted the very conservative assumption that only the youngest 30\% of satellite galaxies are spirals (i.e. have retained their spiral morphology), and have accordingly limited the maximum time since infall in our modelling to $t_{\mathrm{infall}} \le 3.77\,$Gyr. The opposite extreme assumption, is that the group environment has no impact on the probability of a spiral to transform its morphology, in which case it would be appropriate to sample the full distribution of infall times. While this is almost certainly not the case, the true distribution of infall times will lie between these two extremes, albeit likely more on the side of the conservative estimate. Therefore, in order to gain an understanding of the importance of the infall time distribution to the performance of our models, it is informative to consider their performance using the extreme assumption that the full distribution of $t_{\mathrm{infall}}$ is sampled. Figs.~\ref{fig_append_2pmod_FD_lm}, \ref{fig_append_2pmod_FD_hm}, and \ref{fig_append_Qm} which are analogous to Figs.~\ref{fig_2pmod_lm} \& \ref{fig_2pmod_hm}, show the results of the models using this extreme assumption.\newline 

In the high stellar mass range, we find that all three models are formally capable of reproducing the observed distributions even under the extreme infall time distribution. For the delayed quenching model, the preferred parameter values are $\tau_{\mathrm{quench}} =1.5\,$Gyr and $t_{\mathrm{delay}} = 4.9\,$Gyr, while the preferred values for the stochastic delayed quenching model are $\tau_{\mathrm{quench}} =1.5\,$Gyr and $P_{\mathrm{quench}}=0.1\,$Gyr$^{-1}$. For both models, the quenching timescales are longer than when assuming the conservative infall time distribution, notably, however, the delay time is much longer, respectively the quenching probability is lower, expanding beyond realistic estimates of the depletion timescale as discussed in Section~\ref{DEPENDINT}.
In contrast, the refuelling model prefers parameter values $\tau_{\mathrm{fuel}} = 0.98\,$Gyr and$P_{\mathrm{quench}} = 0.5\,$Gyr$^{-1}$,  comparable to those previously found, albeit that the refuelling is slower and the occurrence of quenching is lower. However, as a comparison of Figs~\ref{fig_Qm} \& \ref{fig_append_Qm} reveals, the degenerate parameter space is largely the same under both extreme infall time distributions.\newline

In the low stellar mass range, the refuelling model best reproduces the observed distribution of $\Delta \mathrm{log}(\psi_*)$, still attaining a value of $Q = 0.79$ for the preferred parameter combination ($\tau_{\mathrm{fuel}} = 0.45\,$Gyr, $P_{\mathrm{quench}} = 0.5\,$Gyr$^{-1}$), albeit $0.14$ lower than that found for the conservative distribution of infall times and requiring a high rate of recovery. 
In contrast the stochastic delayed quenching model ($\tau_{\mathrm{quench}} = 3.1\,$Gyr,$P_{\mathrm{quench}}=0.1\,$Gyr$^{-1}$)  and the delayed quenching model ($\tau_{\mathrm{quench}}=3.7\,$Gyr, $t_{\mathrm{delay}}=4.3\,$Gyr) struggle to reproduce the observed distribution under the adopted infall time distribution, both over-predicting the relative number of largely unquenched galaxies, and in the case of the delayed quenching model, markedly under-predicting the number of strongly quenched galaxies. This is also evident from Fig.~\ref{fig_append_Qm} and table~\ref{tab_fitpar}, where there attained values of $Q$ are noticeably lower than for the conservative infall time distribution. Furthermore, the long delay time/low quenching probability are difficult to reconcile with realistic gas exhaustion timescales.\newline

\bibliography{grootes_SPGasfuelling_I_resubmission} 

\end{document}